\documentstyle[12pt]{article}
\newcommand{\beq}{\begin{equation}}   
\newcommand{\eeq}{\end{equation}}
\newcommand{\ra}{\rightarrow}

\newcommand{\gsim}{\lower.7ex\hbox{$
\;\stackrel{\textstyle>}{\sim}\;$}}
\newcommand{\lsim}{\lower.7ex\hbox{$
\;\stackrel{\textstyle<}{\sim}\;$}}

\def\lsim{\mathrel{\rlap{\lower3pt\hbox{\hskip0pt$\sim$}}
    \raise1pt\hbox{$<$}}}         
\def\gsim{\mathrel{\rlap{\lower4pt\hbox{\hskip1pt$\sim$}}
    \raise1pt\hbox{$>$}}}         

\renewcommand{\Im}{{\rm Im}\,}

\newcommand{\aver}[1]{\langle #1\rangle}

\newcommand{\La}{\overline{\Lambda}}
\newcommand{\Lam}{\Lambda_{\rm QCD}}

\newcommand{\al}{\alpha}
\newcommand{\as}{\alpha_s}
\newcommand{\GeV}{\,\mbox{GeV}}
\newcommand{\MeV}{\,\mbox{MeV}}
\newcommand{\matel}[3]{\langle #1|#2|#3\rangle}
\newcommand{\state}[1]{|#1\rangle}

\newcommand{\eq}[1]{Eq.\hspace*{.15em}(\ref{#1})\hspace*{-.3em} 
}

\newcommand{\re}[1]{Ref.~\cite{#1}}
\newcommand{\res}[1]{Refs.~\cite{#1}}

\begin{document}
\begin{titlepage}
\renewcommand{\thefootnote}{\fnsymbol{footnote}}

\begin{center} \Large
{\bf Theoretical Physics Institute}\\
{\bf University of Minnesota}
\end{center}
\begin{flushright}
TPI-MINN-97/02-T\\
UMN-TH-1528-97\\
UND-HEP-97-BIG01 \\ 
hep-ph/9703290\\ 

\end{flushright}
\vspace{.3cm}
\begin{center}

   {\LARGE Aspects Of Heavy Quark Theory}

\vspace*{0.5cm}
\end{center}

\begin{center} {\Large 
I. Bigi} \vspace*{0.1cm}\\
{\it   Physics Department, Univ. of Notre Dame,
Notre Dame, IN 46556}
\vspace*{0.3cm}

and
\vspace*{0.3cm}

{\Large 
M. Shifman and N. Uraltsev$^*$}
\vspace*{0.1cm}\\
{\it  Theoretical Physics Institute, Univ. of Minnesota,
Minneapolis, MN 55455}
\vspace*{1.3cm}

{\Large{\bf Abstract}}\vspace*{.3cm}\\
\end{center}

\noindent
Recent achievements in the heavy quark theory are critically  reviewed. The
emphasis is put on those aspects which 
either did not attract enough attention or cause
heated debates in the current literature. Among other topics  we discuss (i)
basic parameters of the heavy  quark theory; (ii) a class of exact QCD
inequalities; (iii) new heavy quark sum rules; (iv) virial  theorem; 
(v) applications ($|V_{cb}|$ from the
total semileptonic width and from the  $B\ra D^*$ transition at zero
recoil). In some instances new  derivations of the  previously known results
are given, or new aspects  addressed.  In particular, we dwell on the exact
QCD inequalities. Furthermore,  a toy model is considered that may shed light
on the controversy regarding the value of the kinetic energy of  heavy quarks
obtained  by different methods.

\vspace{1.5cm}

\begin{flushleft}
\rule{2.4in}{.25mm} \\
$^*$ Permanent address: Petersburg Nuclear Physics Institute,
Gatchina, St.~Petersburg 188350 Russia
\end{flushleft}

\end{titlepage}

\section{Introduction}
\renewcommand{\theequation}{1.\arabic{equation}}
\setcounter{equation}{0}

Quark-gluon dynamics are governed by the QCD Lagrangian
$$
{\cal L} =-\frac{1}{4} G_{\mu\nu}^a G_{\mu\nu}^a
+\sum_{q}\bar q i\not\!\!{D} q  
 +\sum_{Q}\bar Q (i\not\!\!{D} - m_Q) Q =
$$ 
\begin{equation}
{\cal L}_{\rm light}
+ \sum_{Q}\bar Q (i\not\!\!{D} - m_Q) Q
\label{lagr}
\eeq
where $G_{\mu\nu}^a$ is the gluon field strength tensor,
the light quark fields ($u, d$ and $s$) are generically denoted by $q$ 
and are 
assumed, for simplicity, to be massless, and the heavy quark fields 
are 
generically denoted by $Q$. To qualify as a heavy quark $Q$ the 
corresponding
mass term $m_Q$ must be much larger than $\Lambda_{\rm QCD}$. 
The charmed quark $c$ can be called heavy only with 
some 
reservations and, in discussing the heavy quark theory, it is more 
appropriate 
to keep in mind $b$ quarks. The hadrons to be considered are 
composed from 
one heavy quark $Q$, a light antiquark $\bar q$, or diquark $qq$, 
and 
a gluon 
cloud which also contains light quark-antiquark pairs. The role of
the 
cloud is, of course, to keep all these objects together, in a colorless 
bound state 
which  will be generically denoted by $H_Q$. 

The light component of $H_Q$, its light cloud,\footnote{In some 
papers  devoted to the subject the light cloud is referred to as 
``brown
muck",  which seems to be an unfair name for the soft 
components
of the quark and gluon fields -- perhaps we have not yet unveiled 
their 
beauty.}
has a complicated structure -- the soft modes of the light fields are 
strongly 
coupled and strongly fluctuate. Basically, the only fact which we 
know for sure 
is that the light cloud is indeed light; typical frequencies are of order 
of
$\Lambda_{\rm QCD}$. One can try to visualize the light cloud as a 
soft 
medium. The heavy quark $Q$ is then  submerged in this medium. If 
the hard 
gluon exchanges are discarded, the momentum which the heavy 
quark 
can borrow from the light cloud is of order of $\Lambda_{\rm QCD}$.
Since 
it is much smaller than $m_Q$, the heavy quark-antiquark pairs can  
not
play a role. In other  words, the field-theoretic (second-quantized)
description of the heavy  quark  becomes redundant, and under the
circumstances it is perfectly  sufficient to  treat the heavy
quark quantum-mechanically.  This is  clearly  infinitely 
simpler  than any field theory. Moreover, one can  systematically 
expand in $1/m_Q$. Thus, in the limit $m_Q/\Lambda_{\rm
QCD}\rightarrow\infty$ the heavy quark  component of $H_Q$ 
becomes
easily manageable, allowing one to use the heavy  quark as a probe 
of 
the
light cloud dynamics.   Treating the heavy quark $Q$ in $H_Q$ as a
non-relativistic object submerged in a soft gluon background field of
the light cloud, we open the door for  the use of a large variety of
methods developed in Quantum  Electrodynamics long ago 
\cite{BD,LL}, for
instance, the Pauli  expansion, the Foldy-Wouthuysen technique and 
so on. 

The special advantages of the
limit $m_Q\ra\infty$ in 
QCD were first emphasized by Shuryak \cite{Shuryak}. The next 
logical step was the observation of the heavy quark symmetry
\cite{HQS1,HQS2,HQS3}. The heavy quark theory in QCD was finally 
formalized in 
Refs. \cite{HQET} where the systematic $1/m_Q$ expansion of the 
$H_Q$ dynamics was cast in the form of the effective Largangians. 
Applications of the heavy quark theory are so numerous that they 
comprise now a significant and, perhaps, the most active part of QCD,
with a well-known  record of recent successes and breakthroughs. 
Many  exhaustive  reviews are devoted to the subject
\cite{HQETRev}, and we feel there is no need to repeat what has 
already 
become 
common knowledge in our community. Our task, as we see it, is to 
concentrate on several aspects of the heavy quark theory which are 
still
controversial.  These aspects, as a rule, are related to a specific 
structure of QCD where the gluon field has two faces.
The gluon degrees of freedom play a role of a soft background 
medium, and, simultaneously, they are responsible for the
radiative $\alpha_s$ corrections reflecting the presence of the hard 
component in  the gluon field. These two components -- soft and 
hard -- are tightly intertwined, and in many instances it is hard to 
disentangle them.
Without untangling the hard component one can not  build consistent 
$1/m_Q$  and $\alpha_s$ expansions.  This is in sharp distinction 
with,
say, the Pauli expansion in QED where this problem simply does not 
arise. The   background electromagnetic field can be taken  as soft as 
we want,  the radiative
corrections are fully calculable and small (i.e. they  can be neglected 
in most of the  problems). 

The most remarkable example of the problem mentioned above is 
the definition of the heavy quark mass -- 
what quantity appears in the $1/m_Q$ expansion after all?
In electrodynamics the electron mass is a quantity that can be  
measured. Even in this case, though, one encounters some (mild) 
problems in theoretical description,  due to the masslessness of  
photons. As a matter of fact, the electron never appears ``alone"; it is 
always surrounded by a photonic cloud, manifesting itself in  
infrared
divergences. The  strategy allowing one to treat these divergences 
and
define the mass  of an ``isolated" electron is well-known. 
The situation becomes much more complicated in QCD, where  
isolated quarks simply do  not exist. In defining the heavy quark 
mass one can peel off the gluonic cloud only up to a point. Namely, 
we can eliminate from the definition the contribution coming from 
hard gluons. The contribution 
associated with soft gluons has to be included 
in the definition of the heavy quark mass simply because we do not 
know
the precise structure of the soft-gluon component. How exactly could 
one separate the hard and soft-gluon components in defining the 
heavy quark mass? This question causes 
debate in the 
current literature, and will be one of the focal points of the present
review. 

Another key parameter of the heavy quark theory is the
kinetic operator.  Formally its expectation value can be 
defined as  
\beq
\mu_\pi^2 = \frac{1}{2M_{H_Q}}
\langle H_Q|\bar Q {\vec\pi}^2
Q|H_Q\rangle \; ,\;\;\;\;\vec \pi = - i \vec D \; ,
\label{mupid}
\eeq
where $D$ stands for the covariant derivative 
(note: here and elsewhere, 
we always assume, unless explicitly stated to the contrary, that 
we are in the rest frame of the heavy hadron).

The physical meaning of $\mu_\pi^2$ is quite evident: 
the heavy quark inside $H_Q$ experiences a {\em zitterbewegung}
due to its coupling to other degrees of freedom inside the hadron.
Its average spatial momentum squared is $\mu_\pi^2$. 
The kinetic energy operator appears in the effective 
Lagrangian in the 
next-to-leading order, and basically determines the magnitude of 
$1/m_Q$ corrections in a large number of expressions of practical 
interest --
from the simplest formula for the heavy hadron masses to $B\ra 
D^{(*)}$  formfactor 
at zero recoil. When one tries to proceed from 
formal expansions to practical analysis including perturbative 
${\cal O}(\alpha_s^k)$ effects, the same question 
-- what contribution is to be included in $\mu_\pi^2$ --
immediately resurfaces. One can find in the literature a 
spectrum of answers. 
The numerical value of $\mu_\pi^2$ remains rather uncertain.
Various determinations of this crucial parameter contradict each 
other. We will present a balanced discussion of competing   
definitions trying to  outline a procedure which seems fully 
consistent in  QCD.

A related topic we would like to consider is a QCD analog of the
virial theorem. The issue was raised recently by Neubert  
\cite{MatNeu},
who noted that $\mu_\pi^2$ is related to a certain transition 
amplitude induced by the
chromoelectric field. This observation was exploited for an 
alternative
determination of  $\mu_\pi^2$, which brought a surprise. A  
numerical
value of  $\mu_\pi^2$ emerging in this way was significantly lower 
than the  results of some other determinations. The discrepancy is a 
subject of ongoing discussions. We present a transparent derivation
of the  virial theorem in QCD, and comment on the possible causes of 
the
discrepancy.

This review is organized as follows. First, 
 a brief survey is given of 
the operator product expansion (OPE) as it is used in the heavy quark
theory. This survey allows us to introduce all basic tools  and 
formulate theoretical problems to be 
dealt with in the remainder of the review as they appear in their 
natural environment. We then proceed to 
relatively recent 
applications of the heavy quark theory in the inclusive decays of 
heavy hadrons. The following topics are discussed in some detail:
(i) basic parameters of the heavy quark expansion; (ii) a class of 
exact QCD inequalities; (iii) the virial theorem and (iv) applications 
($|V_{cb}|$
from the total semileptonic width and from the exclusive $B\ra D^*$
transition at zero recoil).
Rather than attempting a comprehensive coverage of a variety of 
results obtained in the last two or three years, we dwell on these 
selected 
topics which 
either have not attracted sufficient attention so far, or are 
subject 
to ongoing debate. A significant part of our presentation is based on 
the results of Ref. \cite{optical} that provides  a consistent 
field-theoretic framework  for treating various aspects of the heavy 
quark theory that are under intense investigation at present.

Even limiting ourselves to this  rather narrow task we had to 
leave aside, with  heavy heart, many relevant issues, e.g.  duality,
inclusive non-leptonic  decays of heavy flavors, etc. We  
apologize to the authors whose results, due to space 
limitations, are left outside the scope of the present review. 
The interested readers are referred to recent talks
\cite{recent}.

\section{Operator Product Expansion; Effective Hamiltonian}

\renewcommand{\theequation}{2.\arabic{equation}}
\setcounter{equation}{0}

The basic theoretical tool of the heavy quark theory is the Wilson  
operator product expansion  \cite{WILSON}. It is used in two 
different 
directions. First, one can restrict oneself to the case of a fixed 
number of heavy quarks $Q$ since $\bar QQ$ fluctuations can be 
ignored. Most often one considers the sector with a single 
heavy quark $Q$ which is treated as being submerged into a 
gluonic medium acting as the sole source of its interactions. 
At this stage we disregard any other possible interaction of the 
heavy quark, e.g. electromagnetic, weak and so on. 
The original QCD Lagrangian (\ref{lagr}) is formulated at very short 
distances, or, which is the same, at a very high normalization point 
$\mu =  
M_0$, where $M_0$ is the mass of an ultraviolet 
regulator; i.e., the 
normalization point is assumed to be much 
higher 
than all mass scales in the theory, in particular, $\mu\gg m_Q$. 
An  
effective theory for  describing the 
low-energy
properties of heavy flavor hadrons is obtained by evolving the
Lagrangian from the high scale $M_0$ down to a 
normalization point
$\mu$ lying below  the heavy quark masses $m_Q$.  
This means that we  integrate out, step by step,
all high-frequency modes in the theory thus calculating the 
Lagrangian ${\cal L}(\mu )$. The latter is a full-fledged  
Lagrangian  with respect to  the soft modes 
with characteristic frequencies less than $\mu$.  The hard 
(high-frequency) modes determine the coefficient  functions
in ${\cal L}(\mu )$, while the contribution of the soft modes
is hidden in the matrix elements of (an infinite set) of 
operators appearing in ${\cal L}(\mu )$. 
The value of this approach,  outlined by Wilson long ago 
\cite{WILSON}, has become  
widely recognized and exploited in countless
applications. The peculiarity 
of the heavy quark theory lies in the fact that the {\em in} and 
{\em out} states contain heavy quarks. Although we  
integrate out the  field fluctuations with the frequencies down to
$\mu \ll m_Q$, the heavy quark fields themselves are not integrated 
out
since we consider the sector with heavy-flavor  
charge
$\neq 0$. The effective Lagrangian ${\cal L}(\mu )$ acts in this
sector. Since the heavy quarks are neither produced nor annihilated, 
any sector with the given $Q$-number can be treated separately
from all others, as a ``vacuum". 

If QCD were solved we could include all modes down to $\mu =0$  
in our explicit evaluation 
of the effective Lagrangian ${\cal L}(\mu )$. 
The resulting Lagrangian would be built in terms of the
fields of physical mesons and baryons rather than quarks and 
gluons --  the latter become irrelevant degrees of freedom in the 
infrared limit $\mu\rightarrow 0$.  All conceivable amplitudes 
could be read off directly from such an effective Lagrangian and 
could be compared with experimental data. 

This picture is of course Utopian: real QCD is not solved
in closed form,
and in explicit calculations of the
coefficients in the effective Lagrangian one cannot put $\mu = 0$.
For decreasing values of $\mu$ a larger and larger 
part of the dynamics has to be 
accounted for in the explicit calculation. 
We would like to have $\mu$ as low as possible, definitely,
$\mu \ll m_Q$. The heavy quark can be treated as a non-relativistic 
object moving in a soft background field {\em only if} the latter 
condition is met. On the other hand,
to keep computational control over the explicit calculations of the
coefficient functions we must stop
at some $\mu\gg \Lambda_{\rm QCD}$, so that 
$\alpha_s (\mu )/\pi$ is still a sufficiently small expansion 
parameter. In practice 
this means
that the best choice (which we will always adopt) is $\mu\sim$ 
several units times $\Lambda_{\rm QCD}$, i.e. $0.7$ to $1 \GeV$.  All 
coefficients in
the effective Lagrangian obtained in this way will be functions of
$\mu$.

Since $\mu$ is an auxiliary parameter, 
predictions for physical quantities cannot depend on $\mu$, of 
course. The $\mu$ dependence of the coefficients must be canceled
by that coming from the physical matrix elements
of the operators in ${\cal L}(\mu )$. 
However, in calculating in the hard and soft domains (i.e.
above $\mu$ and below $\mu$) we make different approximations,
so that the exact $\mu$ independence of the physical quantities does 
not take
place.
Since the transition from hard to soft physics is very steep, one 
may hope that our predictions will be insensitive to the 
precise choice of $\mu$ provided that 
$\mu\sim$ several units times $\Lambda_{\rm QCD}$. 

In descending from $M_0$ to $\mu$, the
form of the 
Lagrangian (\ref{lagr}) changes, and a series of operators of higher 
dimension  appears. 
 For instance, the heavy quark part of the
Lagrangian takes the form
\begin{equation}
{\cal L}_{\rm heavy}=\sum_Q \left\{\bar Q (i\not\!\!D 
-m_Q)Q +
\frac{c_G}{2m_Q}\bar Q \frac{i}{2}\sigma_{\mu\nu}G_{\mu\nu} Q\; +
\sum_{\Gamma ,\;q} 
\frac{d_{Qq}^{(\Gamma )}}{m_Q^2} \bar Q \Gamma Q \bar q \Gamma 
q \right\} + {\cal O}\left(\frac{1}{m_Q^3}\right)
\label{N2}
\end{equation}
where $c_G$ and $d_{Qq}^{(\Gamma )}$ are coefficient functions,
$G_{\mu\nu}\equiv g
G_{\mu\nu}^a t^a $ and $t^a$ is the color generator   (the 
coupling
constant is included into $G$). We 
often use the short-hand notation 
$\sigma G=\sigma_{\mu\nu}G_{\mu\nu}=\gamma_{\mu} 
\gamma_{\nu} G_{\mu\nu}$.
The sum over the light quark flavors $q$ is shown explicitly as well 
as 
the 
sum over
possible structures $\Gamma$ of the four-fermion operators. All 
masses and 
couplings, as well as the  coefficient functions $c_G$ and 
$d^{(\Gamma )}$,
depend on the normalization point. 

The operators of dimension five and higher in Eq. (\ref{N2})
 are due to the  contribution of hard gluons,
with virtual momenta from $\mu$ up to $M_0$.  Here the $1/m_Q$
expansion is explicit.  Implicitly, a $1/m_Q$ expansion is 
generated also by the first (tree-level) term in the Lagrangian 
(\ref{N2}),  
\begin{equation}
{\cal L}_{\rm heavy}^0 = \bar Q (\not\!\!{\cal P} -m_Q)Q\, .
\label{lagrzero}
\end{equation}
Although the field $Q$ in this Lagrangian is normalized at a low point 
$\mu$, it carries a hidden large parameter, $m_Q$. Indeed,
the interaction of the heavy quark with the light degrees of freedom 
enters through ${\cal P}_\mu = i D_\mu$.
The background gluon field $A_\mu$ is weak if 
measured at  
the scale $m_Q$, which means, of course,  that there is a large 
``mechanical"
part
in the $x$ dependence of $Q(x)$, known from the very beginning,
\beq
Q(x) = {\rm e}\,^{-im_Qt}{\tilde Q} (x) \; ; 
\label{tildeq}
\eeq
${\tilde Q} (x)$ is a rescaled bispinor field which, in the  leading
approximation, carries no  information  about the heavy quark mass. 
It
describes  a residual motion  of the heavy quark 
inside the heavy hadron 
\cite{HQET} 
 with   typical  momenta  of order $\Lambda_{\rm QCD}$. 
Remnants of the
heavy quark mass appear in $\tilde Q$ only at the level of $1/m_Q$
corrections. 

Equation (\ref{tildeq}) is written in the rest frame of $H_Q$. In an
arbitrary frame one singles out the factor $\exp (-im_Qv_\mu x_\mu 
)$
\beq
Q(x) = {\rm e}\,^{-im_Qv_\mu x_\mu}{\tilde Q} (x) \; , \,\,\,\,  
v_\mu =
p_\mu/M_{H_Q}\;,
\label{tildeq1}
\eeq
where $v_\mu$ is the four-velocity of the heavy hadron.

The covariant momentum operator ${\cal P}_\mu$ acting on the 
original field
$Q$, when applied to 
the rescaled field $\tilde Q$, is replaced by the operator 
$m_Qv_\mu + \pi_\mu$, 
\beq
iD_\mu Q(x) = e^{-im_Qv_\mu x_\mu}\left( m_Q v_\mu + i 
D_\mu\right) \tilde Q (x)
\equiv e^{-im_Qv_\mu x_\mu}\left( m_Q v_\mu + \pi_\mu
\right) \tilde Q (x)
\, .
\label{pi}
\eeq
If not stated otherwise, we use the rescaled 
field
$\tilde Q$, {\em omitting} the  tilde in all expressions where there is 
no risk of 
confusion. The rescaled field $\tilde Q$
is a {\em four-component} Dirac bispinor, not a two component 
non-relativistic spinor which is usually introduced in the heavy 
quark effective theory (HQET) \cite{HQET}. The Dirac equation 
$(\not\!\!{\cal P} -m_Q)Q = 0$ can be rewritten as follows in terms of 
$\tilde Q$:
\beq
\frac{1-\gamma_0}{2} Q = \frac{\not\!\!{\pi} }{2m_Q} Q\, ,
\label{DEt1}
\eeq
\beq
\pi_0 Q = -\frac{\pi^2 +\frac{i}{2}\sigma G}{2m_Q} Q\, .
\label{DEt2}
\eeq

Armed with this knowledge one can easily expand 
${\cal L}_{\rm heavy}^0$, at the tree level, through order  
$1/m_Q^2$,
$$
{\cal L}_{\rm heavy}^0 =
\bar Q (i\not\!\!D -m_Q)Q\;=\;\bar Q
\frac{1+\gamma_0}{2}\left(1+\frac{(\vec\sigma\vec\pi)^2}{8m_Q^2}
\right)\left[
\pi_0-\frac{1}{2m_Q}(\vec\pi\vec\sigma)^2\;-\right.
$$
\beq
\left.
\frac{1}{8 m_Q^2}\,\left(-(\vec D\vec E)+\vec\sigma\cdot 
\{\vec E\times\vec\pi-\vec\pi\times\vec E\} \right)\,
\right]\left(1+\frac{(\vec\sigma\vec\pi)^2}{8m_Q^2}
\right)\frac{1+\gamma_0}{2}\,Q
\;+\;{\cal O}\left(\frac{1}{m_Q^{3}}\right)\, ,
\label{HQLm}
\eeq
where $\vec\sigma$ denote the Pauli matrices and
$
(\vec\pi\vec\sigma )^2 ={\vec\pi}^2 +\vec\sigma \vec B\,
$, 
$\;\vec E$
and $\vec B$ denote the background chromoelectric and 
chromomagnetic fields, respectively, with the coupling constant $g$ 
and 
the color matrix $t^a$ included in the definition of these fields. 
There is nothing new in this Lagrangian; at the tree level it is the 
same as in QED. The  non-relativistic 
expansions in QED have been known since the thirties, see e.g. 
 Chapter 4 of Bjorken and Drell \cite{BD} or Sect. 33
of the Landau and Lifshitz \cite{LL}.  It is worth noting 
that
\beq
{\cal L}_{\rm heavy}^0 \equiv
\varphi^+(\pi_0-{\cal H}_Q\,)
\varphi 
\label{ham}
\end{equation}
where 
\begin{equation}
\varphi=\left(1+\frac{(\vec\sigma\vec\pi)^2}{8m_Q^2}
\right)\frac{1+\gamma_0}{2}\,Q
\label{18a}
\end{equation}
and ${\cal H}_Q$ is a non-relativistic Hamiltonian,   
\beq
{\cal H}_Q\,=\,\frac{1}{2m_Q}\,({\vec\pi}^2 +
\vec\sigma \vec B)\,+\,
\frac{1}{8m_Q^2}\,\left(-(\vec D\vec E)+ 
\{\vec E\times\vec\pi-\vec\pi\times\vec E\} \right) +{\cal O}(1/m_Q^3)
\label{hamil}
\end{equation}
well-known (in the Abelian case) from \cite{BD,LL}.  
The first term in the $1/m_Q^2$ part is called the  Darwin term and 
the 
second one is
the convection current (spin-orbital) interaction.
Equation (\ref{18a}) is the Foldy-Wouthuysen transformation which 
is
necessary to keep the term linear in $\pi_0$ in its canonic form. For 
unclear reasons  the Foldy-Wouthuysen transformation in the 
context of the heavy quark theory is sometimes
 referred to  as 
``casting the Lagrangian in the  NRQCD form''. For a dedicated 
discussion of  the Foldy-Wouthuysen transformation in the heavy 
quark theory  see Ref.~\cite{korner}.

At the one-loop level the coefficients in the effective Hamiltonian 
 (\ref{hamil}) get some corrections which  are specific for QCD and 
cannot be read off the text-book QED expansion. 
 The work on calculating the one-loop coefficients for all terms 
 through ${\cal O}(1/m_Q^3)$ was recently
 completed  \cite{Mano}.  

The careful reader might have noticed that we consistently avoid 
using
the term HQET. This is done deliberately. 
Although, in principle, HQET is one of a few convenient technical tools
for the heavy quark expansion (an alternative useful approach is
provided by NRQCD \cite{NRQCD}), in many standard presentations  
an additional assumption is made which makes the ``folklore" HQET 
incompatible with the OPE-based expansions. What is missing
in the  commonly accepted  version 
is appreciation of the fact that all 
coefficients and all operators in the Lagrangian, 
individually, are $\mu$ dependent -- as is 
quite obvious from the consideration above. 
The Lagrangian  can be applied to the perturbative  
calculation of the 
Feynman graphs with heavy quarks where the  characteristic virtual 
momenta flowing through  all lines
in the graphs are less than $\mu$. The contribution of all virtual 
momenta {\em above} $\mu$ is explicitly included in the 
coefficients of the 
effective Lagrangian. 

To avoid confusion we suggest, from now on, to use distinct 
notations for the fundamental parameters of the Wilsonian  
Hamiltonian, on the one hand, and (perturbatively defined)
HQET parameters, on the other. The latter correspond
to tending $\mu\ra 0$, after ``appropriate subtraction of 
perturbation theory". It is  clear that the subtraction cannot be 
carried out consistently in all orders, so that the parameters defined 
in this way are to be used with caution and reservations.
Since they are widely exploited in the current literature it is 
convenient to introduce special notations. In particular,
following Ref. \cite{FN}, we will denote the HQET version of 
the kinetic expectation value 
by $-\lambda_1$. Formally, $-\lambda_1$
is given by the same Eq.~(\ref{mupid}) as $\mu_\pi^2$;
the perturbative contribution below $\mu$ is subtracted, however,
from $-\lambda_1$, which results in ambiguities. 
A pragmatically oriented reader, who is uninterested in the 
discussion of the subtraction ambiguities, to be presented below,
may just assume that the subtraction is done, say, at one-loop order 
(or at two-loop order, if calculations are carried out at the
level ${\cal O}(\alpha_s^2 ))$. Even so, one should realize, that all 
general results obtained from the QCD equations of motion, valid 
for $\mu_\pi^2$ and other parameters in the Wilsonian approach,
are, generally speaking, inapplicable to the parameters of HQET. 
Let us parenthetically note that another key parameter appearing in 
the heavy quark theory, the chromomagnetic operator,
\beq
\mu_G^2 = \frac{1}{2M_{H_Q}} {\langle H_Q|\bar Q \frac{i}{2}\sigma 
G 
Q|H_Q\rangle}=  -\frac{1}{2M_{H_Q}}
\langle H_Q|\bar Q\vec\sigma\vec B Q |H_Q\rangle 
+ {\cal O}(1/m_Q)
\,  ,
\label{muG}
\eeq
depends on $\mu$ logarithmically, {\em via} its hybrid anomalous 
dimension \cite{Falkad}; the limit $\mu\ra 0$ is never attempted in 
this case, of course. In  the nomenclature of Ref. \cite{FN} the 
parameter $\mu_G^2$ is  $3\lambda_2$.

The mass formula  relating $m_Q$ to the hadronic mass
is a useful and, perhaps, the simplest application of the expansion 
outlined  above. 
 The $1/m_Q$ corrections to the hadron mass is nothing but  the
expectation value of the effective Hamiltonian (\ref{hamil})
\beq 
M_{H_Q} = m_Q +\bar\Lambda +
\frac{1}{2m_Q} 
\frac{\langle H_Q|{\vec\pi}^2 +\vec\sigma\vec B |H_Q\rangle }
{2M_{H_Q}}+ ... = 
m_Q +\bar\Lambda + \frac{(\mu_\pi^2 - \mu_G^2)_{H_Q}}{2m_Q} 
+ ...
\label{massf2}
\eeq
If we keep only the 
terms up to  $1/m_Q$, it does not matter whether the state $H_Q$ we
average over is the asymptotic state (corresponding to $m_Q=
\infty$) or the actual physical heavy-flavor state. 
The  difference becomes  noticeable only at the level 
$1/m_Q^2$;
it is expressed in terms of a few non-local correlation functions, see
Ref.~\cite{optical} for further details.

The  parameter $\bar\Lambda$ appearing in 
Eq.~(\ref{massf2}) 
was  introduced as a  HQET constant   in 
\cite{Luke}; it 
is associated with those terms in the effective Lagrangian  ${\cal
L}(\mu )$ (disregarded so far) which are entirely due to the light
degrees of freedom. Needless to say that in the Wilsonian approach 
$\bar\Lambda$ is actually 
$\mu$ dependent, $\bar\Lambda (\mu )$. 
Wherever there is the menace of confusion
the $\mu$ dependence of $\bar\Lambda $ will be indicated 
explicitly.

There exist a few 
alternative
expressions  for this parameter. Let us quote the one relating 
$\bar\Lambda$ to the gluon anomaly in the trace of the 
energy-momentum
tensor, 
\beq
\bar\Lambda = \frac{1}{2M_{H_Q}}\langle H_Q |
\frac{\beta (\alpha_s)}{4\alpha_s}G^2
|H_Q\rangle_{m_Q\rightarrow\infty} .
\label{defl}
\eeq
This expression 
was obtained in Ref.~\cite{optical}. Some subtleties left aside   here 
are discussed in detail in that paper. It is
an obvious counterpart of a similar relation 
for the nucleon mass \cite{GAN},
$$
M_N = \frac{1}{2M_N}\langle N |
\frac{\beta (\alpha_s)}{4\alpha_s}G^2
|N\rangle\, .
$$
The renormalization properties of the operator are quite 
different,
however, in the sector of QCD with the heavy quark.
For  simplicity above the light quarks are taken as 
massless;
introduction of the  light quark masses changes only technical details.
If the mass term of the light
quarks is set equal to zero the light quark fields do not appear
explicitly in the trace of the energy-momentum tensor. 

\section{Basic Parameters  of the Heavy Quark Expansion}

\subsection{The Heavy Quark Mass}
\renewcommand{\theequation}{3.\arabic{equation}}
\setcounter{equation}{0}

This section is a brief review of  the the present
status of determination of the heavy quark masses.
An internally consistent definition of the heavy quark mass is 
of utmost importance for $1/m_Q$ expansions.  
Numerical reliability is essential if one 
wants to extract accurate values of the Cabibbo-Kobayashi-Maskawa
(CKM)  parameters  
$|V_{cb}|$ and $|V_{ub}|$. 
While these remarks are obvious 
(in hindsight), the theoretical implications were at first not 
fully appreciated. 

For quarks -- confined degrees of freedom --  there exists no natural
mass definition unlike with asymptotic states. Computational 
convenience 
and
theoretical consistency are the only guiding principles. Convenience
suggests employing the pole mass. However, as will be
explained below, the pole mass does not allow the consistent
inclusion of nonperturbative (power-suppressed) corrections. 

\subsubsection{What, after all,  is the heavy quark mass?}

In quantum field theory the object we begin our work with is the
Lagrangian formulated at some high scale $M_{ 0}$. 
The mass $m_0$  is a parameter in this 
Lagrangian; it enters on the same footing as, say,   
 the coupling constant  $\alpha_s$ with the   only difference being 
that it 
carries dimension. As with any other 
coupling, $m_0$ enters in all  observable quantities in a certain 
universal
combination with  the
ultraviolet cutoff $M_{0}$ (in renormalizable theories). 
Although this combination is universal, its particular form  is 
scheme
and scale-dependent.

The mass parameter $m_0$ by itself is not observable. For  
calculating 
observable quantities   at the scale $\mu \ll  M_{0}$ it is
usually convenient to relate $m_0$ to some  mass  parameter 
relevant to
the scale $\mu$. For instance, in quantum  electrodynamics at low
energies (i.e. $E\ll m_e$) there is  an obvious ``best"
candidate: the actual mass of an  isolated  electron, $m_e$. In  the
perturbative calculations it is determined  as the position of  the
pole in the electron Green function (more exactly, the beginning of 
the cut). The advantages are evident:  $m_e$ is  gauge-invariant and
experimentally measurable. 

The analogous  parameter for heavy quarks in QCD is 
referred
to as the pole quark mass, the position  of the pole of the quark 
Green
function. Like $m_e$ it is gauge  invariant. The idea of using this
parameter in the quark-gluon  perturbation theory apparently 
dates back
to Ref. \cite{NOSVOZ}. Unlike QED, however, the quarks 
do not exist as
isolated objects (there are no states with the quark quantum 
numbers in
the  observable spectrum, and the quark Green function beyond a
given order has neither a pole nor a cut).  Hence, $m_{\rm pole}$ 
cannot 
be directly
measured;  $m_{\rm pole}$ exists only as a theoretical construction.

In principle, there is nothing wrong with using $m_{\rm pole}$
{\em in perturbation theory} where it naturally appears  
in  the Feynman graphs 
for the quark Green functions, scattering amplitudes and so on. It 
may or may not be convenient,
depending on concrete goals.

The pole mass in QCD is perturbatively infrared stable, order by 
order, like in QED.
It is well-defined to every given 
order in perturbation theory. One cannot define it to all orders, 
however.
Nonperturbatively, the pole mass is {\em not} infrared-stable. 
Intuitively this is clear: since the quarks are confined in the full 
theory, the best one can do  is
to define the would-be pole position with  an intrinsic uncertainty
of order $\sim \Lam$ \cite{gurman}. 

The issue  of the appropriate quark mass has two aspects. At 
the perturbative level it is desirable to define the running mass 
$m_Q(\mu)$ at a scale relevant to the scale in the processes at hand, 
$\mu$,
which would incorporate contributions of all virtual momenta
(of quarks and gluons) 
from $M_{0}$ down to $\mu$. This would allow eliminating 
unwanted large coefficients which always emerge under 
unsuitable choice of
expansion parameters. 

If analysis is aimed at higher (power in $m_Q^{-1}$) accuracy,    
the use of mass parameters that can be defined to that accuracy 
becomes mandatory.
Whatever parameter we choose, it 
will not coincide  with the mass of any observable particle,
and is bound to remain a theoretical construction. If the construction 
is consistent, however, and allows us to achieve the desired degree of 
accuracy in the predictions for the observable quantities,
that is all we need.
The most straightforward choice in the theory with 
the confined quarks is as follows. We start from the original 
Lagrangian at $M_{0}$, and evolve it down to the scale $\mu$.
The original mass parameter $m_0$ evolves accordingly. The mass 
parameter appearing in the effective Lagrangian ${\cal L}(\mu )$ 
is $m_Q(\mu )$. 
 
To any order $k$ the relation between the pole mass and the running 
short-distance one $m_Q(\mu )$ takes the form~\footnote{In fact, 
 $m_{\rm pole}^{(k)}$  depends on $\mu$ since the $\as$  series
is truncated at the order $k$. This dependence shows up, however,
only at the level
${\cal O}\left( \as^{k+1}\right)$.}
\beq
m_{\rm pole}^{(k)} \;=\; m_Q(\mu )\;\sum_{n=0}^{k}\,
C_n\left(\frac{\mu}{m}\right) \,\left(\frac{\as (\mu )}{\pi}\right)^n
\; , \;\;\; C_0=1\;.
\label{m11}
\eeq

Three questions (more exactly,   three facets of one and the same 
question) naturally arise in connection with Eq. (\ref{m11}):

i) Does the series (\ref{m11}) converge to a reasonable number?

ii) How well does it behave numerically in low orders, say, in the 
second 
or third order in $\as$ which we usually deal with  in  actual 
calculations?

iii) What happens when we start analyzing nonperturbative 
corrections suppressed by powers $1/m_Q$?

The answer to the first question is negative, which entails negative 
consequences in the second and the third question, as we
will see shortly. 

\begin{figure}
\vspace{2.4cm}
\includegraphics{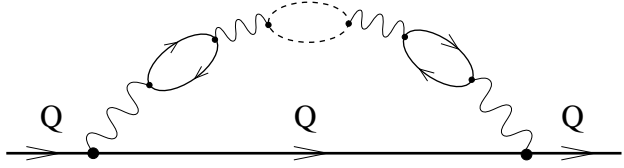}
\caption{
Perturbative diagrams leading to the IR renormalon
uncertainty in $m_Q^{\rm pole}$ of order $\Lam$. 
The number of bubble insertions in the gluon propagator
can be arbitrary.}
\end{figure}

Let us begin  with the  divergence of the
perturbative series in Eq.~(\ref{m11}). Although more than one 
 reason for the factorial divergence may exist, it is easy
to identify a particular source, 
the so-called infrared renormalon singularity in the perturbative
expansion of the pole mass \cite{pole,bbpole}, see
diagrams in Fig.~1. The bubble chain generates the
running of the strong coupling $\alpha_s$. To leading
order, it can be accounted for by inserting the running coupling 
constant 
$\alpha_s(k^2)$ in the integrand corresponding to the one-loop 
expression.
In  the non-relativistic regime, when   the internal momentum $|k| 
\ll 
m_Q$,   the expression is simple, 
\begin{equation}
\delta m_Q \sim - \frac{4}{3}
\int \frac{d^4k}{(2\pi )^4 i k_0} \frac{4\pi \alpha_s(-k^2)}{k^2} =
\frac{4}{3}
\int \frac{d^3\vec k}{4\pi ^2} \frac{\alpha_s (\vec k^2)}{\vec k^2}\, .
\label{DELTAMQ}
\end{equation}
Now, expressing the running $\alpha_s(k^2)$ in terms of 
$\alpha_s(\mu 
^2)$,  $k^2 <\mu ^2$, 
\begin{equation}
\alpha_s(k^2) =
{\alpha_s(\mu ^2)}\left\{1 + \frac{\alpha_s(\mu ^2)}{4\pi}
b \,\ln{\frac{k^2}{\mu ^2}}\right\}^{-1} , \; \;
b = \frac{11}{3} N_c - \frac{2}{3}n_f \, ,
\label{run}
\end{equation}
and expanding $\alpha_s(k^2)$ in a power series in
$\alpha_s(\mu ^2)$, it is easy to  
 find  the $(n+1)$-th order contribution to
$\delta m_Q$,
\begin{equation}
\frac{\delta m_Q^{(n+1)}}{m_Q} \sim \frac{4}{3}
\frac{\alpha_s(\mu ^2)}{\pi} n!
\left( \frac{b \alpha_s(\mu ^2)}{2\pi}\right) ^n \; .
\label{DELTAMQN}
\end{equation}
Observe that the coefficients grow factorially and contribute
with the same sign. Therefore, one cannot define
the sum of these contributions even using the Borel transformation. 
The best one
can do is to truncate the series judiciously. An optimal truncation   
leaves us with an irreducible
uncertainty $\sim {\cal O}(\Lambda _{QCD})$ \cite{pole,bbpole}. 

Thus, the perturbative
expansion {\em per se} anticipates the onset of the nonperturbative 
regime (the impossibility of locating the 
would-be quark pole to accuracy better than $\Lam$).  Certainly, the 
concrete numerical value of the uncertainty
in $m_{\rm pole}$
obtained through renormalons is not trustworthy.
The renormalons do not represent the dominant 
component of the infrared dynamics. 
However, they are a clear indicator of the presence of 
the
power-suppressed nonperturbative effects, or infrared instability of
$m_{\rm pole}$; the very fact that there is a correction ${\cal 
O}(\Lam
/m_Q)$ is beyond any doubt. 

One can illustrate the emergence of this correction in the following
transparent way.  Consider the energy stored in the
chromoelectric field in a sphere of radius $R \gg 1/m_Q$
around a static color source of mass $m_Q$,
\begin{equation}
\delta {\cal E}_{\rm Coul} (R) \propto
\int _{1/m_b \leq |x| < R} d^3x {\vec E}^{\,2}_{\rm Coul}
\propto {\rm const} - \frac{\as (R)}{\pi} \frac{1}{R}\, .
\label{COLOURCOUL}
\end{equation}
The definition of the pole mass amounts to setting 
$R \ra \infty$; i.e.,
in evaluating the pole mass one undertakes to integrate the
energy density associated with the color source over
{\em all} space assuming that it has the Coulomb form.
In real life the color interaction becomes strong at
$R_0 \sim 1/\Lam$; at such distances the 
chromoelectric field
has nothing to do with the Coulomb tail. Thus, one cannot include the
region beyond $R_0$ in a meaningful way. Its contribution
which is of order $\Lam$, thus, has to be considered
as an irreducible uncertainty which is power-suppressed relative
to $m_Q$, 
\begin{equation}
\frac{\delta_{\rm IR}m_Q^{\rm pole}}{m_Q}\; = \; 
{\cal O}\left(\frac{\Lam}{m_Q}\right)\; .
\label{POLEUNC}
\end{equation}

 It is worth noting that the pole mass was the first example 
where  a quantity which is perturbatively infrared-stable was shown  
not to be stable  nonperturbatively at the level $\Lam^1$. 
The observation of Refs.~\cite{pole,bbpole} gave impetus 
to dedicated analyses of other perturbatively infrared-stable 
observables 
in numerous hard processes  without OPE, in particular, in  jet 
physics. Such nonperturbative infrared contributions, linear in
$\Lam /Q$ were indeed   found shortly after 
in thrust and many other jet characteristics (for a review and a 
representative list of references see e.g. \cite{VB}).

To demonstrate that the problem of divergence of the
$\alpha_s$ series for $m_Q^{\rm pole}$ is far from being academic,
let us examine how the perturbative contribution to the $b$ quark 
mass looks numerically:
$$
m_b^{\rm pole} = m_b (1\GeV) +
\delta m_{\rm pert}(\leq 1\GeV) \simeq
$$
\begin{equation}
4.55 \, {\rm GeV}
+ 0.25\, {\rm GeV} + 0.22\, {\rm GeV} + 0.38\, {\rm GeV} +
1\, {\rm GeV} + 3.3\, {\rm GeV}  + ...  ,
\label{POLEMASSUNC}
\end{equation}
where $m_b (1\, {\rm GeV}) $ is the running mass at $\mu = 1 \GeV$,
and $\delta m_{\rm pert}(\leq 1\,  {\rm GeV})$ is the perturbative
series taking account of the loop momenta from $1\GeV$ down to zero.
It is quite obvious that  the corrections start to blow up already in
rather low orders! 
Expressing observable infrared-stable quantities (e.g.
the inclusive semileptonic width  $B\ra X_u \ell\nu $)
in terms of the pole mass
will necessarily entail large coefficients in the
$\alpha_s$ corrections compensating for the explosion of the 
coefficients in $m_b^{\rm pole}$. We will return to this point later
(Sect.~8.1.3). 

Summarizing, the pole mass {\em per se} does  not appear in  OPE for
infrared-stable quantities.  Such expansions  operate with the
short-distance (running)  mass. Any  attempt to
express the  OPE-based results in terms of the pole mass creates a 
problem   making the Wilson coefficients ill-defined theoretically and
poorly convergent numerically. 

Concluding this section, we make a side remark 
concerning
the $t$-quark mass. The peculiarity of the $t$ quark is that it has a
significant width $\Gamma_t \sim 1 \GeV$ due to its weak decay. 
The 
perturbative position of the pole in the propagator is, thus, shifted
into the complex plane by $-\frac{i}{2} \Gamma_t$. 
The finite decay 
width of the $t$ quark introduces a {\em  physical} infrared cutoff 
for
the infrared QCD effects \cite{khoze}. In particular, 
the observable decay
characteristics do not have ambiguity  associated with the 
uncertainty 
in the pole mass discussed above. The uncertainty cancels in any
physical quantity  that can be measured. That is not the case, 
however,
in the position of the pole of the $t$-quark propagator in the complex
plane (more exactly, its real part). The quark Green function is not 
observable there, and one would encounter the very same infrared 
problem
and the same infrared renormalon. The latter does not 
depend on
the absolute value of the quark mass (and whether it is real or have
an imaginary part). Thus, in the case of top, one would observe an 
intrinsic uncertainty due to the renormalon, of several hundred 
$\MeV$, 
in
relating the peak in the physical decay distributions to the position
of the propagator singularity in the complex plane.

While this fact will pose no practical problem any time soon, it will 
do
so in the future, in particular when analyzing top production at linear
${\rm e^+e^-}$ colliders.  The observables can be conveniently 
expressed
in terms of a mass $m_t$ defined at the scale $\mu \approx 
\Gamma_t$
that is  free from the  renormalon ambiguities. It can be defined and 
measured
without intrinsic limitations. It is worth trying to 
come
to an agreement in advance what kind of top quark mass should be  
listed
in the PDG tables, and thus avoid the problems which accompanied 
attempts to pinpoint the masses of $c$ and $b$ quarks.

\subsubsection{Short distance masses}

Since the pole mass is theoretically ill-defined, it is preferable to
use a short-distance mass. 
The
most popular choice is the so-called $\overline{\rm MS}$ mass $\bar
m(\mu)$.

The $\overline{\rm MS}$ mass is not  a parameter in the effective 
Lagrangian;
rather it is a certain {\em ad hoc} combination of the  parameters  
which is
particularly convenient in the perturbative calculations using 
dimensional
regularization. Its relation to the perturbative pole mass is known to 
two loops \cite{gray}:
\beq
m_Q^{\rm pole}\;=\;\bar{m}_Q(\bar m_Q)\left\{1+\frac{4}{3} 
\frac{\as(\bar m_Q)}{\pi} + (1.56\,b 
-3.73)\left(\frac{\as}{\pi}\right)^2
+...\, .
\right\}
\label{m15}
\eeq
At $\mu \gsim m_Q$ the $\overline{\rm MS}$ mass coincides,
roughly  
speaking,
 with the running Lagrangian  mass seen at the scale $\sim
\mu$. However, it becomes rather meaningless at  $\mu  \ll  m_Q$. 
To
elucidate the point let us note that, by definition, it adds to  the bare
mass $m_0$ the same logarithmic  contribution $\sim \as\, m_Q\,
dk^2/k^2$  both at $k^2 > m_Q^2$, where it is indeed present, and
 below $m_Q$ where  such a contribution is nonsensical.
$\bar{m}_Q(\mu)$  logarithmically  diverges when $\mu \ra 
0$. The
actual contribution coming from   domain below $m_Q$ is essentially 
smaller,
$\sim \as\, dk^2/k$.  For this reason $\bar m(\mu)$ is not 
appropriate
in  the  heavy quark theory where  the possibility of
evolving down to a low normalization  point,  $\mu \ll  m_Q$,
is crucial.

The genuine short-distance low-scale mass $m_Q(\mu)$ at $\mu \ll  
m_Q$
can be introduced in different ways, and it always exhibits an 
explicit
linear $\mu$ dependence,
\beq
\frac{{\rm d}m_Q(\mu)}{{\rm d}\mu}\; =\; -c_m 
\frac{\as(\mu)}{\pi}\; 
- ...\, .
\, .
\label{m16}
\eeq
The coefficient $c_m$ is definition-dependent. 
For the most natural choices \cite{volopt,pole,upset}, $c_m$ varies
between $4/3$ and $16/9$. The difference in the definition of the 
running mass in the heavy quark theory can, thus, 
constitute $\sim 50$ to $100 \MeV$ at $\mu \sim 1\GeV$. Once a 
particular definition is chosen, no ambiguity remains. A convenient 
physical (gauge-invariant) 
definition of the  short-distance heavy quark mass with $\mu \lsim 
m_Q$
was outlined in \cite{optical} (Sects.~III.D and V.A). Technical aspects
are discussed in detail in the dedicated paper \cite{five}, where
$c_m=16/9$. 

\subsubsection{Which mass is to be used?}

It is legitimate to use any  short-distance mass. The normalization 
point $\mu$ can be arbitrary as long as $\mu \gg \Lam$. It does not 
mean, however, that all masses  are equally
practical,  since the  perturbative series are necessarily truncated
 after a few  first  terms. Using an inappropriate scale makes
numerical approximations bad. In particular, relying on 
$\bar{m}_Q(m_Q)$
in treating the low-scale observables can be awkward. The 
following
toy example illustrates this point.

Suppose, we would like to exploit QCD-like methods in solving the 
textbook
problem of the positronium mass, calculating the standard Coulomb
binding energy. To use  the result written in terms  of the
$\overline{\rm MS}$ mass one would, first, need  to evaluate $\bar
m_e(m_e)$. This  immediately leads to technical problems: $\bar
m_e(m_e)$ is known only to the order $\alpha^2  m_e$; therefore,  
the
``theoretical uncertainty"  in $\bar m_e(m_e)$ would generate the 
error bars $\sim \alpha^3 m_e$ in the binding energy, i.e.  at 
least
$0.01\,{\rm  eV}$. Moreover, without cumbersome two-loop 
calculations
one would not know the binding  energy  even at a few ${\rm eV}$ 
level
-- although, obviously, it is known to a much better
accuracy (up to $\al^4 m_e$ without a dedicated field-theory 
analysis),
and the result
\beq
M_{P}\;=\; 2m_e(0)\; - \; \frac{\al^2m_e}{4}
\label{m18}
\eeq
is obtained without actual loop calculations!

Thus, working with the  $\overline{\rm MS}$ mass
we would have to deal here with the ``hand-made" disaster. 
The reason is obvious. The relevant momentum scale in this 
problem is the inverse Bohr radius, 
$\mu_B = r_B^{-1}\sim \alpha m_e$. Whatever contributions emerge 
at   much shorter distances, say, at $\mu^{-1}\sim m_e^{-1}$,  
their sole role is
to  renormalize  the low-scale parameters $\al$ and $m_e$. If the
binding energy is  expressed in terms of these parameters,  further
corrections are  suppressed by powers  of $\mu_B/\mu$. Exploiting  
the
proper  low-energy  parameters $\alpha(0)$, $m_e(0)$ one 
automatically 
resums chains of  corrections.  In essence, this  is  the basic  idea
of the effective Lagrangian  approach,  which, unfortunately,  is often
forgotten.

Needless to say, that in the processes at  high energies it is the 
high-scale  masses that appear directly. In particular, the inclusive 
width $Z\ra b \bar b$ is sensitive to $m_b(M_Z)$;  using 
$\overline{\rm
MS}$  mass normalized at $\mu\sim M_Z$ is appropriate here. 
On the
contrary, the inclusive semileptonic decays $b\ra c\,\ell \nu$ are
rather low-energy in this respect \cite{five}, and, to some extent, 
that is true even for $b\ra u$.  

\subsubsection{The numerical values of $m_c$ and $m_b$}

Accurate phenomenological determination of  $m_b$ and  $m_c$ at 
large
$\mu$  (higher or comparable to the quark masses themselves)  
is not
easy since it requires very precise data and a high degree of control 
over the
perturbative and  nonperturbative corrections. The  terms $\sim
(\as/\pi)^2 m_b$ by themselves constitute typically $\sim  
200\MeV$,
which, thus,  sets the scale of accuracy one can expect.

The mass of the $c$ quark at the scale $\sim m_c \sim 1\GeV$ can 
be 
obtained from
the charmonium sum rules \cite{SVVZ}, $
m_c(m_c)\; \simeq\; 1.25 \GeV$. The result to some extent 
is  affected by the value of the gluon condensate. 
To be safe, we conservatively ascribe a rather large  
uncertainty,
$$
m_c(m_c) = 1.25 \pm 0.10 \, \mbox{GeV}\, .
$$
One can argue that the precision 
charmonium sum rules
\cite{SVVZ} actually fine-tune the charmed quark mass to a much 
better accuracy. The argument would lead us far astray,
and is irrelevant for our present purposes since the convergence of 
the $1/m_c$ expansion is not good anyway. 

The desire to get rid of  potentially large perturbative corrections in 
the
$b$ quark mass,  scaling as $m_b$, suggests examining the 
threshold
region. 
The process ${\rm e^+e^-} \ra b \bar b$ in the threshold domain 
provides us with the  opportunity  of determining  the 
low-scale running 
mass, along the lines suggested in Ref. \cite{SVVZ}.  Using 
dispersion relations 
\beq
\Pi_b(q^2) \; =\; \Pi_b(0)\;+\; \frac{q^2}{2\pi^2} 
\,\int\,\frac{ds\, R_b(s)}{s(s-q^2)}
\label{m25}
\eeq
one evaluates the polarization operator $\Pi_b(q^2)$ (and its
derivatives) induced by the vector currents $\bar b\gamma_\mu b$,  
in
the  complex $q^2$  plane at an adjustable distance $\Delta$ from the
threshold. Such  quantities are proportional to weighted integrals 
over
the experimental cross  section; the integrals are saturated  in the
interval $\sim \Delta$ near threshold, and are very sensitive to the 
mass $m_b(\Delta)$.

A dedicated analysis was carried out  by  Voloshin \cite{volmb} who 
considered a
set of relatively high derivatives of $\Pi_b$ at $q^2=0$. On the 
phenomenological side they  are  expressed through moments of 
$R_b(s)$, 
$$
\frac{2\pi^2}{n!}\, \Pi_b^{(n)}(0) \; =\; I_n\; = \;
\int\,\frac{ds\,R_b(s)}{s^{n+1}}
\simeq
$$
\beq
M_{\Upsilon(1S)}^{-2(n+1)}\, \int \: ds\:R_b(s)\,
\exp\left\{ -(n+1)\left(s-4M_{\Upsilon(1S)}^2\right)\right\}\, ,
\label{m26}
\eeq
while the theoretical expressions for the very same moments
are given in terms of the running quark mass and $\alpha_s$.
The relevant momentum scale here is $\mu \sim m_b^2/\sqrt{n}$.
Considering small-$n$ moments $I_n$ one
would determine $m_b$ at the scale of the order $m_b$.
The small-$n$ moments  are contaminated by the contribution of 
$R_b$ above the open beauty threshold where 
experimental data  are poor. Increasing $n$ 
shrinks the interval of saturation and, thus, lowers the effective
scale. On the other hand, we 
can not go to too high values of $n$ where  infrared 
effects
(given, first, by the gluon condensate) explode. There is still enough
room to keep the gluon condensate small and, simultaneously, 
suppress
the domain above the open beauty  threshold.  In the fiducial 
window, 
$\mu$ must be large enough to ensure control over the QCD 
corrections. The latter requires a nontrivial summation of
 enhanced Coulomb terms unavoidable   in  the non-relativistic
situation. As known from textbooks, the part of the perturbative
corrections to the polarization operator, associated with the potential
interaction, is governed by the parameter    $\alpha/|\vec{v}\,|$ 
rather 
than by  $\alpha$ {\em per se}. In \cite{volmb} a typical scale is
$\mu\sim 
0.7$ to $1.5\GeV$. A sophisticated resummation technique was used 
to access the moments as high as $16$ to $20$. 

The crucial 
question is the estimate of uncertainties in determination of $m_b$. 
The  
uncertainty of the fit for 
$m_b$ is a meagre few $\MeV$. This is not surprising since the 
theoretical 
expression for
the moments very sharply depends on the value of $m_b$, 
\beq
I_n\; \propto \; 
{\rm e}\,^{-2n\delta m_b/m_b}\;.
\label{m27}
\eeq
It is clear that the uncertainty of the fit {\em per se} 
is overshadowed  by a systematic theoretical uncertainty. The 
calculation of the
Coulomb effects in Ref. \cite{volmb} was not genuinely two-loop: the 
effect of 
running of
$\alpha_s$ was accounted for only through  the 
Brodsky-Lepage-Mackenzie (BLM) scale fixing \cite{BLM}.  
Considering the fact that 
the dominant,
BLM $\as^2$ effects were accounted for, it seems  safe to assume that 
the 
actual
accuracy of the determination of $m_b(\mu)$ with $\mu\sim 1 
\GeV$ is 
not worse than $(\alpha_s (\mu )/\pi)^2 \mu \sim 30\MeV$  
\cite{volmb}. To 
be 
on the safe side, we increase it up to $50\MeV$. This number 
includes, 
additionally, possible scattering
 in the definitions of $m_b(\mu)$ at the level 
$\sim
(\as/\pi) \mu^2/(2m_b)$ which can constitute  $10$ to 
$20\MeV$. It is
worth  mentioning  that the most labor-consuming analysis of  the 
next-to-leading  
$\as^{n+1}/|\vec{v}\,|^n$ effects in \cite{volmb}, although absolutely
necessary for determination of $\as$,  led to a modest change in 
the numerical value of $m_b$, 
$\sim 15\MeV$,  as compared to the ten-year old analysis of Ref. 
\cite{old}
\footnote{After this review was essentially finished it
was reported \cite{Pich} that the  analysis of the $\bar b b$ sum 
rules had been repeated, with some rather surprising conclusions. 
Resummation of the full Coulomb ladder is not addressed, however, 
in 
Ref. \cite{Pich}. We choose to rely on the previous results
 until the matter is more closely looked
at.}.

This method could be refined in several respects to yield the overall
precision in the ballpark of $10\MeV$.  At this level  the 
accurate definition of the renormalization  procedure is mandatory.
If the experimental cross section ${\rm e^+e^-} \ra b \bar b$ above 
the
threshold were better measured in a reasonably wide interval, one 
could
study  lower moments $I_n$ and  get an accurate determination of 
$m_b$
without tedious  analysis of the perturbative Coulomb 
corrections.

Numerically, the $b$ quark mass turns out to be 
\cite{volmb}
\beq
m_b(1\GeV)\;=\;4.64\GeV \pm 0.05\GeV\, .
\label{m28}
\eeq
The definition  corresponds to the scheme in which  $c_m=16/9$. In 
other words, the ``one-loop" pole mass is
\beq
m_b^{(1)}\;=\;4.64\GeV \;+\frac{16}{9}\frac{\as}{\pi}\cdot 1\GeV 
\;\simeq \; 4.83\GeV
\label{m29}
\eeq
at $\as=0.336$. The result is, of course, sensitive to the choice of 
$\mu$. The sensitivity to the uncertainty in $\alpha_s$ is around
$10\MeV$. The corresponding value of $\La(1\GeV) \approx
0.6\,{\rm GeV}$.

The heavy quark masses can be measured, in principle,  by studying  
the distributions in 
the
semileptonic $B$ decays  \cite{prl,volspec}. Such analyses 
were undertaken recently \cite{chern,gremm}. Unfortunately, the 
data are 
not
good enough yet to yield a competitive determination. On the 
theoretical
side, there are  potential problems with higher-order corrections due 
to
not too large energy release $m_b-m_c\simeq 3.5\GeV$ and/or 
relatively
small mass of  the $c$ quark. In particular, the effect of higher-order
power corrections can be noticeable. The result of analysis reported  
in Refs.~\cite{chern,gremm} is compatible with
the value (\ref{m28}), within the quoted uncertainties. 
This approach seems to be most natural.

In many applications one needs to know the difference between 
$m_b$ and
$m_c$. If both masses are normalized at the same low-scale point 
below
both masses, $m_b(\mu)-m_c(\mu)$ depends on $\mu$ weakly. This
difference is well constrained in the heavy quark expansion. For
example,
\beq
m_b-m_c=\overline{M}_B-\overline{M}_D +
\mu_\pi^2\left(\frac{1}{2m_c}-\frac{1}{2m_b}\right) +
\frac{\rho_D^3-\bar\rho^3}{4}
\left(\frac{1}{m_c^2}-\frac{1}{m_b^2}\right) \;+\; {\cal
O}\left(\frac{1}{m^3}\right), 
\label{m30}
\eeq
$$
\overline{M}_B=\frac{M_B+3M_{B^*}}{4}\;,
\;\;\;\overline{M}_D=\frac{M_D+3M_{D^*}}{4}\;,\;\;\;\bar\rho^3
\equiv
\rho_{\pi\pi}^3+\rho_S^3
$$
where $\mu_\pi^2$ is the asymptotic expectation value of the kinetic
operator, $\rho_D^3$ is the expectation value of the Darwin term 
given
by the local four-fermion operator and
$\rho^3_{\pi\pi}$, $\rho^3_{S}$  are positive non-local correlators
\cite{optical}. All quantities in \eq{m30}, except the  meson masses,
depend on the normalization point $\mu$ which can be arbitrary.
In this way we arrive at
\beq
m_b-m_c \simeq 3.50\GeV \;+\; 40\MeV\,
\frac{\mu_\pi^2-0.5\, {\rm GeV}^2}{0.1\,{\rm GeV^2}}
\;+ \; \Delta M_2\;\;,\;\;\;\; |\Delta M_2|\lsim 0.015\GeV \; .
\label{m30a}
\eeq
The estimate at $\mu_\pi^2=0.5\GeV^2$ appears to be in  
good agreement with 
the separate determinations of $m_b$ and 
$m_c$  
from the sum  rules in charmonia and $\Upsilon$.

The main uncertainty in $m_b-m_c$ is due to that in the value of 
$\mu_\pi^2$.  
The
Darwin term can be reasonably estimated relying on  factorization
\cite{motion}; it is of order $0.1\GeV^3$. The non-local 
positive
matrix elements  $\rho^3_{\pi\pi}$ and $\rho^3_{S}$ are 
expected,
generally speaking, to be of the same order. Altogether, assuming 
$|\rho_D^3-\bar\rho^3| \lsim 0.1\GeV^3$, one arrives at the 
uncertainty
in $m_b-m_c$ due to the higher-order terms $\Delta M_2$ quoted 
above. 

\subsection{The chromomagnetic and kinetic operators }

The non-relativistic Hamiltonian of the heavy quark (\ref{hamil}) 
in the order $1/m_Q$ contains two
operators (\ref{mupid}) and (\ref{muG}).
Their expectation values in the heavy meson $B$ are  the key 
players in many applications.
Let us note that the expectation values of the chromomagnetic 
operator in $B$ and $B^*$ are related, 
\beq
\mu_G^2 = \frac{1}{2M_B}\matel{B}{\bar b \frac{i}{2} 
\sigma_{\mu\nu}
G_{\mu\nu} b}{B} \simeq -\frac{3\;\;}{2M_B}\matel{B^*}{\bar b 
\frac{i}{2}
\sigma_{\mu\nu} G_{\mu\nu} b}{B^*}\, .
\label{p6}
\eeq

The value of $\mu_G^2$ is known: since 
$$
\frac{1}{2m_Q} \bar{Q} 
\frac{i}{2}
\sigma_{\mu\nu} G_{\mu\nu} Q
$$ 
describes the interaction of the 
heavy
quark spin
with the light cloud and causes the hyperfine splitting between 
$B$ and
$B^*$,
it is easy to see that
\beq
\mu_G^2\; \simeq\;\frac{3}{4}\:2m_b(M_{B^*}-M_B) \;\simeq\;
\frac{3}{4}\,(M^2_{B^*}-M^2_B)\;\approx\; 0.36\GeV^2\;.
\label{p7}
\eeq
In the above relations  the operators are normalized at the scale
$\sim m_b$. Evolving them perturbatively to the  
normalization 
scale $\mu \sim 1\GeV$ slightly enhances the chromomagnetic 
operator 
and the  value of $\mu_G^2(\mu)$, but this effect is numerically 
insignificant, and
we will  neglect it.

In Sect. 3.1.4 we mentioned that the heavy 
quark mass can be extracted from the data on semileptonic $B$ 
decays. By the same token, 
the  value of the kinetic operator is, in principle, directly 
measurable
 in  experiment.
However, it remains rather  uncertain at the
moment. First  attempts at extracting  it from the semileptonic 
distributions were reported
\cite{chern,gremm}. Although the method seems to 
be the most promising, so far the outcome is inconclusive.
Typically, the results fall in the interval $0.2$ to $0.3\GeV^2$. 
We hasten to add, however, that the difference between
$\mu_\pi^2$ and $-\lambda_1$ was not fully untangled. 
The normalization point $\mu$ should be explicitly introduced in the 
perturbative corrections to reveal the distinction between these two 
parameters. Until this is done it is unclear to which particular 
definition of $-\lambda_1$ the number 
quoted refers.\footnote{This remark applies also 
to other numerical determinations discussed below in this section, 
with the exception of Eq.~(\ref{p11}).} 
As we will 
see shortly, numerically the difference 
between $\mu_\pi^2$ and $-\lambda_1$ is of order 
$0.1\GeV^2$ ($-\lambda_1$ is smaller than $\mu_\pi^2$).
Note also that the numerical
analysis  of the higher-order power corrections \cite{grekap} led 
the authors to 
the
conclusion that these effects are  significant in the 
application to the semileptonic spectra.

Historically, the first attempt to determine the average of the kinetic 
operator from the 
QCD sum rules was Ref. \cite{neubold} where  a negative  
value  $\sim -1\GeV^2$ was obtained! (This result was later 
revised by the author.) Shortly after,  a more 
 thoughtful
application of the QCD sum rules yielded  \cite{pp}  
$\mu_\pi^2\approx 0.6\GeV^2$. The prediction was later 
refined by the authors
 \cite{ppnew}
\beq
\mu_\pi^2\;=\; (0.5\pm 0.15) \GeV^2 \; .
\label{p10}
\eeq

 Meanwhile a model-independent lower bound was established
\cite{motion,volpp,vcb,optical}
\beq
\mu_\pi^2\;>\;\mu_G^2\approx 0.4 \mbox{GeV}^2
\label{p11}
\eeq
which  constrained  possible values of $\mu_\pi^2$.
It is worth emphasizing that the inequality takes place for 
$\mu_\pi^2$ normalized at any point $\mu$, provided 
$\mu_G^2$ is normalized at the same point. 
For large $\mu$ it becomes uninformative; so, it is in our best 
interests to use it at $\mu$ = several units $\times \Lambda_{\rm 
QCD} \lsim 1 \GeV$. 
The inequality  does not necessarily 
hold for
$-\lambda_1$. 
 Exact QCD inequalities of this type will be discussed in the next 
section. 

Recently, an  approach combining the QCD sum rules and the virial 
theorem was exploited \cite{neubp} for determination of 
$\mu_\pi^2$.
The corresponding analysis is claimed to produce a surprisingly  
precise value 
\beq
0.1\pm
0.05\GeV^2\, ,
\label{NNR}
\eeq
noticeably below the range obtained in other analysis. We 
will dwell
on various aspects of this approach, in an attempt to reveal
 reasons explaining the discrepancy, in the  
Sect. 6.  
 
The expectation value of the kinetic operator was also estimated in 
the quark
models, with a spectrum of predictions: the relativistic quark 
model
\cite{rel} gave about $0.6\GeV^2$  or even slightly larger; the
estimates of Ref.~\cite{hwang} yield a close value $0.5\GeV^2$. 
Attempts of extracting $\mu_\pi^2$ on the lattices are also reported
\cite{lat}.

The scattering of the numerical results above should not surprise the 
reader  too much.  As was mentioned above, although the formal 
definition of $\mu_\pi^2$ and $-\lambda_1$ is the same, see 
Eq.~(\ref{mupid}), these quantities are not identical, and one should 
expect noticeable differences. What is surprising is the discrepancy 
obtained under one and the same definition and within basically the 
same method. The most notable example of this type is
the sum rule rule analysis, Ball{\em et al.} \cite{pp} versus Neubert
\cite{neubp}. We will dwell on this point in Sect.~6.2. 
Now we would like  to make several remarks regarding the 
difference 
between $\mu_\pi^2$ and $-\lambda_1$. Conceptually the problem 
with $\lambda_1$ is similar to the problem  encountered in the  
attempt
to define the pole mass (more exactly, $\La_{\rm HQET}=M_B-
(m_b)_{\rm pole})$.

If $\mu_\pi^2$ is defined in the context of the Wilsonian OPE,
and incorporates all soft fluctuations, including perturbative 
fluctuations
from 0 to $\mu$,
HQET assumes  the existence of a ``genuinely nonperturbative" 
parameter
\beq
-\lambda_1 \,=\, \mu_\pi^2(\mu) \,- \, 
\mbox{``$\left(\mu_\pi^2(\mu)\right)_{\rm pert}$''}
\label{p13}
\eeq
where $\left(\mu_\pi^2(\mu)\right)_{\rm pert}$ is a ``perturbative 
part" of
$\mu_\pi^2$ coming from the domain below $\mu$. 
As was emphasized long ago \cite{fail}, the procedure of
separating out ``perturbative 
parts" of the condensates cannot be carried out in general,
simply because the notion of perturbation theory below $\mu$ is 
non-existent.  The best one can do is to construct $-\lambda_1$,
starting from $\mu_\pi^2 (\mu)$, in the given (say, the first or the 
second) order in $\alpha_s (\mu)$. 

The Wilsonian kinetic average 
$\mu_\pi^2$ is $\mu$ dependent. For example, to the first order in 
$\as$ one
has 
\beq
\frac{{\rm d}\mu_\pi^2(\mu)}{{\rm d}\mu^2}\; = \; c_\pi\, 
\frac{\as}{\pi}
\label{p12}
\eeq
where under some natural choice of the cut-off (i.e. the 
normalization point) 
$c_\pi=4/3$ \cite{optical}.

Equation (\ref{p12}) implies that to the first order in
$\alpha_s$
\beq
-\lambda_1^{(1)} = \mu_\pi^2(\mu) - 
\,c_\pi\, \frac{\as}{\pi}
\,\mu^2
\label{p16}
\eeq
where $\as$ is the  coupling constant (which does not run in
one-loop calculations) and the superscript {\small (1)} indicates
the one-loop nature of $\lambda_1$ introduced in this way.

The first observation is obvious: $-\lambda_1^{(1)}$ is smaller than
$\mu_\pi^2(\mu)$. How much smaller? The answer 
depends on the choice of $\mu$; it is also affected by the 
value of $\alpha_s$ in a particular one-loop calculation.
If $\mu=0.5\GeV$ 
the values of the kinetic operator quoted in \cite{pp,neubp} 
must be increased by 
approximately
$0.05\GeV^2$ to translate them from $-\lambda_1$
to $\mu_\pi^2$; at $\mu=0.7\GeV$ the shift constitutes $\sim 
0.1\GeV^2$. A similar (probably, somewhat larger) adjustment 
applies to empiric determinations from
the semileptonic spectra \cite{chern,gremm}. These shifts 
lie within the error bars of the corresponding analyses.

Constructing a  general ``ready-to-use" cut off procedure for the 
Wilsonian operators normalized at $\mu$  is  a
non-trivial problem going   beyond the scope of this review.
A brief discussion in heavy quark theory can be found in 
\cite{blmope,Ji1} and on lattices in Ref.~\cite{Ji2}; the issue will 
be further elaborated  in a dedicated publication \cite{future}.
Here we mention  a few basic points relevant to the issue of 
$\mu_\pi^2$.

A consistent field-theoretic definition of the operator $\bar Q
(i\vec D)^{\,2} Q $ was given  using the small velocity (SV) sum rules
\cite{optical,blmope}. Limiting oneself to the  purely perturbative 
level, 
one can readily 
outline  other cut-off schemes allowing to  
calculate the perturbative $\mu$ dependence of $\mu_\pi^2(\mu)$.
In doing so one must keep in mind that some purely perturbative 
cut-off schemes that have no transparent physical meaning  violate 
the general quantum-mechanical properties of 
the operator in question. Therefore, one must exercise extreme
caution in connecting various results regarding the condensates
obtained in different approaches with each other. 
Moreover, perturbative cutoff procedures do not define the operators 
at
the nonperturbative level. 
 
Keeping in mind all these  problems, one may ask whether 
the  controversy in the current literature is to be taken seriously
at all. 
The numerical 
comparison 
given above seems to show that the existing scattering in the value 
of
the kinetic operator 
goes
beyond the  
mismatches of different  definitions. A better understanding of 
theoretical uncertainties in the sum rule and empiric determinations
is badly needed. This task obviously must  be the subject of a 
special investigation. In the absence of such an investigation
we will focus in this review on a class of results which seems to be 
theoretically clean: exact QCD inequalities. 

\section{Exact Inequalities of the Heavy Quark Theory in the Limit
$m_Q\ra \infty$ }

\renewcommand{\theequation}{4.\arabic{equation}}
\setcounter{equation}{0}

Exact inequalities reflecting the most general features of QCD
(e.g. the vector-like nature of the quark-gluon interaction) have 
been with us since the early eighties \cite{EIQCD}.
The advent of the heavy quark theory paved the way to a totally 
new class of inequalities among the fundamental parameters.
As with the old ones, they are based on the equations of motion of 
QCD 
and certain positivity properties. All technical details of the 
derivation are different, however, as well as the sphere of 
applications.

The most well-known example is Eq. (\ref{p11}) which was  subject  to
intense scrutiny.   As a matter of fact, this expression is just one
representative of a large class, including,  among others, the Bjorken
and Voloshin relations, and the so-called third (BGSUV) sum rule. Since
the corresponding derivations are scattered in the literature,  it makes
sense to give here a coherent presentation of the subject.

The starting point for all inequalities is the set of the sum rules
\beq
\varrho^2(\mu)-\frac{1}{4} =
 \sum_n|\tau_{1/2}^{(n)}|^2 +
2\sum_m|\tau_{3/2}^{(m)}|^2 \, ,
\label{bj}
\eeq
\beq
\La(\mu) = 2\left(
\sum_n \epsilon_n|\tau_{1/2}^{(n)}|^2 +
2\sum_m \epsilon_m|\tau_{3/2}^{(m)}|^2
\right)\, , 
\label{vol}
\eeq
\beq
\frac{\mu_\pi^2(\mu)}{3} =
\sum_n \epsilon_n^2|\tau_{1/2}^{(n)}|^2 +
2\sum_m \epsilon_m^2|\tau_{3/2}^{(m)}|^2\, ,
\label{srmupi} 
\eeq
\beq
\frac{\mu_G^2(\mu)}{3} =
-2 \sum_n \epsilon_n^2|\tau_{1/2}^{(n)}|^2 + 
2 \sum_m \epsilon_m^2|\tau_{3/2}^{(m)}|^2 \, ,
\label{pig}
\eeq
\beq
\frac{\rho_D^3(\mu)}{3}=
\sum_n \epsilon_n^3|\tau_{1/2}^{(n)}|^2 +
2\sum_m \epsilon_m^3|\tau_{3/2}^{(m)}|^2\, , 
\label{asr1}
\eeq
\beq
-\frac{\rho_{LS}^3(\mu)}{3}= 
-2 \sum_n \epsilon_n^3|\tau_{1/2}^{(n)}|^2 +
2 \sum_m\epsilon_m^3|\tau_{3/2}^{(m)}|^2 \, ,
\label{fourth}
\eeq
a sequence which can be readily continued further.\footnote{Higher 
moments than those presented above are not  very 
useful in practice since they are saturated at higher $\epsilon$, 
the $\mu$ dependence becomes too essential, etc.} 
Here $\epsilon_k$ is the 
excitation energy of the $k$-th intermediate state (``$P$-wave 
states"
in the quark-model language), 
$$
\epsilon_k=M_{H_Q^{(k)}}-M_{P_Q}\,  ,
$$
while the functions $\tau_{1/2}^{(n)}$
and $\tau_{3/2}^{(m)}$ describe the transition amplitudes of the 
ground state $B$ meson to
these intermediate states. We follow the 
notations of  Ref.~\cite{NIMW},
\beq
\frac{1}{2M_{H_Q}}\langle H_Q^{(1/2)} |A_\mu |P_Q\rangle
= - \tau_{1/2} (v_1 - v_2)_\mu\, ,
\eeq
and
\beq
\frac{1}{2M_{H_Q}}\langle H_Q^{(3/2)} |A_\mu |P_Q\rangle
= -\frac{1}{\sqrt{2}}i\tau_{3/2} \epsilon_{\mu\alpha\beta\gamma}
\varepsilon^{*\alpha}(v_1 + v_2)^\beta (v_1 - v_2)^\gamma\, ,
\eeq
where 1/2 and 3/2 mark the quantum numbers of the light cloud in 
the intermediate states, $j^\pi = 1/2^+$ and $3/2^+$, respectively,
and $A_\mu$ is 
the axial current. Furthermore, the slope parameter $\rho^2$ of 
the Isgur-Wise function is defined as 
\beq
\frac{1}{2M_{P_Q}}
\matel{P_Q(\vec v)}{\left(\bar Q \gamma_0 Q\right)_\mu}{P_Q}=
1-\varrho^2(\mu)\frac{\vec v^{\,2}}{2} +{\cal O}(\vec v^{4})\, .
\label{slope}
\eeq
 All quantities on the left-hand side of Eqs. 
(\ref{bj}) -- (\ref{fourth}) depend on the normalization point $\mu$,
since they are defined in the Wilsonian sense. 
The right-hand side also depends on this point. 
Thus, all sums over the excited states appearing in 
Eqs. 
(\ref{bj}) -- (\ref{fourth}) imply a cut-off at the excitation energy 
equal to $\mu$,
\beq
\epsilon_k \leq \mu\, .
\label{smartnp}
\eeq

Equation (\ref{bj}) is nothing but the Bjorken sum rule \cite{BJSR}.
It was discussed in great detail in Ref.  \cite{NIMW} where, 
among other things, the cut-off was introduced in the form 
(\ref{smartnp})
almost a decade ago!
Equation (\ref{vol}) was obtained by Voloshin \cite{MVSR}. The 
expression 
for $\mu_\pi^2$ is the BGSUV sum rule \cite{third}.
The next one is new. 
The last two sum rules are obtained along the same lines; the one for 
the Darwin term 
$\rho_D^3$ 
was presented in \cite{pirjol}. 
A unified field-theoretic derivation of all these relations was 
discussed 
in Ref.~\cite{optical} where the transition amplitude in the SV limit 
was considered (see also Sect. 8.2.1). Since the presentation in 
that paper is sufficiently 
pedagogical, the interested reader is referred to it directly. In this 
section we will give an alternative (new) derivation of Eqs.~
(\ref{srmupi})--(\ref{fourth}). A new derivation of Eq. (\ref{vol})
will be sketched in Sect. 6.1.

At first, however, we pause to make a few explanatory remarks 
regarding $\mu$ dependence.  To avoid inconsistencies it must 
be done in the same 
way in all theoretical calculations, including perturbation theory, the 
definition of the heavy quark current (and, hence, 
the Isgur-Wise function) and so on. This cut-off scheme does not 
coincide with the one implied by dimensional 
regularization,
although they are related in the perturbative domain $\mu\gg 
\Lam$. In
particular, the logarithmic scale dependence of $\varrho^2(\mu)$ 
is the same. The bound 
$\varrho^2 > 1/4$ \cite{BJSR} holds only for this definition of the 
renormalized 
Isgur-Wise
function and is not necessarily true for other definitions. 
Without the cut-off at $\epsilon_k =\mu$ the 
sums over the excited states in the above relations diverge at high
 excitation energies. The high-energy tails of the sums  are dual 
to the perturbative
transition probabilities which, thus,  determine the $\mu$ 
dependence of 
the
operators at hand. Cutting  the integrals over the
excitation energies (i.e. limiting  the sums by $(\epsilon_k)_{\rm 
max} 
= \mu$)  is a physical way of  introducing the normalization point 
without
violating the general properties of QCD and not endangering the 
quantum-mechanical
aspects of the field 
theory. It is often convenient to use the 
exponential
weight ${\rm e}^{-\epsilon/\mu}$ in the sums. As long as the 
weight function is universal,
all relations are the same as with a step-like cut-off.

Now, we proceed to a consideration which 
illustrates the physical meaning of the sum rules presented above. 

Since in the heavy quark limit, $m_Q \gg\Lam$, $\bar QQ$ pairs are 
not produced, 
we will use the quantum-mechanical language with respect to 
$Q$  (but not the  light cloud, of course).  Moreover, it is 
convenient, at the first stage to assume $Q$  to be spinless. 
The $Q$ spin effects are 
trivially included later. 
Then the  lowest-lying 
states, the  $S$-wave configurations corresponding  to $B$ and $B^*$, 
are  spin-1/2 fermions, with  two spin orientations of the 
 light cloud. We shall denote them  $|\Omega _0\rangle $; the  spinor 
wavefunction of this state is 
$\Psi _0$. It is obvious that  
\beq
\matel{\Omega _0}{\bar Q (i D_j) (i D_k) Q}{\Omega _0} \equiv 
\frac{\mu _{\pi}^2 }{3} \delta_{jk}
\Psi_0^\dagger \Psi_0 - \frac{\mu_G^2}{6} 
\Psi_0^\dagger \sigma_{jk} \Psi_0 = 
\sum _n \matel{\Omega _0}{ \pi_j}{n}
\matel{n}{\pi_k}{\Omega _0}, 
\label{p14}
\end{equation}
where  a complete set of intermediate states is  inserted.
They are spin-1/2 states (of the opposite parity with respect  to 
$|\Omega _0\rangle $), generically denoted by 
$\phi ^{(n)}$, and spin-3/2 states $\chi ^{(n)}$. 
We will use the Rarita-Schwinger wavefunctions for the latter, i.e. a 
set 
of three
spinors $\chi_l$ obeying the constraint $\sigma_l\chi_l=0$. The 
normalization  
of these spinors is fixed by the sum over polarizations $\lambda$ 
\begin{equation} 
\sum_{\rm \lambda} \chi _i (\lambda)\chi _j^{\dagger}(\lambda) = 
\delta _{ij} - 
\frac{1}{3} \sigma _i \sigma _j \;\;.
\end{equation} 
Defining the reduced matrix elements $a_n$ and $b_m$ as 
\begin{equation} 
\matel{\phi ^{(n)}}{\pi_j}{\Omega _0} \equiv 
a_n \phi ^{(n)\dagger}\sigma _j \Psi _0 \; ,\;  \; \; 
\matel{\chi^{(m)}}{\pi_j}{\Omega _0}\equiv 
b_m \chi_j^{(m)\dagger} \Psi _0\, ,
\end{equation} 
where $\phi ^{(n)}$ and $\chi ^{(m)}$ stand for the states as well as 
for their wavefunctions, we get 
\begin{equation} 
\mu _G^2 \;=\; 
-6 \sum_n |a_n|^2 + 2 \sum_m |b_m|^2 \; , 
\label{3.7a}
\eeq
and 
\beq
\mu _{\pi}^2 \;=\; 
3 \sum_n |a_n|^2 + 2 \sum_m |b_m|^2 \; .  
\label{3.7b}
\end{equation} 

These expressions can be immediately  generalized to the actual 
case of the spin-1/2 quarks $Q$. The quantities $a_n$ and $b_m$ 
are  to be understood as the matrix elements of $\bar b i\vec D 
b$ 
between the $B$ meson  and higher even-parity states. 
They are related to $\tau_{1/2}^{(n)}$ and 
$\tau_{3/2}^{(m)}$ as follows:
\beq
\tau_{1/2}^{(n)}\;=\; \frac{a_n}{\epsilon_n}\;\;,\;\;\;\; 
\tau_{3/2}^{(m)}\;=\;
\frac{1}{\sqrt{3}} \frac{b_m}{\epsilon_m}\;,
\label{p30}
\eeq
and, therefore, 
$$
\mu _\pi^2 =
3\left( \sum _n \epsilon_n^2|\tau_{1/2}^{(n)}|^2 + 
2\sum_m \epsilon_m^2|\tau_{3/2}^{(m)}|^2 \right)
\; ,\;\; 
\mu_G^2 =
3 \left( -2 \sum _n \epsilon_n^2|\tau_{1/2}^{(n)}|^2 +
2 \sum_m \epsilon_m^2|\tau_{3/2}^{(m)}|^2 \right)
\, .
$$

Relations (\ref{p30}) are most easily obtained from  
the fact
that the SV amplitudes of transitions to the corresponding states are 
given by the 
overlap 
$\langle n(\vec{v})|B(\vec{v}=0)\rangle $ and by exploiting  Eq. 
(\ref{v7}) for $|B(\vec{v})\rangle $ from Sect. 6.1.

The sum rules (\ref{bj})--(\ref{fourth}) 
obviously entail a set of  exact QCD inequalities. The first one
is nothing but the Bjorken inequality $\varrho^2 > 1/4$  \cite{BJSR} 
which was already mentioned above. Others from this sequence are:
\beq
\La(\mu)  \geq 2 \Delta_1\left( \varrho^2(\mu)-\frac{1}{4}\right)\, ,
\label{IQ1}
\eeq
\beq
\mu_\pi^2(\mu) \geq \frac{3}{2} \Delta_1\La (\mu )\,\,\,\,
\mbox{and} \,\,\, \,  \mu_\pi^2(\mu) \geq 3\Delta_1^2 
\left( \varrho^2(\mu)-\frac{1}{4}\right)\, ,
\label{IQ2}
\eeq
\beq
\mu_\pi^2\;=\; \mu_G^2 + 9 \sum_n \:\epsilon_n^2\, 
|\tau_{1/2}^{(n)}|^2\;\geq\;\mu_G^2\; ,
\label{p31a}
\eeq
\beq
\rho_D^3(\mu) \geq  \Delta_1 \mu_\pi^2(\mu)\,\,\,\,
\mbox{and} \,\,\,\,  \rho_D^3(\mu) \geq \frac{3}{2} \Delta_1^2\La 
(\mu )
\,\,\,\,   \mbox{and} \,\,\,\,  \rho_D^3(\mu) \geq 3 \Delta_1^3
\left( \varrho^2(\mu)-\frac{1}{4}\right)\, ,
\eeq
\beq
\rho_D^3(\mu) = -\rho_{LS}^3(\mu)  + 9 \sum_n \:\epsilon_n^3\, 
|\tau_{1/2}^{(n)}|^2\;\geq \, -\rho_{LS}^3(\mu)\:,\;\; \mbox{and} \; \;
\rho_D^3(\mu) \ge  \frac{|\rho_{LS}^3(\mu)|}{2} \: .
\label{IQ3}
\eeq
Here $\Delta_1$ is the first excitation energy.
Strictly speaking, the lowest excitation of $B$ is $B\pi$;
the pions are coupled rather weakly, however. Moreover,
the $B\pi$ intermediate state becomes irrelevant in the limit
$N_c\ra\infty$. Therefore, for all practical purposes
we can equate $\Delta_1$ to the mass difference of $B^{**}$ and $B$,
i.e. $\Delta_1\approx 0.5\GeV$. 

It is  important that the above inequalities  
are based  only on  general  QCD relations and the heavy-quark 
limit: no recourse has been made to
model assumptions. In particular, dynamics of the the light cloud 
 is included in full and exactly. The inequality $\rho_D^3 -
(1/3)\rho_{LS}^3 > 0$ following
from Eqs. (\ref{asr1}) and (\ref{fourth}) is useful for the zero recoil
sum rule discussed in Sect. 8.2.3.

The values of $\tau_{3/2}$ and, in particular, $\tau_{1/2}$ are not well
known at present. For the lowest states
with $\epsilon \approx 0.5 \GeV$ they were evaluated in QCD sum rules.
According to \cite{bsrho}, $\tau_{1/2}^{(1)}\approx 0.25$;  
the similar estimate for $\tau_{3/2}^{(1)}$ is $\tau_{3/2}^{(1)} =
0.4\pm 0.1$ \cite{tau32}.
A model estimate \cite{NIMW} falls quite close, 
$\tau_{1/2}^{(1)}\approx \tau_{3/2}^{(1)}\approx 0.31$. 
Recently, the value $\tau_{3/2}^{(1)}\approx 0.35$ was derived
\cite{tau32exp} from the experimental information on the yield of the
excited charm states in semileptonic $B$ decays, yet also with sizeable
uncertainty.
Substituting $\tau_{1/2}^{(1)}\approx 0.25$ in Eq. (\ref{p31a}) we 
arrive at a stronger 
bound on $\mu_\pi^2$, quoted in Eq. (\ref{p32}) below. 

All these inequalities are {\em saturated} if it is only the first 
excitation that contributes to the sums. This is explicitly so in
the so-called  two-doublet model of Ref.~\cite{Koyrakh}. If $\mu$ is 
not
too large,  in the ballpark $0.7$ to $1 \GeV$, the actual situation
seems to be not too far from this model. (In the two-doublet model
the higher excited states are exactly dual to the perturbative
corrections).

Assigning to the  estimate of  
$\tau_{1/2}^{(1)}$ above an uncertainty
$\pm 30\%$,   one obtains from Eq. (\ref{p31a})
\beq
\mu_\pi^2(0.7\GeV) = (0.6\pm 0.15)\GeV^2\;. \label{p32} 
\eeq
Moreover, assuming 
that \footnote{The experimentally measured slope is centered around 0.9,
with the errors $\sim \pm 0.2$. To get $\varrho^2(\mu)$ from this 
slope one must eliminate
$1/m_c$ and hard perturbative corrections. According to estimates 
in \cite{neubpr,leib}, this amounts to subtracting $\sim 0.1$ from the 
experimental number.} $\;\varrho^2(\mu) = 0.8\pm 0.2$ 
and $\Delta_1 = 0.5 \GeV$ 
we get 
\beq
\overline\Lambda (\mu ) = 0.35\,\,\, \mbox{to} \,\,\, 0.75\,\,\, 
(0.5\,\,\, \mbox{to} \,\,\, 1)\, \mbox{GeV} \, ,
\eeq
\beq
{\mu_\pi^2}(\mu ) = 0.25\,\,\, \mbox{to} \,\,\, 0.56\,\,\, 
(0.38\,\,\, \mbox{to} \,\,\, 0.8)\, \mbox{GeV}^2\, ,
\eeq
\beq
{\rho_D^3}(\mu ) = 0.13\,\,\, \mbox{to} \,\,\, 0.28\,\,\, 
(0.2\,\,\, \mbox{to} \,\,\, 0.4)\, \mbox{GeV}^3\, .
\eeq
The first interval assumes complete saturation by the first excitation, 
while the interval indicated in parenthesis assumes $70 \%$ 
saturation. 

The exact QCD inequalities presented above take place at any $\mu$,
provided that all parameters refer to the Wilsonian formulation. 
At large $\mu$ they become non-informative 
since all dimensionful parameters (except $\mu_G^2$, $\rho_{LS}^3$) 
are dominated by the large 
perturbative 
pieces scaling like the  corresponding power of $\mu$. 

Concluding this section, let us stress again that the information 
encoded in the exact inequalities
is model-independent. 
It makes sense to impose the corresponding 
constraints in any analysis aimed at extracting the set of the 
fundamental parameters, say, from the semileptonic spectra. 
If a particular determination leads to results incompatible with these 
inequalities, one may be sure that something is wrong either with 
data or with the analysis.

In the practical applications discussed in the later sections we will
use, as a central value, $\mu_\pi^2=0.5\GeV^2$ normalized at $0.5$ to
$0.7\GeV$. 

\section{More on  $\mu_\pi^2> \mu_G^2$}

\renewcommand{\theequation}{5.\arabic{equation}}
\setcounter{equation}{0}

In view of importance of this particular 
inequality  a few additional comments are in order. 
Unlike  quantum-mechanical potential problems, in QCD 
the heavy quark
kinetic operator  is expressed in terms of the 
{\em covariant}
derivatives $\pi_j=-i D_j= -i\partial_j-A_j$ which contain the vector 
potential
$A_j$. The presence of $A_j$ 
leads to non-commutativity of different spatial 
components 
of the  momentum operator in the presence of the 
chromomagnetic field
$\vec{B}$,
\beq
[\pi_j,\pi_k]\;=\; iG_{jk}\;=\; - i\epsilon_{jkl} B_l\;.
\label{p20}
\eeq
This non-commutativity immediately leads to the lower bound on 
the 
expectation
value of $\vec{\pi}^{\,2}$ \cite{motion}, in full analogy with the  
uncertainty
principle in quantum mechanics. The simplest quantum-mechanical  
formulation was 
suggested in Ref.~\cite{volpp}, 
\beq
\matel{B}{(\vec\sigma \vec\pi)^2}{B} = 
\matel{B}{\vec{\pi}^{\,2}+\vec\sigma \vec B}{B} = 
\mu_\pi^2-\mu_G^2 \;>\;0\;.
\label{p21}
\eeq

This inequality  has a transparent physical interpretation. We deal 
here with  the Landau
precession of a colored, i.e. ``charged" particle in the 
(chromo)magnetic
field. Hence,  one has $\aver{p^2}\ge |\vec B|$. Literally
speaking, in the $B^*$ meson the 
quantum-mechanical 
expectation value of the chromomagnetic  field is suppressed, 
$\aver{B_z}=-\mu_G^2/3$. It 
completely vanishes  in the $B$ meson. However, the essentially
non-classical nature of $\vec{B}$ 
(e.g.  $\aver{\vec{B}^{\,2}} \ge 3
\aver{\vec{B}}^{2}$), in turn, enhances the bound  which then takes 
the
same form as in the  external classical field.
 
The SV sum rules \cite{vcb,optical}  translate the 
quantum-mechanical derivation of the above inequality to
 the field-theoretical language;
 the difference $\mu_\pi^2(\mu)-\mu_G^2(\mu)$ 
is represented  
as 
the imaginary part of the transition operator which is given by a sum 
of
certain transition probabilities. 

\section{Virial Theorem}

\renewcommand{\theequation}{6.\arabic{equation}}
\setcounter{equation}{0}

In this section we continue the discussion of the kinetic operator 
from 
a different perspective. 
As was mentioned, the expectation value of $\vec\pi^2$ in the $B$ 
meson was estimated in the QCD sum rules 
directly  \cite{pp}
and  \cite{neubp}
{\em via} the virial theorem. A strong discrepancy between the two 
determinations, which do not overlap within their respective error 
bars,
prompts us to revisit the issue. 
Two questions arise. A practical one is what result is more 
trustworthy 
and why. A  more theoretical problem is what went wrong with the 
evaluation 
of the uncertainties in the competing calculations? 

Clearly, only the authors themselves could answer these questions in
full. We would like to suggest, however, some qualitative 
insights
which, hopefully, may shed  light on the discrepancy and be helpful 
in
future refined treatments of the issue. We must start, however, with 
the
brief description of the virial relation itself.

\subsection{Alternative  derivations}

The kinetic and chromomagnetic matrix elements are defined via the
expectation values of the corresponding operators in $B$ meson.  In 
Ref.
\cite{MatNeu} a certain relation for the matrix elements  of $\bar
Q{\vec \pi}^2 Q$ was obtained involving a new, 
chromoelectric
operator  $\bar Qg\vec E Q$. The    transition matrix
element of  $\bar Qg\vec E Q$ between the states with {\em 
different} 
velocities was considered, and it was found that
$$
\frac{1}{2M_{H_Q}}\langle H_Q(v')| \bar Q i G^{\mu\nu} 
Q|H_Q(v)\rangle
=
$$
\beq
\frac{1}{3}\left( v^\mu v'^\nu - v^\nu v'^\mu \right) 
\left\{ \frac{1}{2M_{H_Q}}\langle H_Q(v)| \bar Q {\vec \pi}^2 
Q|H_Q(v)\rangle 
+{\cal O}(vv'-1)\right\} \, .
\label{virialth}
\eeq
This equation is valid for the spinless $H_Q$; for 
non-vanishing  spins it is valid after averaging over polarizations of
$H_Q$. In the rest frame of the initial hadron ($\vec{v}=0$) the
relation is nontrivial for $\{\mu,\nu\} = \{0,k\}$; then $G_{0k}=
- E_k^a t^a$ where $\vec E$ is the chromoelectric field. 
Then Eq.~(\ref{virialth}) takes  the form
\beq
\frac{1}{2M_{H_Q}}\langle H_Q(v')| \bar Q (i E^i) Q|H_Q(0)\rangle
=\frac{v'^i}{3}
\mu_\pi^2 + ...
\label{viriale}
\eeq
where the ellipses denote ${\cal O}({\vec v}'^2)$ terms
which will be systematically omitted.

Below several complementary pedagogical derivations will be 
presented. First, however, 
 a few words as to why Eq. (\ref{virialth})
is sometimes referred to as a virial theorem \cite{MatNeu}.
Assume that, instead of QCD with its messy dynamics of infinite 
number of degrees of freedom residing in the light cloud,
we consider a non-relativistic bound-state problem:
a heavy quark interacting with a lighter one {\em via}
a potential. Using the formalism of non-relativistic quantum 
mechanics
it is easy to rewrite the left-hand side of Eq. (\ref{viriale}) as follows:
\beq
\langle e^{-im_Q\vec v^{\prime}\vec x } ig  E^i  (\vec x)\rangle \;\ra\; 
 m_Q v'^j \langle  g  x^j E^i (\vec x)\rangle
=
m_Q v'^j \langle  g  x^j \: \nabla_i 
V(\vec x)\rangle\, ,
\label{virder}
\eeq
where the angle brackets denote averaging over the given 
bound state (in the rest frame), 
$V$ is the  binding  potential and, we restore 
the coupling constant contained in $G_{\mu\nu}$.
Equation (\ref{virder}) now 
takes the text-book form of the virial theorem, 
\beq
\langle\pi_k^2 \rangle = m \langle x_k \partial_k V(x) \, .
\rangle
\eeq
No sum over $k$ is implied here. (Note that in quantum 
mechanics $\aver{mx_k \partial_k V(x)}$ remains  the same being 
calculated for the heavy ``quark" $Q$ or the lighter one $q$.)

Equation  (\ref{virialth})  might  be helpful in various applications. It  
was  exploited, in particular,  in  studying the properties of the 
kinetic 
energy
operator under mixing  \cite{MN1,MN2}. 

The most elementary derivation of \eq{virialth} is as follows. 
If $\vec v = 0$ and $|\vec v '|\ll 1$ then 
$$
\matel{H_Q(v')}{\bar Q (\pi_0\pi_i - \pi_i \pi_0 )Q}{H_Q(v)}= 
\matel{H_Q(v')}{\bar Q (\pi_0-\vec\pi \vec v ' + \vec\pi \vec v ' ) 
\pi_i Q}{H_Q(v)},
$$
where all $\pi$'s act to the right, and we have used the equation
of motion neglecting the terms suppressed by $1/m_Q$. Moreover,
to the first order in $\vec v '$ the operators $\pi$ on the right-hand 
side can be considered as acting to  the left (the difference is the full 
derivative which is obviously ${\cal O}(|\vec v '|^2)$). 
When $\pi_0-\vec\pi \vec v ' $ acts to the left
it again produces zero due to the equation of motion and we are left 
with 
\beq
\matel{H_Q(v')}{\bar Q (\vec\pi \vec v ')\pi_i
 Q}{H_Q(v)}
 \label{der1}
\eeq
which is equivalent to 
Eq. (\ref{virialth}).

It is instructive to give a different derivation,
more directly related to the quantum-mechanical aspect of the 
problem. 
Consider
$$
\matel{H_Q(0)}{\pi_j\pi_k}{H_Q(0)} = \matel{H_Q(0)}{\pi_j\pi_0^{-
1}\, 
\pi_0\pi_k}{H_Q(0)} = 
$$
\beq
\matel{H_Q(0)}{\pi_j\pi_0^{-1}\,
(\pi_0\pi_k-\pi_k\pi_0)}{H_Q(0)}\;.
\label{v6}
\eeq
The operator $\pi_0^{-1}(\vec v\vec\pi)$ acting on $\state{H_Q}$ is 
nothing but
the generator of the boost along direction $\vec v$:
\beq
\state{H_Q(\vec v)}\;=\; \state{H_Q(0)} + 
\pi_0^{-1}\vec{v}\vec{\pi}\state{H_Q(0)}\;+\;{\cal O}(\vec v^{\,2})\;.
\label{v7}
\eeq
This is a useful relation nicely  elucidating the meaning of the 
small
velocity sum rules. It represents the  first-order perturbation theory 
in 
$\delta {\cal H}= \vec
v\vec \pi$; the unperturbed Hamiltonian is ${\cal H}_0=\pi_0$ 
(further  details can be found in \cite{optical}, Eq.\,(178) and 
Sect.~VI). 
In the second-quantized notations the very same relation  takes the 
form
\beq
\state{H_Q(\vec v)}\;=\; \state{H_Q(0)}\; +\;
\int\;d^3\vec{x}\:
\bar Q\pi_0^{-1}\vec{v}\vec{\pi}Q(x)\state{H_Q(0)}\;+\; 
{\cal O}(\vec v^{\,2})\;.
\label{v8}
\eeq
Keeping in mind that $\matel{H_Q(0)}{\pi_i}{H_Q(0)}$ vanishes and 
invoking the
definition 
\beq
[\pi^0,\pi^k]\;=\;iG^{0k}\;=\; - iE^k,
\label{E}
\eeq
we immediately rewrite \eq{v6} in the form 
\beq
\matel{H_Q(0)}{\bar Q
\vec v\vec \pi\, \pi_k Q}{H_Q(0)}\;=\; 
\matel{H_Q(\vec v)}{\bar Q iG_{0k} Q}{H_Q(0)} 
\label{v9}
\eeq
which is again the virial relation. 

The simple derivations above are formulated in terms of the 
on-mass-shell states. In some applications (e.g. the QCD sum rules),
rather than dealing with the on-mass-shell states, one works with 
the Green functions and amplitudes induced by  interpolating 
currents. For future applications we find it useful to formulate
yet another  derivation of the virial theorem based on the technique
of the Green functions. In this approach we consider
 the  three-point functions induced by appropriately chosen 
currents, and  then use the reduction formula. The relevant
correlators are treated in the background field method (for a
review see \cite{NSVZ}), 
although actually we use very little of the  background field 
formalism.
The advantage of this approach is its 
manifest field-theoretic nature. One operates with full QCD and takes 
the limit $m_Q\ra\infty$, as in Ref. \cite{Shuryak}. 

Consider the three-point function depicted in Fig.~2. 
The central vertex in this triangle is the operator $G^{0i}$. Two other 
vertices
are generated by the interpolating currents
\beq
J = \bar Q i\gamma_5 q \,\,\, 
\mbox{and}\,\,\, J^\dagger = \bar q i\gamma_5 Q\, ,
\label{J}
\eeq
where the pseudoscalar current is chosen merely  for definiteness. 
The 
sides of the triangle are  Green's functions of the quarks in the
background  gluon field. The reduction theorem tells us that  to get 
the 
transition amplitude  $\langle H_Q | \bar Q G^{0i} Q | H_Q \rangle $ 
from this  three-point function we  amputate it:   
multiply
by $(p^2-M_{H_Q}^2)$  and $(p'^2-M_{H_Q}^2)$ and  tend $p^2$  and 
$p'^2$
to $M_{H_Q}^2$.  This singles  out the double pole whose residue is
proportional to  the  meson-to-meson  transition on $G^{0i}$. 

\begin{figure}
\vspace{3.75cm}
\includegraphics{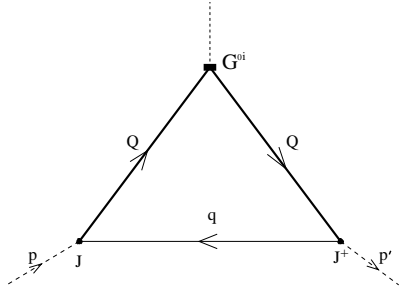}
\caption{
Three-point function relevant to the derivation of the virial 
theorem in QCD.}
\end{figure}

The expression for the three-point function of Fig.~2 has the form
\beq
i\mbox{Tr}
\left\{ i\gamma_5 \, \frac{1}{\not\! p' +\not\!\!{\pi} -m_Q} G^{0i}
\frac{1}{{\not\! p} +\not\!{\pi} -m_Q}
\,  i\gamma_5 \,  
\frac{1}{\not\!{\pi} - m_q}\right\}\, ,
\label{triangle}
\eeq
where $\pi$ is the momentum operator,  Tr implies taking trace
in the color, Lorentz and (abstract) momentum spaces, and $p$ and 
$p'$ are
external momenta flowing through the currents.   In the 
non-relativistic 
limit, keeping only terms linear in $\vec v'$, one can readily 
rewrite Eq.
(\ref{triangle}) as follows:
\beq
i\mbox{Tr} \left\{\left[
\frac{1}{(\varepsilon  +{\pi_0}) - \vec\pi\vec v'}\, G^{0i} \,
\frac{1}{(\varepsilon  +{\pi_0})}\right]
\, {i\gamma_5}\, \frac{\gamma_0+1 -\vec v'\vec\gamma}{2} \,
 \frac{1}{\not\!{\pi} - m_q} \,  
\frac{\gamma_0+1 
}{2}\, {i\gamma_5}
\right\}\,  ,
\label{nrtriangle}
\eeq
where $\varepsilon+m_Q$ is the time component of the external 
momentum
$p$ (see Fig.~2), and we set $\varepsilon ' =\varepsilon$,
since only the double pole of the type
\beq
\frac{1}{(\varepsilon -\bar\Lambda)^2}
\label{dpdeg}
\eeq
is of interest for  singling out the ground state
$H_Q$ from the tower of states created by the interpolating currents 
$J$.

Now, let us use the fact that 
\beq
G^{0i} = -i [\pi^0, \pi^i ] = -i \left\{ (\varepsilon +\pi _0)\pi^i
-
\pi^i (\varepsilon +\pi _0)\right\}\, .
\label{com}
\eeq
Substituting Eq. (\ref{com}) in Eq. (\ref{nrtriangle}) 
and expanding the denominator in $\vec v\,'$
we observe that the 
square bracket reduces to
$$
\left[
\frac{1}{(\varepsilon  +{\pi_0}) - \vec\pi\vec v\,'}\, \left( 
(\varepsilon 
+\pi_0)\pi^i
-
\pi^i (\varepsilon +\pi _0)\right)
\,
\frac{1}{(\varepsilon  +{\pi_0})}\right] \ra\,\,\, \mbox{terms with 
single pole} 
+
$$
\beq
\frac{1}{(\varepsilon  +{\pi_0})}\vec\pi\vec v\,'\, \pi^i
\frac{1}{(\varepsilon  +{\pi_0})}  +{\cal O}(\vec v\,'^2)\, .
\label{dp}
\eeq
In the limit $\varepsilon \ra \bar\Lambda$, the residue of the 
double pole 
(in the second line in Eq. (\ref{dp})) is just
$$
\frac{1}{3} v'^i \, \frac{1}{2M_{H_Q}} \langle H_Q | \bar Q {\vec \pi}^2 
Q|H_Q\rangle\, ,
$$
which proves Eq. (\ref{viriale}). Note that it is crucial to single
out the double-pole contribution.

By the same token, a similar consideration would lead us to Eq. (94) 
from Ref. \cite{optical}, and, 
eventually, to the sum rule (\ref{vol}). Consider, instead of the 
three-point function (\ref{triangle}), the two-point function
\beq
i\mbox{Tr}
\left\{ i\gamma_5 \, \frac{1}{{\not\! p} +\not\!{\pi} -m_Q}
\,  i\gamma_5 \,  
\frac{1}{\not\!{\pi} - m_q}\right\}\, ,
\label{diangle}
\eeq
with the external momentum 
$ p = m_Q v_\mu + \epsilon_\mu$. It is assumed that
$v_\mu = \{v_0, \vec v\}$ where $|\vec v | \neq 0$ but
$|\vec v | \ll 1$ and $\epsilon_\mu = \{\epsilon , \vec 0\}$. 
We then take the limit $m_Q\ra\infty$, expand in $\vec v$
keeping the terms quadratic in this parameter,  isolate
the double pole (\ref{dpdeg}) and compare the result with the 
phenomenological representation for the same two-point function. 
Equation (94) from Ref. \cite{optical} is immediately reproduced. 

\subsection{Virial theorem and QCD sum rules}

In this section we dwell on the problem of determining the kinetic 
energy 
from different QCD sum rules.

The technical reason of the existing discrepancy is quite evident: in 
the 
sum rule analyzed  by Ball {\em et al.}  \cite{pp} the free-quark
diagram  gives  a 
contribution in the theoretical side of the sum rule. This contribution  
is then   ``corrected" by ${\cal O}(\alpha_s)$ terms 
and
nonperturbative condensates, a  standard situation in the 
QCD sum
rules technology. On the other hand, Neubert suggests \cite{neubp} 
considering 
the virial partner of $\vec\pi^2$, and analyzing
the QCD sum rule for $\bar Q \vec E Q$ rather than for
$\bar Q \vec\pi^2 Q$. In this case 
 the free-quark contribution on the right-hand side is absent.
The theoretical side of the sum rules is solely due to interaction 
terms which must be kept  relatively
small, by necessity. Although such a regime is much less studied than
the standard one, some 
experience is still available. 
 
To get an idea of how  the QCD sum rules work in such problems
it is instructive to consider a toy model which was used for this 
purpose previously more than once \cite{MSO}. We mean   
illustrating 
the sum-rule approach in the three-dimensional harmonic oscillator. 

Assume 
that we  have an infinitely heavy ``quark" $Q$ and a 
lighter but still non-relativistic
``quark" $q$ with  mass $m\ll m_Q$, interacting 
{\em via} the harmonic oscillator potential,
\beq
V(\vec r ) = \frac{m\omega^2 {\vec r}^{\,2} }{2}
\eeq
where $\vec r$ is the distance between $Q$ and $q$.

Needless to say, this model is exactly solvable. We will pretend,
however, that we do not know it, and will calculate the matrix  
element
of interest using the sum-rule approach. Parallelizing  QCD we will
consider the expectation value of the kinetic energy and its virial
partner $\frac{1}{2}x_i\partial_iV(x)$.  Since the potential is 
quadratic, the virial
partner 
is the potential energy operator itself,
\beq
E_{\rm pot} = \frac{m\omega^2{\vec r}^{\,2}}{2} \, .
\eeq
The kinetic energy operator  in the model at hand has the form
\beq
E_{\rm kin} = \frac{{\vec p}^2}{2m} = -\frac{1}{2m}\Delta\, .
\eeq
As is well-known, the expectation values of the $E_{\rm pot}$
and $E_{\rm kin}$ are indeed always equal to each other in the 
harmonic
oscillator; for the ground state
\beq
\langle E_{\rm kin}\rangle_0 =
\langle E_{\rm pot}\rangle_0 = \frac{3\omega}{4}\, .
\eeq
Note that 
\beq
E_{\rm kin} + E_{\rm pot} = H
\eeq
where $H$ is the Hamiltonian of the system. 

The important element is that, like in QCD, the potential energy
vanishes for free particles and appears only when the interaction is
included. Thus,  its calculation is technically similar to the QCD sum 
rule for $\bar Q \vec E Q$. The kinetic energy, clearly, is nonzero 
even for 
free -- but moving -- quarks and, in this respect, is the counterpart
of $\bar Q \vec\pi^2 Q$.

Now, in the spirit of the sum-rule approach, we pretend that
the only quantities we are able to calculate reliably are the 
correlation functions at short (Euclidean) times. Namely, we consider 
the following amplitudes: 
at the initial moment of time the quarks are at zero separation,
at final (Euclidean) time $T=\tau_1 +\tau_2$ they are at zero 
separation again; at $\tau_1 $  the insertion of the operator
$E_{\rm kin}$ or $E_{\rm pot}$ is made. These amplitudes are 
perfect analogues of the Borel-transformed two- and three-point 
functions in the QCD sum rules.  

All relevant formulae are collected in Ref. \cite{bsrho}
where a toy-model calculation in a related problem is carried out.
The two-point function is
\beq
S(T) = K(\vec 0, T |\vec 0, 0)  = \left( 
\frac{m\omega}{2\pi}\right)^{3/2}
\frac{1}{(\sinh \omega T)^{3/2}}\, ,
\label{TPF}
\eeq
where $K(\vec r, \tau |\vec 0, 0)$ is the amplitude
(time-dependent Green function) of the propagation from the point
$(\vec 0, 0)$ to the point $(\vec r, \tau )$ in the Euclidean time,
$$
K(\vec r, \tau |\vec 0, 0) = \sum_n \psi^*_n (\vec 0)\psi_n (\vec r)
e^{-E_n\tau}\, =
$$
\beq
\left( \frac{m\omega}{2\pi\sinh \omega T}\right)^{3/2}
\exp\left(-\frac{m\omega}{2\sinh (\omega\tau )}{\vec r}^{\,2}\cosh
(\omega\tau )\right) \, .
\label{KHO}
\eeq
The three-point functions with the insertion of
$E_{\rm pot}$ and $E_{\rm kin}$ are
\beq
S_{\rm kin} = \int d^3\vec r K(\vec 0, \tau_1 +\tau_2  |\vec r, 
\tau_1)\, E_{\rm kin}\,  K(\vec r, \tau_1 |\vec 0, 0)\, ,
\eeq
plus the same expression with the insertion of $E_{\rm pot}$.
Following the standard routine of the QCD sum-rule practitioners we 
will analyze the sum rules at the symmetric point,
$\tau_1 =\tau_2 = T/2$. 
Using Eq. (\ref{KHO}) above it is easy to find that
\beq
S_{\rm kin} = \frac{3}{4}\omega S \frac{\cosh (\omega T) +1}{\sinh 
(\omega T) }\;\; , \;\;\;\;\;
S_{\rm pot} = \frac{3}{4}\omega S \frac{\cosh (\omega T) -1}{\sinh 
(\omega T) }\; ,
\label{QQ}
\eeq
where $S$ is defined in Eq.~(\ref{TPF}). 

Let us review how the sum rule technology works. As a classic 
example, we determine the ground state energy of the 
three-dimensional
oscillator.
To this end we will analyze two sum rules
 \beq
S = \sum_{n=0}^\infty \psi_n (0) \psi_n^* (0) e^{-E_n T} = \left 
(\frac{mT^{-1}}{2\pi}\right)^{3/2} \left( 1 -\frac{\omega^2 T^2}{4}
+ \frac{19 \omega^4 T^4}{480} + ...\right)\, ,
\label{2PFSR}
\eeq
and
$$
S_{\rm kin}+ S_{\rm pot} =\sum_{n=0}^\infty \psi_n (0) \psi_n^* 
(0)E_n  e^{-E_n T} =
$$
\beq
 \frac{3T^{-1}}{2}\left (\frac{mT^{-
1}}{2\pi}\right)^{3/2} 
\left( 1 +\frac{\omega^2 T^2}{12}
- \frac{19 \omega^4 T^4}{288} + ...\right)\, .
\label{3PFSR}
\eeq
Equation (\ref{2PFSR}) is a two-point function sum rule, Eq. 
(\ref{3PFSR}) is a three-point function sum rule with the insertion of 
the Hamiltonian. The right-hand sides in both expressions are given 
in 
the small-$T$ expansion since we pretend not to know the exact 
solutions. Note, that the leading terms in these expansions
are $\omega$-free; they correspond to the propagation of free 
``quarks"
and are analogous to the free-quark diangle and triangle diagrams in 
QCD. The subsequent terms are proportional to powers of $\omega$
and are analogous to the condensate corrections in the sum-rule
approach, and, indeed, can be easily calculated without the exact
solution at hand.

To single out the ground state contribution it is desirable to go to as 
high values of $T$ as possible since all higher states are 
exponentially suppressed at large $T$. We cannot go to the 
asymptotically large values of $T$, however, since the power 
corrections blow up. Upon  inspecting Eqs. (\ref{2PFSR}) and 
(\ref{3PFSR}) it becomes clear that if we want to keep the power
corrections under control (i.e.  the last correction kept 
to be less than $\sim 30\%$)
we can not go higher than
$T\sim 1.5 \omega^{-1}$. At such values of $T$ the suppression of 
the
higher states in the sum rules, although quite visible, is still not good 
enough.  Thus, the relative contribution of the first excited state in 
Eq. (\ref{2PFSR}) is $(3/2) e^{-2\omega T}$, while in Eq. 
(\ref{3PFSR}) its relative weight is $(7/2) e^{-2\omega T}$. 
We can see it in a different way: taking the ratio of Eqs.
(\ref{3PFSR}) to (\ref{2PFSR}) would produce $3\omega /2$
if all higher excitations were fully suppressed. Instead, we get
$\sim 0.75(3\omega /2)$ at $T\sim 1.5 \omega^{-1}$.

If we aim at better accuracy we have to have at least a rough idea of 
the contribution coming from the excited states. The standard 
strategy in the sum rule analyses 
is representing  all states higher than the ground state
by the free-quark approximation, starting from some effective 
continuum threshold $\epsilon_c$. The particular form of the 
corresponding 
spectral densities is determined by the free-quark term in the power 
expansion. In the sum rules (\ref{2PFSR}) and (\ref{3PFSR})
we have, respectively,
$$
\sum_{n=1}^\infty \psi_n (0) \psi_n^* (0) e^{-E_n T}
\approx \left( \frac{m}{2\pi}\right)^{3/2}
\frac{2}{\sqrt{\pi}} \int_{\epsilon_c}^\infty 
d{\cal E} {\cal E}^{1/2} e^{-{\cal 
E} T} 
$$
and
\beq
\sum_{n=1}^\infty \psi_n (0) \psi_n^* (0) E_ne^{-E_n T}
\approx \frac{3}{2} \left( \frac{m}{2\pi}\right)^{3/2} 
\frac{4}{3\sqrt{\pi}} \int_{\epsilon_c}^\infty 
d{\cal E} {\cal E}^{3/2} e^{-{\cal 
E} T} \, .
\eeq
Transferring the ``continuum" contribution to the right-hand side of 
the sum rules we arrive at 
$$
\psi_0 (0) \psi_0^* (0) e^{-E_0 T} = \left (\frac{m}{2\pi}\right)^{3/2} 
\left( \frac{2}{\sqrt{\pi}} \int_0^{\epsilon_c} 
d{\cal E} {\cal E}^{1/2} e^{-{\cal 
E} T}  -\frac{\omega^2 T^2}{4}
+ \frac{19 \omega^4 T^4}{480} + ...\right)
$$
and 
\beq
\psi_0 (0) \psi_0^* (0) E_0e^{-E_0 T} = \frac{3}{2}\left 
(\frac{m}{2\pi}\right)^{3/2} \left(
\frac{4}{3\sqrt{\pi}}\int_0^{\epsilon_c} 
d{\cal E} {\cal E}^{3/2} e^{-{\cal E} T}  +\frac{\omega^2 T^2}{12}
- \frac{19 \omega^4 T^4}{288} + ...\right)\,\, .
\eeq

Assume that our idea of the excited state contributions is roughly 
correct. If $T$ can be chosen large enough so that the higher states 
are sufficiently suppressed, on the one hand, and 
small enough to keep the power expansion in the right-hand side 
under control, on the other,  the ratio of the two
expressions must be close to $E_0= (3/2\omega)$. The interval of $T$ 
satisfying the above conditions is called window 
(fiducial domain). Of course, the constancy ($T$ independence) of the 
ratio 
is valid only to the extent one can neglect the higher-order power 
corrections and the error of  the continuum approximation. Thus, we 
expect the ratio to be approximately constant. The solid curve in 
Fig.~3 shows
the ratio, as a function of $T$ in the fiducial domain (the plot 
corresponds to $\epsilon_c =2.5 
\omega$; the value of the ``continuum threshold" can be fitted in its 
turn). We see a clear-cut plateau, and the height of the
plateau differs from the exact ground state energy ($E_0=3\omega 
/2$) by at most
$\pm 5\%$. 

\begin{figure}
\vspace{4.0cm}
\includegraphics{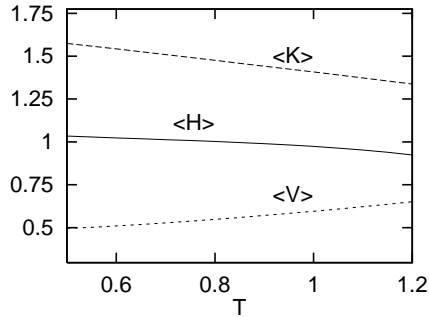}
\caption{
Sum rules for the three-dimensional harmonic oscillator.
The solid curve presents $\langle H\rangle$ (in units of 
$\frac{3}{2}\omega$)  vs. 
$T$. The dashed curve gives $\langle E_{\rm kin} \rangle$ in 
units of $\frac{3}{4}\omega$, the dotted curve  $\langle E_{\rm pot} \rangle$ 
in units of $\frac{3}{4}\omega$. The horizontal axis is $T$ in units of
$\omega^{-1}$. The exact results for the ground state energy, kinetic 
energy and potential energy are $\frac{3}{2}\omega$, $\frac{3}{4}\omega$,
$\frac{3}{4}\omega$, respectively.}
\end{figure}

So far the application of the sum rule technique is successful.
What happens, if instead of the sum of the kinetic and potential 
energy operators we would analyze them separately?

From the exact solution of the problem we know that the most
drastic change is the occurrence of the off-diagonal transitions in the 
three-point sum rules for $S_{\rm kin}$ and $S_{\rm pot}$. These 
off-diagonal
contributions are absent, by construction, in $S$ and are obviously 
absent in the sum rule for
$\langle H\rangle$, too, since the off-diagonal matrix elements of the 
Hamiltonian vanish. On the contrary, the off-diagonal matrix 
elements
of the operators $E_{\rm kin}$ and $E_{\rm pot}$ do not vanish and 
are numerically relatively large compared to the diagonal matrix 
elements.
For instance, $\langle {\vec r}^{\,2}\rangle_{01} = -2\langle {\vec 
r}^{\,2}\rangle_{00}$.  Moreover, if the first excited state appearing 
in 
$S$ or $\langle H\rangle$ has the excitation energy $2\omega$
(above the ground state),  the first excited state in $\langle E_{\rm 
kin}\rangle$ and $\langle E_{\rm pot}\rangle$ is separated from the 
ground state by only $\omega$, 
as explained in detail in Ref.~\cite{bsrho}.
The fact that the excited states lie closer to the ground state and
their relative weight is relatively large makes it harder to suppress 
them sufficiently/and or approximate them by the free-quark 
continuum.
The fiducial domain is bound to shrink.

A closer inspection shows that the situation is especially bad
in the sum rule for $\langle E_{\rm pot}\rangle$, since in this case 
the continuum simply vanishes in the free quark approximation.
This is obvious from the definition of the operator, and is clearly 
seen from the small-$T$ expansion of the second expression in Eq.
(\ref{QQ}). Following the standard strategy we would be forced
merely to neglect all higher states. The exact solution 
 shows \cite{bsrho} that the excited states contribution consists
of two components -- the diagonal transitions and off-diagonal ones 
-- and each component separately does not vanish in the free quark 
approximation and is large, but they have opposite relative signs.
For highly excited states the sign-alternating terms, after summation,
are smeared to zero, so this approximation of 
``no continuum" 
is not  bad. 
The first excitation, however, is too conspicuous
and is not annihilated by the second one. For values of $T$ up to 
$\sim 1.5\omega^{-1}$ the first excitation plays an important role in 
the sum 
rule, and effectively screens the ground state contribution. Accepting 
the ``standard" continuum model (i.e. no continuum in the case at 
hand) we significantly underestimate the expectation value
of $E_{\rm pot}$ in the ground state. Figure 3 illustrates the trend.
Assuming that the fiducial domain is the same as in the sum rule for
$S$ or $\langle H\rangle $ we get around 0.5 of the exact value.
As a matter of fact this means that the fiducial domain shrinks to 
zero in this case.

Analysis of the kinetic energy operator is better, but not much 
better. The diagonal and off-diagonal transitions are the same in the 
absolute value as above (this is obvious from Eq. (\ref{QQ})),
but the relative sign is now  positive.  The free quark approximation 
gives a non-vanishing continuum contribution. This is always good 
news.
The bad news is that the lowest lying excited state (the off-diagonal 
transition), even after smearing, is not well-described by
the free-quark curve.  On average, the free-quark curve falls below
the actual value of the (smeared) spectral density. To make a 
relatively precise prediction we would have to go to such high values 
of $T$ where the power expansion, truncated at a couple of the first 
terms, fails. If we stay inside the former fiducial domain, we with 
necessity overestimate the value of the kinetic energy in the ground 
state. The outcome of the sum-rule calculation includes a positive 
and rather significant contamination from the first excitation.
The extent of this contamination is clear from Fig.~3. Instead of
$3\omega /4$ we get roughly $1.5 \cdot 3\omega /4$. The sum is very
close to the exact result, $3\omega /2$, but since the divergences go 
in the opposite directions, the difference, instead of being zero in 
accord with  the virial theorem,  is quite large.

What lessons can be inferred for the QCD calculations? 
Although the toy model discussed above is far too simple to closely 
follow the QCD pattern, some features are still rather close.
The virial partner of ${\vec\pi}^2$ in QCD is $\vec E$.
It is quite obvious that the sum rule for $\vec E$
does not have the free-quark graph. In other words,
just like in the case of $\langle E_{\rm pot}\rangle$
the free-quark continuum vanishes \cite{neubp}.
This was believed to be a positive feature of the analysis. 
We see now that it is not. We expect a large physical spectral density
in close proximity to the ground state contribution
(caused by off-diagonal transitions), and of the 
opposite sign. To suppress it one would need to go to such high 
values of the Borel parameter where -- alas -- we cannot go without 
losing control over the power expansion. The screening of the 
ground 
state contribution by  the off-diagonal transitions results in
an underestimated value of $\langle g\vec E\rangle$,
which through the virial theorem leads to an underestimated value 
of ${\vec \pi}^2$.

At the same time, and for the same reason, the calculation of 
Ref.~\cite{pp}, which follows the standard pattern, is expected to give
an overestimated prediction for ${\vec \pi}^2$.
Whether the genuine value of $\mu_\pi^2$ is closer to the one edge 
of the interval or to the other -- we do not know. It seems clear, 
however, 
that the error bars in Eqs. (\ref{p10}) and (\ref{NNR})
reflect only  uncertainties in various parameters entering the sum 
sum-rule calculations and  do {\em not} reflect the fact that
the continuum saturation in the problem at hand is likely to 
deviate from the standard pattern.\footnote{The authors of 
Ref.~\cite{pp} estimated the uncertainty of the continuum model
reflected in Eq.~(\ref{p10}) at the 
$30\%$ level, i.e. $0.15\GeV^2$.}

The actual value of
${\vec \pi}^2$ lies, most probably, somewhere in between the two  
estimates.
Unfortunately,  the potential interaction is strictly speaking 
non-existent in QCD, let alone the quadratic potential of the  
oscillator model. 
Therefore,  mechanically averaging the estimates
(\ref{p10}) and (\ref{NNR}) is unlikely to  suppress the 
off-diagonal
transitions and yield a more reliable approximation to the genuine
value of $\mu_\pi^2$. 
In no way do we suggest such an averaging.
It is clear that more sophisticated  models for the off-diagonal 
transitions 
are needed in both calculations. Once we realize that
the standard continuum approximation is inadequate, developing a 
better one does not seem to be a hopeless problem.

\vspace*{.2cm}

In summary, 
it seems safe to say, that the result obtained in Ref. \cite{neubp}
significantly underestimates the kinetic operator.
The sum rule technology for 
$\vec\pi^2$ seems to operate in a somewhat better environment 
than that in the case of $\vec E$. This conclusion is indirectly 
supported by the fact that  the alternative 
analysis relying on the exact QCD inequalities (plus the data on the 
slope of the Isgur-Wise function and the observed spectrum of 
the
excited states) favors larger values of $\mu_\pi^2$. 

\section{OPE for Inclusive Weak Decays}
\renewcommand{\theequation}{7.\arabic{equation}}
\setcounter{equation}{0}

The number of applications of the heavy quark expansion 
dramatically grows
when one includes external (non-QCD) interactions of the heavy 
quark, e.g. electromagnetic, weak and so on. A variety of problems 
can be formulated in the language of the OPE -- inclusive heavy 
quark decays are the best-studied example of this kind. 
In this case, the operator product expansion is built in a slightly 
different way. The expansion parameter is not necessarily
$1/m_Q$, but rather it is regulated by the energy release in the 
problem at hand, or by other external parameters. 
The general consideration runs parallel to the treatment of  $\sigma
({\rm e^+e^-}\ra \mbox{ hadrons})$. One describes the decay rate 
into an 
inclusive final state $f$ in terms of the imaginary part of a  forward
scattering operator (the so-called transition operator)  evaluated to
second order in the  weak interactions \cite{SV,CHAY}
\begin{equation} 
\Im \hat T(Q\rightarrow f\rightarrow Q)\;= \;
\, \Im \int d^4x\ i\,T \left({\cal L}_W(x){\cal 
L}_W^{\dagger}(0)\right)\ 
\label{OPTICAL} 
\end{equation} 
where $T$ denotes the time ordered product and 
${\cal L}_W$ is the relevant weak Lagrangian at the normalization 
point higher or about $m_Q$.  The space-time separation 
$x$ in
Eq.~(\ref{OPTICAL}) is fixed by the inverse energy release. If the 
latter
is 
sufficiently large in the decay, one can express the {\em non-local} 
operator
product in  Eq. (\ref{OPTICAL})  as an infinite sum of {\em local}
operators $O_i$ of increasing  dimensions.  
The width for $H_Q\rightarrow f$ is then  obtained by  averaging  
$\Im \hat T$ over the heavy-flavor hadron $H_Q$,
$$
\frac{\matel{H_Q}{\Im \hat T (Q\rightarrow f\ra 
Q)}{H_Q}}{2M_{H_Q}} \propto 
\Gamma (H_Q\rightarrow f) = 
$$
\beq
G_F^2 |V_{\rm CKM}|^2m_Q^5   
\sum _i  \tilde c_i^{(f)}(\mu ) 
\frac{\matel{H_Q}{O_i}{H_Q}_{\mu }}{2M_{H_Q}}    
\label{OPE}
\end{equation} 
with $V_{\rm CKM}$ denoting the 
appropriate combination of the CKM parameters. A few comments 
are 
in order to elucidate the content of Eq. (\ref{OPE}).

(i) The parameter $\mu$ in Eq. (\ref{OPE}) is the normalization point, 
indicating that we explicitly evolved from $m_Q$ down to $\mu$. 
The
effects of momenta {\em below}  $\mu$  are lumped into the matrix 
elements of the operators $O_i$. 

(ii)  
The coefficients $\tilde c_i^{(f)}(\mu )$ are dimensionful, they 
contain powers of 
$1/m_Q$ that go up with the dimension of the operator $O_i$.  
Sometimes it is convenient to introduce dimensionless coefficients 
$c_i^{(f)}(\mu ) = m_Q^{d_i-3}\tilde c_i^{(f)}(\mu )$ where $d_i$ 
denotes the dimension of the operator $O_i$.
The dimensionless coefficients 
$c_i^{(f)}$ depend on 
the ratio of final- to initial-state quark masses.
Using the normalization introduced in Eq.~(\ref{OPE}), one obtains 
on dimensional grounds 
$$
\tilde c_i^{(f)}(\mu ) \frac{1}{2M_{H_Q}} 
\matel{H_Q}{O_i}{H_Q}_{(\mu )} 
\sim 
{\cal O} \left( \frac{\Lam^{d_i-3}}{m_Q^{d_i-3}},\;\frac{\as
\mu^{d_i-3}}{m_Q^{d_i-3}} \right) 
$$
with $d_i$ denoting the dimension of operator $O_i$.
The contribution from the  lowest-dimensional operator obviously 
dominates  in the limit $m_Q \to \infty$.   

(iii)  
It seems natural then that the expansion of total rates can be given 
in powers of $1/m_Q$. 
The master formula (\ref{OPE}) holds for a host of different  
integrated 
heavy-flavor decays: semileptonic, nonleptonic and radiative 
transitions, 
CKM-favored or suppressed, etc. For semileptonic and nonleptonic 
decays, treated through order $1/m_Q^3$, it takes the 
following form: 
$$
\Gamma (H_Q\ra f)=\frac{G_F^2m_Q^5}{192\pi ^3}|V_{\rm 
CKM}|^2\times
$$
$$
\left[ c_3^{(f)}(\mu )\frac{\matel{H_Q}{\bar 
QQ}{H_Q}_{(\mu)}}{2M_{H_Q}}
+ c_5^{(f)}(\mu ) m_Q^{-2}
\frac{
\matel{H_Q}{\bar Q\frac{i}{2}\sigma G Q}{H_Q}_{(\mu )}}
{2M_{H_Q}\;\;}+ \right. 
$$
\begin{equation}  
\left. +\sum _i c_{6,i}^{(f)}(\mu )m_Q^{-3}\frac{\matel{H_Q}
{(\bar Q\Gamma _iq)(\bar q\Gamma _iQ)}{H_Q}_{(\mu )}}
{2M_{H_Q}\;} + {\cal O}(1/m_Q^4)\right]  \, .
\label{WIDTH} 
\end{equation} 

We pause here to make a few  explanatory remarks on   this 
particular   
expression.
First, the main statement of OPE is that there is {\em no} 
correction of order $1/m_Q$ \cite{buv}.  This is particularly  
noteworthy because the hadron masses, which control the phase 
space, 
do contain such a correction:   
$M_{H_Q} = m_Q \left( 1 + \bar \Lambda /m_Q + 
{\cal O}(1/m_Q^2)\right)$; the parameter $\La$, different for
different hadrons, does not enter the width!
The reason for the absence of the  $1/m_Q$ correction in the total 
widths is two-fold: the corrections to the expectation value of the
leading QCD operator $\bar Q Q_{(\mu)}$ is only $\sim 
\mu^2/m_Q^2$, and 
there is no independent QCD operator of dimension $4$ for forward 
matrix
elements. Since the coefficients functions are purely short-distance,
infrared effects neither can penetrate  into them. The absence of the 
dimension-4 operator in  HQET was noted in \cite{CHAY,luke}.

A physically more illuminating way to realize the absence of 
corrections of
order $1/m_Q$ is to realize that the bound-state 
effects in the {\em initial} state (mass shifts, etc.) do 
generate corrections of order $1/m_Q$ to the total width -- 
as does hadronization in the final state. Yet local 
color symmetry demands that they cancel against each other, 
as can explicitly be demonstrated in simple models. 
It is worth realizing that this is a peculiar feature of QCD
interactions -- other dynamical realizations of strong confining forces
would, generally, destroy the exact cancelation.

Second, the  leading nonperturbative  corrections are $\sim {\cal
O}(1/m_Q^2)$, i.e. small in  the total decay rates for beauty hadrons: 
$(\mu /m_b)^2 \sim {\rm few} \, \% $ if $\mu \lsim 1 \GeV$. The 
first
calculation of the leading nonperturbative corrections in the decays 
of
heavy flavors was done in \cite{buv,bs,dpf,prl}.

Third, the four-quark operators  $(\bar Q\Gamma _iq)(\bar 
q\Gamma 
_iQ)$ depend
explicitly on the light-quark flavors denoted by $q$. They, 
therefore, 
generate differences in the weak transition rates for the  different
hadrons of a given heavy flavor.\footnote{Expanding
$\langle H_Q|\bar Q i \sigma G Q|H_Q\rangle /m_Q^2$
also yields contributions of order $1/m_Q^3$; those are, however,
practically insensitive to the light quark flavors.}
Their effects were calculated already in
mid-eighties \cite{vsold}. 

Fourth, the  short-distance coefficients $c_i^{(f)}(\mu )$ 
in practice are 
 calculated in perturbation theory. However it is quite 
conceivable that certain nonperturbative effects arise also in the 
short-distance regime. They are believed to be rather small in 
beauty decays \cite{inst}.
 
Fifth, a new matrix element appearing in  OPE, not discussed so far,  
is  the scalar heavy quark density. Its expansion originally 
established in Ref. \cite{buv} takes the form 
\beq 
\matel{H_Q}{\bar QQ}{H_Q} = \matel{H_Q}{\bar Q\gamma_0 Q}{H_Q} + 
\frac{\matel{H_Q}{\bar Q\left(\pi^2+\frac{i}{2}\sigma G \right) 
Q}{H_Q}}{2m_Q^2}+
{\cal O}(1/m_Q^4)\;\;.
\label{QQBAR} 
\eeq
Since $\matel{H_Q}{\bar Q\gamma _0Q}{H_Q}=2M_{H_Q}$, the 
spectator
ansatz indeed emerges  as the asymptotic scenario universal for all 
types
of hadrons,  and holds up to $1/m_Q^2$ corrections. In addition to
$\bar Q \vec\pi^{\,2} Q$, the second 
dimension-five operator is  the chromomagnetic 
operator
$\bar Q i\sigma G Q$. 
Since $\bar Q \vec D^{\,2} Q$  is  not a  Lorentz scalar, it does not
appear independently in   Eq.~(\ref{WIDTH}).

Equation (\ref{QQBAR}) is readily obtained in  the heavy quark 
expansion
if one uses  proper non-relativistic heavy-quark spinors 
incorporating
the Foldy-Wouthuysen  transformation. In the context of HQET this  
was
suggested   in Ref. \cite{korner}. When the Foldy-Wouthuysen
transformation is ignored, the  straightforward evaluation of the 
scalar
density in  HQET leads to  incorrect $1/m_Q^2$ terms.   Recovering 
the
additional terms required  a certain revision of the standard HQET 
strategy \cite{falkls,mawise}. At present, the
full-QCD derivation and that based on HQET are in perfect agreement. 
We thus have, for example, for the pseudoscalar mesons, 
\begin{equation}
\frac{1}{2M_{P_Q}}\matel{P_Q}{\bar QQ}{P_Q}=1 - 
\frac{\mu _{\pi}^2}{2m_Q^2}+ 
\frac{3}{8} \frac{M_{V_Q}^2-M_{P_Q}^2}{m_Q^2}+ 
{\cal O}(1/m_Q^3)
\label{QQBARPQ}
\end{equation}   
The reason for the kinetic operator term to appear is quite 
transparent.
The  first two quantities on the right-hand side of the equation  
represent the
mean value of the factor  $\sqrt{1-\vec v^2}$ reflecting the time
dilation which slows  down the  decay of the quark $Q$ moving 
inside
$H_Q$ \cite{prl,WA,Ds}. 

Finally,  Eqs. (\ref{WIDTH})--( \ref{QQBAR}) show that the two   {\em
dimension-five} operators do produce differences in $B$ versus
$\Lambda_b/\Xi_b$ versus $\Omega_b$ decays of order  $1/m_Q^2$. 
To a small extent they can also differentiate $B$ and $B_s$ via the
$SU(3)$ breaking in their expectation values. Differences in the 
transition rates inside the  meson family  are generated at order 
$1/m_Q^3$ by 
{\em
dimension-six} four-quark operators. They are usually estimated in 
the
vacuum saturation approximation which -- although cannot be exact 
--
represents a reasonable starting approximation. There is an 
intriguing way to
check factorization  experimentally \cite{WA}: similar four-fermion 
operators 
enter
semileptonic $b\ra u$ transition rates. Moreover, in the heavy quark 
limit the four-fermion operators  populate mainly the transitions into 
the hadronic states 
with
low energy, and thus show up, for example, in the end-point domain 
of the 
lepton spectrum where their relative effect is enhanced. Considering 
the
difference of the decay characteristics of  the  charged and neutral 
$B$'s in
this domain, one can measure these matrix elements and even feel 
their
scale dependence. Further details regarding the ``flavor-dependent"
preasymptotic effects can be found in \cite{BELLINI}.

\begin{figure}
\vspace{3.2cm}
\includegraphics{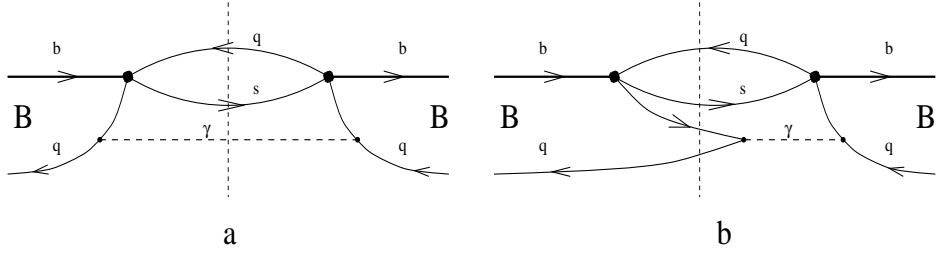}
\caption{
{\bf 
a)} Spectator effect leading to $1/m_b$ correction in the decay 
width $b\rightarrow  s +\gamma $.
{\bf b)} A different cut of the same diagram leads to an electromagnetic 
correction in the hadronic decay width. Terms $1/m_b$ and  $1/m_b^2$ 
cancel out in the sum of two decay widths. The solid dot in the vertex denotes
the penguin-induced $ b\rightarrow s \bar q q$ interaction.  }
\end{figure}

It is important to keep in mind that the OPE approach discussed in 
this
section implies that {\em all} decay channels induced by 
a
given term in the short-distance Lagrangian in Eq.~(\ref{OPTICAL}) 
are
included. It
is not  enough to consider the states that  appear at the
free-quark
level. The final-state interactions can annihilate, for example, the
$\bar{c} c$ quark pair into light hadrons; electromagnetic interaction,
if considered, can do the same. Disregarding the channels that can
emerge due to such final state interactions can violate the general
theorems: even $1/m_Q$ terms can appear in such incomplete
``inclusive" widths. For example, the correction to 
the $b\ra s+\gamma$ width does have nonperturbative 
corrections 
scaling like $1/m_b$ due to the effect of weak annihilation (in 
mesons,
or weak scattering in baryons) of the
light quarks with emission of a hard photon in the 
penguin-induced weak
decay $b\ra s\bar{q}q$, see Fig.~4a. This effect would only cancel
against the (virtual) {\em electromagnetic} correction to the hadronic
penguin-induced width, Fig.~4b. \cite{mirage,WA}.

\section{Applications of Heavy Quark Expansion}

\renewcommand{\theequation}{8.\arabic{equation}}
\setcounter{equation}{0}

If in the first part of the review we were more concerned with the 
theoretical foundations, here we pass to applied problems. 
The number of applications is quite large, and continues to grow. 
We will limit ourselves to those applications
where the heavy quark expansion is exploited as a tool
for extracting such fundamental parameter as 
$|V_{cb}|$, the element of the CKM matrix. Below we dwell on two 
methods often briefly called inclusive and exclusive approaches.

\subsection{$|V_{cb}|$ from the total semileptonic $B$ width}

The idea of extracting $|V_{cb}|$ from the inclusive semileptonic
transition rate $b\ra c \ell\nu$ is quite evident.
If $m_b\ra\infty$ the parton formula becomes exact.
Comparing the decay rate in this limit to the theoretical expression 
one could determine $|V_{cb}|$.
In the real world, where $m_b$ is finite,
analysis of various pre-asymptotic corrections, which became 
possible after the advent of the heavy quark expansion, is needed.

Dimensional arguments tell us that the total decay width scales as
the 
fifth power of  mass, but not 
which mass it is. {\em A priori} possible choices would be the heavy 
flavor  quark or  hadron masses which agree 
to leading order,
$(M_{H_Q} - m_Q)/m_Q \sim {\cal O}(\bar \Lambda /m_Q)$. 
However, in the 
case of beauty $M_B^5 \approx 1.5  m_b^5$. This difference, 
which would 
represent a power-suppressed nonperturbative correction,  
has to be brought under theoretical control if one wants to 
extract reliable numerical values for 
the CKM parameters from the measurement of the total 
semileptonic $B$ width.  In particular, the theoretical 
uncertainties
intrinsic to the pole masses must be
exterminated.

The  heavy quark expansion yields an unambiguous answer. Not 
surprisingly, everything is expressed in terms of the {\em quark} 
masses \cite{buv,dpf,prl}:
\beq
\Gamma_{\rm sl}(B) \! = \! \frac{G_F^2 m_b^5 |V_{cb}|^2 }
{192\pi^3} \left\{z_0
\left(1\!-\!\frac{\mu_\pi^2\!-\!\mu_G^2}{2m_b^2} \right)\!
- \! 2\left(1\!-\!\frac{m_c^2}{m_b^2} \right)^4
\! \frac{\mu_G^2}{m_b^2}
\!- \! \frac{2\as}{3\pi} z_0^{(1)} \!+ \!...\!
\right\}
\label{SLGAMMA}
\eeq
where ellipses stand for higher order perturbative and/or power 
corrections;
$z_0,\, z_0^{(1)}$ are known parton phase space factors depending on 
$m_c^2/m_b^2$ (see, e.g. \cite{bbbsl}). The $1/m_b^2$ power 
corrections to $\Gamma_{\rm sl}$ are rather small, about $-5\%$, 
and thus 
the
direct impact of the higher-order power corrections is negligible. 
One obviously has to be careful 
in dealing with the quark mass  in Eq. (\ref{SLGAMMA})  
since it  enters in a high power. Our strategy is to first 
present a complete theoretical formula which explicitly exhibits
theoretical uncertainties. Then we inspect 
the  extracted value   for $|V_{cb}|$.  
Finally, we  detail the criticism that sometimes is raised
against such an analysis. 

The first-order perturbative correction term
$z_0^{(1)}\left(m_c^2/m_b^2\right)$ is essentially the same as for 
$\mu\ra e\bar{\nu} \nu$ known since the mid-fifties \cite{muon}. 
Recently the QCD calculation was improved by computing the 
second-order \cite{wise} and all-order \cite{bbbsl} corrections 
arising
due to the running of $\as$ in the first-order diagrams (i.e. the
BLM part of $\as^k$  corrections). These are presumably dominant 
terms. They have a small numerical impact on the width, though. For 
completeness, 
the
second-order BLM terms are included below.      

Evaluating Eq. (\ref{SLGAMMA}) we find
$$
|V_{cb}|=0.0419\left(\frac{{\rm BR}(B\rightarrow 
X_c\ell\nu)}{0.105}
\right)^{\frac{1}{2}}\left(\frac{1.55\,\rm 
ps}{\tau_B}\right)^{\frac{1}{2}}
\cdot \left(1-0.012\frac{(\mu_\pi^2-0.5\GeV^2)}{0.1\,\rm
GeV^2}\right)
\times
$$
\beq
\left(1-0.01\frac{\delta m_b(\mu)}{50\,\rm MeV}\right)
\left(1+0.006 \frac{\as^{\overline{\rm MS}} (1\GeV)-
0.336}{0.02}\right)
\left(1+0.007\frac{\bar\rho^3}{0.1\,{\rm GeV}^3}\right).
\label{w12}
\end{equation}
Here we have relied on the mass $m_b(\mu)$, see  Eqs.~(\ref{m28}) 
--
(\ref{m29}),  
and on the value of the low-energy coupling 
$\as^{\overline{\rm MS}} (1\GeV)\simeq \as^{V}(2.3\GeV) = 0.336$ 
obtained in the dedicated analysis of
\cite{volmb}. The charmed quark mass $m_c(\mu)$ is then obtained 
using the relation 
(\ref{m30}) which introduces also the parameter $\bar \rho$. 
Furthermore, 
$\bar \rho ^3$ reflects the dependence on the $1/m_Q^3$ terms.
Electroweak corrections are still neglected.
It is worth noting that all expressions for 
the perturbative
coefficients depend on the concrete definition of mass and
relevant  operators. This dependence propagates into numerical 
expressions, 
which,
however, give the same final result for observables for 
commensurate input. The numbers shown above correspond to the 
definitions 
discussed at length  in the preceding sections.

 The proper evaluation of  
the theoretical uncertainties is a real challenge which we now 
discuss. 
The last four terms in parentheses in Eq. (\ref{w12})
exhibit some of the uncertainties; others will be considered shortly. 

(i)     As was mentioned, the 
perturbative corrections are known exactly to first order in 
$\as$.
Their impact on $|V_{cb}|  \sim \left(\Gamma_{\rm sl}^{-
1/2}\right)_{\rm
theor}$ is only 
about
$5\%$. The (presumably  dominant) BLM part of higher-order 
corrections
is now known to all orders; already in  the  second order the impact 
is
less than $1\%$. The only remaining uncertainty,  thus, is  the 
genuine,
non-BLM second-order corrections. They are not known completely, 
but
there are good reasons to believe that  they are indeed small. 

There 
exist 
some enhanced higher-order corrections not related to the running of
$\as$, that are specific to the inclusive widths \cite{five}. They
appear if one uses the masses  normalized at a high scale $\mu 
\gsim
m_b$.  They have been accounted for in the analyses 
\cite{vcb,upset},
but were not included  in Refs. \cite{bn,bbbsl}. 

In the SV limit, 
when 
$$
\xi\equiv (m_b-m_c)/(m_b+m_c) \ll 1 \, ,
$$
 the perturbative 
corrections to
the widths can be related \cite{look} to  the zero-recoil
renormalization of the axial current, which was recently calculated 
by
Czarnecki \cite{czar} to two loops. In this way one obtains for the 
second-order
non-BLM coefficient $a_2^{0}$ 
\beq
a_2^{0}(\xi)\;=\; \frac{85}{36}-\left(\frac{5}{2}-
\ln{2}\right)\frac{\pi^2}{9}-
\frac{\zeta(3)}{6}\; +\; {\cal O}\left(\frac{\mu^2}{m_c m_b},\, \xi^2 
\right)\; 
\simeq\; 0.179\; +\; {\cal O}\left(\frac{\mu^2}{m_c m_b},\, \xi^2 
\right)
\label{w10} 
\eeq 
where $a_i$ are defined in \eq{w31} below. If $\xi =0$ it is clear 
that  
$a_2^{0}(\xi)$  is rather small numerically. At the
actual value of $\xi  \approx 0.5$ the coefficient $a_2^{0}$ is
different, of course, but hardly can be large if the  appropriate value 
of $\mu$ 
is used. For example, the 
second-order non-BLM term in ${\rm
d}\Gamma_{\rm sl}/{\rm d} q^2$ at $q^2=q^2_{\rm max}=(m_b-
m_c)^2$
constitutes approximately $0.1 (\as/\pi)^2$ for the actual quark 
masses,
according to \re{czar}. The same correction to ${\rm
d}\Gamma_{\rm sl}/{\rm d} q^2$ at $q^2=0$ was recently evaluated 
 $\approx 1.25 (\as/\pi)^2$ in \re{czarmel}, thus 
demonstrating 
that 
the
non-BLM effects are moderate.
Altogether, we assess a $\pm 2\%$  uncertainty in 
$\Gamma_{\rm
sl}$ due to higher-order perturbative corrections, as a conservative
estimate. There is an additional uncertainty associated  with the
experimental error bars in $\alpha_s$, but it is minor.

(ii)     
The power corrections in $\Gamma_{\rm sl}$ scaling like
$\Lam^3/(m_b-m_c)^3$, are at the $1 \%$ level,  i.e. tiny. They 
emerge 
from several sources. First,  there are $1/m_b$ corrections in the 
 expectation values of the
kinetic and chromomagnetic operators over the actual $B$ meson 
state. They are  small
and, in any case, covered by the existing uncertainties in the 
values of the condensates. 
The ``genuine" corrections are given by the local  Darwin term 
$\rho_D^3$
and the ``spin-orbital" expectation value $\rho^3_{LS}$ \cite{optical}
appearing in Eq.~(\ref{hamil}). The latter, in turn,  appears  also  as
a (local) $1/m_b$ correction in the expectation value  of 
$\matel{B}{\bar b\frac{i}{2}\sigma G b}{B}$ which enters the 
$1/m_b^2$ 
term in  the widths. Thus, this effect is also negligible. It is worth
noting that  $\rho^3_{LS}$ is expected to be suppressed in the
ground-state mesons \cite{optical}.

The largest direct effect is due to the Darwin term; relying 
on the
approximation $m_c^2/m_b^2 \ll 1$ and using the calculations of 
Refs.~\cite{bds,grekap} one arrives at the 
estimate~\footnote{There is a minor disagreement in the relevant
expressions in the two works. Numerically it is negligible.}
\beq
\Gamma_{\rm sl} (b\ra c)\; \propto \; 
\left( z_0 - 0.01\frac{\rho_D^3}{0.1\,{\rm GeV}^3}
\right) \; \simeq \; 
z_0\left( 1 - 0.02\frac{\rho_D^3}{0.1\,{\rm GeV}^3}\right) \, .
\label{w40}
\eeq

Finally, if $m_c$ is related to $m_b$ {\em via} the  mass formula 
(\ref{m30}) -- so far, the most accurate method -- one has an 
indirect 
dependence on $\rho_D^3$ and $\bar\rho^3$ through $m_b-m_c$. 
Even though the
value of $\rho_D^3$ is reasonably well estimated, the overall
dependence on it practically cancels out in $\Gamma_{\rm sl}$
in this method. As a result, the  extracted value of $|V_{cb}|$ is 
sensitive only to non-local  correlators
$\rho^3_{\pi\pi}$ and $\rho^3_S$
as far as the 
$1/m^3_Q$ terms are concerned. 
Both $\rho^3_{\pi\pi}$ and $\rho^3_S$ are positive, see 
\cite{optical}. The corresponding shift in $\Gamma_{\rm sl}$ is thus,
probably,  negative but below $2\%$.

(iii)    
The calculation of the width could be affected by violations of 
duality,
which are conceptually related to the asymptotic nature  of the OPE
expansions. 
This is the least understood ingredient of the theoretical analysis. 
Some model considerations are given in Ref. \cite{inst}.
With  the energy release in the  
semileptonic
widths $\gsim 3.5\GeV$, the duality violation  is expected to be 
negligible, below $1\%$ level, and not to 
exceed the effect of $1/m^3_Q$ corrections.

Assembling all pieces together and 
assuming, conservatively, that at present $|\delta m_b| < 50\MeV$ 
and 
$|\bar\rho| <0.1\GeV^3 $, we arrive at the model-independent 
estimate
$$
|V_{cb}|=0.0419
\left( \frac{{\rm BR}(B\rightarrow X_c\ell\nu)}{0.105}
\right) ^{\frac{1}{2}}
\left( \frac{1.55\,\rm ps}{\tau_B}\right) ^{\frac{1}{2}}
\times
$$
\beq
\left( 1-0.012\frac{(\mu_\pi^2-0.5\,\rm GeV^2)}{0.1\,\rm
GeV^2}\right)
\cdot \left( 1 \pm 0.015_{\rm pert} \pm 0.01_{m_b} \pm 
0.012
\right) \, ,
\label{w20}
\end{equation}
where the last error reflects $m_Q^{-3}$ and higher
power corrections, including possible deviations from duality.

\subsubsection{$\Gamma (B\ra X_u \ell \nu )$}

Similar to the treatment of $\Gamma (B\ra X_c \ell \nu )$  it is
straightforward to relate the value of $|V_{ub}|$ to the total
semileptonic width $\Gamma(B\ra X_u\, \ell\nu)$ 
\cite{upset}:
\beq
|V_{ub}|=0.00465\left(\frac{{\rm BR}(B^0\rightarrow 
X_u\ell\nu)}{0.002}
\right)^{\frac{1}{2}}\left(\frac{1.55\,\rm 
ps}{\tau_B}\right)^{\frac{1}{2}}
\cdot \left(1 \pm 0.025_{\rm pert} \pm 0.03_{m_b} 
\right)
\;.
\label{22}
\eeq
The dependence on $\mu_\pi^2$ is practically absent here.

Accurate measurement of the inclusive $b\ra u\, \ell\nu$ width 
is
difficult and for a long time seemed unfeasible. However, recently 
ALEPH 
announced the first direct measurement \cite{aleph}:
$$
{\rm BR}(B\rightarrow X_u\ell\nu) = 0.0016\pm 0.0004\;\;.
$$
It is not clear how reliable and model-independent are the quoted 
error 
bars in this complicated analysis. It certainly will be clarified in the
future. Taking these numbers at their face value, one would arrive at 
the
model-independent result        
\beq
|V_{ub}|/|V_{cb}|\; = \; 0.098\pm 0.013\;\;.
\label{19}
\eeq
The theoretical uncertainty in translating $\Gamma(B\ra X_u\,  
\ell\nu)$ 
into $|V_{ub}|$ \cite{upset} is a few times smaller than the above 
experimental one.

\subsubsection{Caveat}

Deviation from duality is the most vulnerable element of the 
theoretical prediction. Reliable information on this aspect of QCD
is scarce. The only model specifically designed to address the issue
is presented in Ref. \cite{inst}. If it can be relied upon,  at least for 
orientation, we can safely neglect deviations from duality in the
semileptonic decays of $B$ mesons, but not in
those of $D$ mesons. And indeed, a parallel analysis
of $\Gamma (D\ra X_s \ell\nu )$ \cite{Ds,DSL,bds}, along the lines 
discussed above, shows that the expression for
$\Gamma (D\ra X_s \ell\nu )$, similar to Eq.  (\ref{SLGAMMA}), falls 
short of the experimental number  roughly by a factor of two,
provided that
the modern values of the fundamental parameters ($m_c$, 
$\mu_\pi^2$ and so on) are used. 

\subsubsection{Why has $\Gamma_{\rm sl}(B)$ been sometimes 
discarded in the quest for $|V_{cb}|$?}

The status of the radiative corrections to the widths had been  the
subject of some controversy recently, but finally the issue seems to 
be settled.
The reason for apparently contradicting opinions was related, once 
again,
to the problem of the heavy quark mass in the context of 
the
heavy quark expansion. We will briefly sketch the problem and 
explain the answers. 

The inclusive width can be expressed in terms of any 
{\em well-defined} mass parameter. Its {\em perturbative} 
component  through  a certain order  can conveniently be given in 
terms
of the  pole masses. Yet they are not  defined with the necessary 
accuracy 
so
as to include  power-suppressed corrections.  As discussed 
before, the 
pole mass contains long-distance contributions of order  $\Lam$. 
With the width being a short-distance  quantity we can then
infer that the indefiniteness  $\sim \Lam$ in the quark pole
mass has its  counterpart in an irreducible uncertainty  $\sim \Lam 
/m_Q$ 
in the perturbative corrections  if those are evaluated with
pole masses as input. This conclusion  has been proven in the most 
general 
terms by virtue of  OPE where the emergence of the $1/m_Q$ 
infrared
renormalon was  noted \cite{pole}. It was also  substantiated by the
perturbative calculation. A concrete analysis in the
framework of the BLM  resummation was performed later in 
\cite{bbz,bbbsl}.
Consider the 
perturbative expansion of $\Gamma_{\rm sl}$  in terms of the pole 
masses: 
\beq
\Gamma_{\rm sl} = \frac{G_F^2 m_{b\;{\rm pole}}^5}{192\pi^3} 
|V_{cb}|^2 
\cdot z_0
\left(\frac{m_{c\;{\rm pole}}^2}{m_{b\;{\rm pole}}^2}\right) 
\cdot c_3^{\rm (sl)}\; ,
\;\;\;
c_3^{\rm (sl)}= 
\left(1+
a_1^{(p)} \frac{\as}{\pi}+ a_2^{(p)} \left(\frac{\as}{\pi}\right)^2 +...
\right)
\label{w30}
\eeq
One finds that $a_1^{(p)} \simeq -1.7$ is followed by a huge 
coefficient \cite{wise,bbbsl} $a_2^{(p)} \simeq  - 10$ to $-20$ for 
$b \ra c$ (the exact value depends on the choice of the 
expansion parameter $\as$) or even 
$a_2^{(p)} \simeq - 30$ for $b \ra u$. If this were the end of the 
story, this fact 
would mean that the $\as ^2$ corrections by themselves 
reduce the width significantly, namely, by about $10 \%$, suggesting, 
at first sight,  
 that the unknown third (and higher) order 
corrections are  sizeable. In that case one could argue that 
the theoretical uncertainties in extracting $|V_{cb}|$ from 
$\Gamma_{sl}(B)$ are considerably larger than stated above. It 
would be natural to be 
concerned \cite{wise} whether such an extraction can be 
trusted at all! 

The key 
point that eliminates the concern, is as follows. When one calculates 
the radiative correction factor $c_3^{\rm
(sl)}$  through order $k$, one has to evaluate the pole mass likewise 
 to that order -- since the pole mass is only a formal 
perturbative
construction. 
Then one finds that $m^{(k)}_{\rm pole}$ receives sizeable increase 
as well; $m_{\rm pole}$ breaths as $k$ changes. The shifts are 
correlated so  that the width remains almost 
the same 
numerically. 

In doing so one must be careful always to use the very same
perturbative approximation both in masses $m_{\rm pole}^{(k)}$ 
and in the coefficient $c_3$. 
Speaking theoretically, the perturbative approximations to
the width computed in such a strictly correlated way contain no
uncertainty $\sim {\cal O}(\Lam/m_Q)$.
Uncertainties  
$\sim {\cal O}(\Lam^2 /m^2_Q)$  still remain due to
spurious terms $\sim \left(\delta m_{\rm pole}\,/\,m_Q\right)^2$ 
 having no  OPE interpretation.
On the practical side, the latter  problem emerges at the one-percent
level only --  unless one attempts to continue to too high orders. 

If one, however, exploits the  short-distance 
masses, say the 
masses normalized at the scale $\sim 1\GeV$, all these problems are 
completely avoided and one does not have to rely on what 
seems to be a numerical miracle. Indeed,   one 
can
use a
 fixed  number  for $m_{b,c}$ which, unlike $m_{\rm pole}$,
does not breath.
Moreover,  one finds then that 
\beq
\Gamma_{\rm sl} = \frac{G_F^2 m_b^5(\mu)}{192\pi^3} |V_{cb}|^2 
\cdot z_0
\left(\frac{m_c^2(\mu)}{m_b^2(\mu)}\right) \cdot \left(1+
a_1(\mu) \left(\as/\pi\right) + a_2(\mu) \left(\as/\pi\right)^2 +...
\right)
\label{w31}
\eeq
with $a_1(\mu)\approx -1$, $a_2(\mu)\approx 1$,
etc. \cite{upset}. The exact values of $a_k(\mu)$ depend on 
the scheme 
used. The same is true for  $m_{b,c}(\mu)$. The scheme dependence 
affects the overall result for 
$\Gamma_{\rm sl}$  only at a percent level. Such variations are 
unavoidable since  the perturbative
series are truncated.

To reiterate, the suspicion of  a large and uncontrollable impact of 
the 
perturbative
corrections on the absolute values of the semileptonic widths, quite 
popular in
the literature in the past, was mainly due to 
theoretical
subtleties with the pole mass. In particular, it was tacitly 
assumed  
that the  quark pole mass has an unambiguous value, exact up to a 
hundred $\MeV$, or so.  Then it was observed that:

\vspace{0.2cm}

(a) It is
difficult to accurately extract $m_b^{\rm pole}$ from experiment. In
any particular calculation one can  identify effects left out,
which can change its value by $\sim 200\MeV$. This uncertainty 
leads to
a  theoretical error $\delta_{\rm I}\simeq 10\%$ in 
$\Gamma_{\rm
sl}(B)$.

 (b) When routinely calculating $\Gamma_{\rm sl}(B)$ in  
terms
of the pole masses, there are significant higher order corrections 
$\delta_{\rm II} \simeq 10\%$. 

Thus the conclusion was made:  $\Gamma_{\rm sl}(B)$ cannot be  
calculated 
with accuracy 
better than $\sim 20\%$, and, correspondingly,  at best  $\delta 
|V_{cb}|/|V_{cb}| \sim
10\%$. 

  Both observations (a) and (b) above are correct, beyond any doubt. 
If one more step is taken, however, the conclusion is invalidated -- 
one should take into account the fact 
that the origin of these two uncertainties is actually the same, and
therefore they practically cancel each other. 
\vspace{0.2cm}

Let us mention that the problems associated with the  the pole mass 
sometimes 
surface in a  superficially different form, leading  (in the past)  to the 
same feeling,  
that $\delta |V_{cb}|/|V_{cb}| $ is as large as $\sim
10\%$. Quite often, inflated error bars in $m_b$ are quoted, see 
e.g. \cite{NEUBERT}, where the uncertainty in $m_b$ 
is set  $200$ to  $300\MeV$. One may notice that this variation  
actually reflects the change  in the values of $m_b$  emerging in 
different perturbative definitions of $m_b$. Such  a 
procedure is like comparing a quantity 
renormalized in the ${\rm MS}$ and $\overline{\rm MS}$ schemes at 
the same 
scale without including the non-trivial translation between 
those two definitions. The large perturbative uncertainty $\sim 
20\%$ 
in the value of $\Gamma_{\rm sl}$  in \cite{NEUBERT} was  deduced 
not
from the analysis of the absolute prediction for the semileptonic 
widths.
Instead, it was based on  a comparison of the 
perturbatively-improved
calculation of $\Gamma_{\rm sl}$ to an {\em ad-hoc}  quantity 
dubbed
$\Gamma_{\rm tree}$; the latter was supposed to represent a 
numerical
estimate of $\Gamma_{\rm sl}$ before any radiative corrections are
included. However, $\Gamma_{\rm tree}$ involves the fifth
power of the mass which was presumed to be $m_{\rm pole}$ in
\cite{NEUBERT}. Then everything depends on what is the value of 
$m_{\rm
pole}$ in the {\em tree}-level calculations. The large uncertainty 
stated
in \cite{NEUBERT} can be traced back  to the assumption that this 
{\rm tree}-level
mass is given by the actual all-order sum of the perturbative series
emerging  when one attempts to calculate the would-be pole mass of 
the
$b$ quark in terms of $\bar m_b(m_b)$. This  procedure relies on 
summation of the series 
similar to Eq.~(\ref{POLEMASSUNC}), which is particularly unnatural 
in
the context of the {\it tree}-level evaluation. A finite value for this
tree-level mass, $\simeq 5.05\GeV$, used in \cite{NEUBERT}, was due 
to a
trick with a  non-summable series, a principal-value
prescription for the Borel integral, and so was the ratio
$\Gamma/\Gamma_{\rm tree}=0.77\pm 0.05$. Values other  than 
$0.77$ could have
been obtained as well, invoking other prescriptions. In any case, it is
clear that such definition of $\Gamma_{\rm tree}$ is not appropriate 
for
evaluating the impact of the perturbative corrections.

\vspace*{.2cm}

The epilogue of the long story with the determination of $|V_{cb}|$
from the inclusive width can be summarized as follows.
After all refinements and thorough analysis of the
corrections, we are basically back to square one. 
The expression for $|V_{cb}|$ given in Ref. \cite{vcb}
has changed by less than one percent (if one inputs the same 
values of the experimental parameters  as in \cite{vcb}).
What became clearer in the last two years is the acceptable range of 
the underlying key QCD parameters entering the heavy quark expansion, and 
the
size of theoretical corrections. 

\subsection{$|V_{cb}|$ from 
$\Gamma (B \ra D^* \ell \nu )$ at Zero Recoil}

Introduction of the universal Isgur-Wise function \cite{HQS3}
was a crucial step in the evolution of heavy quark 
theory. One aspect is of a special significance in applications: the fact 
that it is normalized to unity at zero recoil \cite{HQS1,HQS2,HQS3}.
As was emphasized later (see e.g. \cite{Nzr})
one can exploit this 
feature for determinations of $|V_{cb}|$ from the exclusive
$B\ra D^*$ semileptonic transition extrapolated to zero recoil.
To this end  
one measures the  differential 
rate, extrapolates to the  point of zero recoil and 
gets the quantity $|V_{cb}F_{D^*}(0)|$, where $F_{D^*}$ is the axial
$B \ra l \nu D^*$ formfactor. To extract $|V_{cb}|$ it is  necessary
to know $|F_{D^*}(0)|$, which, although close to unity,
still deviates from unity due to various perturbative and 
nonperturbative 
corrections. In the real  world 
$$
F_{D^*}(0) = 1 + {\cal O}\left(\frac{\as}{\pi}\right) + \delta_{1/m^2} + 
\delta_{1/m^3}+ ... = 
$$
\beq
1 + {\cal O}\left(\frac{\as}{\pi}\right)
+ {\cal O}\left(\frac{1}{m_c^2}\right) + 
{\cal O}\left(\frac{1}{m_cm_b}\right)+
{\cal O}\left(\frac{1}{m_b^2}\right)+...
\label{vpc}
\eeq 

The absence of $1/m_Q$ corrections in  $|F_{D^*}(0)|$ was noted in 
passing in Ref.
\cite{HQS2}. This fact was cast in the form of a theorem by Luke
\cite{luke}. This  is nothing but the  heavy-quark analog of the 
Ademollo-Gatto theorem
for the $SU(3)_{\rm fl}$ breaking effects
which is routinely exploited in determinations of
$|V_{us}|$ from $K\ra\pi\ell\nu$ and semileptonic hyperon decays. 

The task of precision determination of $|V_{cb}|$ from the 
exclusive transition requires a
detailed dynamical analysis of various  
pre-asymptotic corrections in Eq.~(\ref{vpc}).
The perturbative part, albeit  technically complicated, is  at least 
conceptually transparent.  The theory of $1/m_Q^2$
corrections is more challenging.

The urgent need in evaluation of the $1/m_Q^2$ corrections in Eq.
(\ref{vpc}) for practical purposes was realized quite early
\cite{FN}. In these days the theory of the power corrections 
in heavy quarks was immature, our knowledge was scarce, so that it 
was hard to
decide even the sign of $\delta_{1/m^2}$.
The general impression was that this deviation is quite small,
$\lsim 2\%$ \cite{neubbc}.  
At the present stage we learned much 
more about the  nonperturbative  $1/m_Q^2$ and perturbative  
corrections. We discuss them in turn, and then proceed to numerical 
analysis and survey of the literature. At first, a brief excursion into 
the theory is undertaken to sketch the main ingredients of the 
direction 
where most of the advances have been achieved --
the sum rules for heavy flavor transitions \cite{vcb,optical}. 
Although related to those presented in Sect. 4, these sum rules  are 
not identical (with a few exceptions, see Sect.~8.2.2) 
since we account for
$1/m_Q$ effects. 
They express the moments of the 
{\em inclusive} probabilities   in terms of the quark 
masses and the expectation values of certain local operators. 
Since the transition probabilities 
are non-negative, the sum rules lead to  constraints on the
{\em exclusive} form factors. 

\subsubsection{Heavy flavor  sum rules: generalities}

The sum rules are derived in QCD using the standard methods of the
short-distance expansion \cite{optical}. One starts the analysis from  
the forward 
transition
amplitude
\beq
T^{(12)}(q_0;\,\vec{q})\;=\; \frac{1}{2M_B} \int\; d^3x\, dx_0 {\rm
e}\,^{i\vec{q}\vec x -iq_0x_0}\, \matel{B}{iT(\bar c \Gamma^{(1)} 
b(x), \:
\bar b \Gamma^{(2)} c(0)\,)}{B}
\label{s2}
\eeq
where $\Gamma^{(1,2)}$ are some spin structures or, 
more generally, local operators. The transition amplitude 
contains a lot of information about the decay probabilities. As usual 
in QCD, 
one cannot calculate it completely in the physical domain of $q$.
The amplitude
(\ref{s2}) 
has
several  cuts corresponding to different physical processes 
\cite{CHAY}. The
discontinuity at the physical cut $q_0 < M_B- 
\sqrt{M_D^2+\vec{q}^{\,2}}$
describes the inclusive decay probabilities at a given energy released 
into
the final hadronic system. The cut continues further than 
the domain 
accessible in
the actual decays, see Fig.~5. 

\begin{figure}
\vspace{3.0cm}
\includegraphics{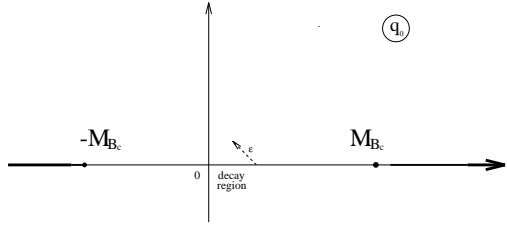}
\caption{ 
Cuts of the transition amplitude in the complex $q_0$ plane. The physical cut
for the weak decay starts at $q_0=M_B-
(M_D^2+\vec{q}^{\,2})^{1/2} $ 
and continues towards $q_0=-\infty$. 
Other physical processes generate cuts starting near 
$q_0=\pm 
(M_{B_c}^2+\vec{q}^{\,2})^{1/2}
$ (one pair); another pair of cuts originates
at close values of $q_0$. 
An additional channel opens at 
$q_0 \gsim 2m_b +m_c$. }
\end{figure}

In QCD we  calculate the amplitude (\ref{s2}) away 
from all its cuts. Essentially, it can be expanded  in the 
inverse
powers of the distance from the physical cut, $\epsilon\,$:
\beq
\epsilon =  M_B-\sqrt{M_D^2+\vec{q}^{\,2}}-q_0 =
m_b-\sqrt{m_c^2+\vec{q}^{\,2}}+\delta(\vec q^{\,2})- q_0\;,
\label{s3}
\eeq
where
$$
\delta(\vec q^{\,2})=
(M_B-m_b)-(\sqrt{M_D^2+\vec{q}^{\,2}}-\sqrt{m_c^2+\vec{q}^{\,2}}) 
\sim
{\cal O}\left(\Lam\frac{\vec{q}^{\,2}}{m_Q^2}\right) + 
{\cal O}\left(\frac{\Lam^2}{m_Q}\right)\;.
$$
The ``off-shellness" $\epsilon$ must be chosen in such a way that 
that $|\epsilon|$ and $|\epsilon \cdot 
\arg{\epsilon}| \gg 
\Lam$ but, 
simultaneously,
$|\epsilon| \ll \sqrt{m_c^2+\vec{q}^{\,2}},\,m_b$. The second 
requirement allows us  to ``resolve" the
contributions of the separate cuts.

How does the deep Euclidean expansion of $T^{(12)}$ help 
constrain the amplitude
in the physical domain of real $\epsilon$ lying just on the cut? The
dispersion relation immediately tells us that the coefficients in the
expansion of the amplitude in powers of $1/\epsilon$ are given by 
the
corresponding moments of the spectral density in $T^{(12)}$,
\beq
T^{(12)}(\epsilon;\,\vec{q})\;=\; \frac{1}{\pi} \,\int\; 
\frac{\Im T^{(12)}(\epsilon';\,\vec{q})}{\epsilon' -\epsilon}\; 
d\epsilon'
\;=\;
-\frac{1}{\pi} \, \sum_{k=0}^\infty \:\frac{1}{\epsilon^{k+1}}\,\int\; 
\Im T^{(12)}(\epsilon';\,\vec{q})\,\epsilon'^k\; d\epsilon'
\;.
\label{s4}
\eeq
Various subtleties going far beyond the scope of the present review, 
are discussed in Refs. \cite{optical} and \cite{inst}. 

On the other hand, we can build the 
large-$\epsilon$ expansion of the transition amplitude {\em per se},
treating it as the propagation of the virtual heavy quark 
submerged into a  soft medium. The expansion takes the general 
form
$$
T^{(12)}(q_0;\,\vec{q})\;=\; 
\frac{1}{2M_B} \int\; d^3x\, dx_0 {\rm
e}\,^{i\vec{q}\vec x -iq_0x_0}\times
$$
$$
\matel{B}{\bar b(x) \Gamma^{(1)} 
\: (m_c+i\not\!{D}-\not\!{q}) 
\frac{1}{m_c^2-(iD-q)^2-\frac{i}{2}\sigma G}
\:\Gamma^{(2)} b(0)}{B}\;=
$$
$$
\aver{\Gamma^{(1)}  (m_c+m_b\gamma_0-\not\!{q} +\not\!\pi ) 
\frac{1}{m_c^2-(m_b-q_0)^2 +\vec q^{\,2} - 2m_b \pi_0+2q\pi
-\pi^2-\frac{i}{2}\sigma G}
\:\Gamma^{(2)}}\;=
$$
\vspace{0.1cm}
$$
\aver{\Gamma^{(1)} 
(m_c+(E_c+\epsilon-\delta(\vec q^{\,2}))\gamma_0+\vec
q\vec\gamma+\not\!\pi) \times
$$
\vspace{0.1cm}
\beq
\sum_{n=0}^\infty
\frac{\left[2(E_c+\epsilon-\delta(\vec q^{\,2}))\pi_0 + 2\vec q \vec 
\pi
+ \pi^2+\frac{i}{2}\sigma G\right]^n} 
{\left[-(\epsilon-\delta(\vec q^{\,2}) )(2E_c + \epsilon-\delta(\vec
q^{\,2}))\right]^{(n+1)}}
\:\Gamma^{(2)}}\;.
\label{s7}
\eeq
\vspace{0.1cm}
Here $E_c=\sqrt{m_c^2+\vec{q}^{\,2}}$, and we use the short-hand 
notation
$$
\aver{...} = \frac{1}{2M_B}\, \int\; d^3x\, dx_0 {\rm
e}\,^{i\vec{q}\vec x -iq_0x_0}\,\matel{B}{\bar b(x) ...b(0)}{B}\;.
$$

In the first equation we used the full QCD fields and then passed  to
the low-energy fields according to Eqs.~(\ref{tildeq})--(\ref{pi}). 
Picking up the corresponding term $1/\epsilon^{k+1}$ in the 
expansion
over $1/\epsilon$ at $\epsilon \gg \Lam $ 
and evaluating the expectation value of the resulting {\em local}
operators (e.g., $\bar b(0)\pi_\mu b(x) \equiv \delta^4(x)\, \bar
b\pi_\mu b(0)$), one gets the  sum rules sought for. Taking $k=0$ 
yields 
the sum rule for the equal-time commutator of the currents $\bar
c \Gamma^{(1)} b $ and $\bar b \Gamma^{(2)} c$; $k=1$ selects the
commutator for the time derivative, etc. After this  brief  general
overview we now outline a few concrete applications 
\cite{vcb,optical}.

\subsubsection{Sum rules for $\mu _{\pi}^2$ and $\mu _G^2$} 

Considering the semileptonic transitions driven by the pseudoscalar 
weak current $J_5= \int \,d^3 x\, (\bar c i\gamma_5 b)(x)$ (i.e., at  
zero recoil, 
$\vec q = 0$) one  obtains the sum rule 
\footnote{Corrections in the
overall  normalization of the current are ignored as 
irrelevant. The structure functions are defined, e.g. in 
\cite{Koyrakhsf}.} 
for the structure function $w^{(5)}$
\beq
\frac{1}{2\pi} \:\int_{0}^{\mu} w^{(5)}(\epsilon)\;  d\epsilon \; = \;
\left( 1/2m_c-1/2m_b\right) ^2
\left( \mu_\pi^2(\mu) -\mu_G^2(\mu)\right) \;.
\label{z5}
\eeq

The normalization point 
$\mu$ of $\mu _{\pi}^2$ and $\mu _G^2$ is introduced through the 
cut-off
of the integral on the left-hand side. Since the structure 
functions are 
non-negative, one arrives at the conclusion that 
$
\mu_\pi ^2(\mu) \; \ge  \mu _G^2(\mu )$, which is 
the  field-theoretic analog of the difference of the sum rules 
(\ref{3.7b}) and (\ref{3.7a}). 

The sum rule
(\ref{3.7a}) for $\mu_G^2$ can be also easily  obtained
in this way. For 
example, 
one can consider the antisymmetric part 
(with respect to  $i$ and $j$)  of the
correlator of the vector currents ($\Gamma^{(1)}= \gamma_i$,
$\Gamma^{(2)}= \gamma_j$). Another possibility is to turn directly 
to 
the
sum rule for the correlator of the non-relativistic currents $\bar c
\pi_j b$ and $\bar b \vec\sigma \vec \pi c$ in the leading order 
in $1/m_Q$. The OPE guarantees that all such relations are 
equivalent.

\subsubsection{$F_{D^*}$ at zero recoil}

The axial current $\bar c \gamma_{i}\gamma _5 b$ produces
$D^*$, $D\pi$  and higher excitations in semileptonic $B$ decays at 
zero
recoil. A straightforward derivation yields the following sum rule 
for this  current, 
\beq
|F_{D^*}|^2 + \frac{1}{2\pi} \:\int_{\epsilon>0}^{\mu} w^A(\epsilon)
\;  d\epsilon
\; = \;
\xi_A(\mu) \;-\; \Delta^A_{1/m^2} \;-\;\Delta^A_{1/m^3}\;+
{\cal O}\left(\frac{1}{m^4}\right) \,,
\label{z4}
\eeq
where
\beq
\Delta^A_{1/m^2} = \frac{\mu_G^2}{3m_c^2} +
\frac{\mu_\pi^2-\mu_G^2}{4}
\left(\frac{1}{m_c^2}+\frac{1}{m_b^2}+\frac{2}{3m_cm_b}
\right)\;,
\label{z4a}
\eeq
and $\matel{D^*(\vec q = 0)}{\bar c \gamma _i \gamma _5 b} {B} =
\sqrt{2M_BM_{D^*}} F_{D^*} \epsilon _i$. Furthermore, $w^A$ denotes 
the 
corresponding
structure function of the heavy  hadron excited  by the  axial current. 
Its
integral describes the contributions from {\em excited} charm states
with mass $M_i = M_{D^*} + \epsilon _i$, up to $\epsilon _i \leq 
\mu$:
\beq 
\frac{1}{2\pi}\:\int_{\epsilon>0}^{\mu} w^A(\epsilon) \;  d\epsilon = 
\sum
_{\epsilon _i < \mu}|F_i|^2 
\eeq
Contributions from excitations with $\epsilon$ higher than $\mu$ 
are
dual to perturbative contributions and get lumped into the 
coefficient
$\xi _A(\mu )$ of the unit operator, the first term on the right-hand 
side of 
Eq. (\ref{z4}). 

The role of $\mu$ is thus 
two-fold:
in the left-hand side it acts as an ultraviolet cutoff in the effective
low-energy theory, and by the same token determines the 
normalization
point for the local operators; simultaneously, it defines the  infrared
cutoff in the Wilson coefficients. 

Equation (\ref{z4}) immediately implies
\beq
|F_{D^*}|^2 = \xi_A(\mu)- \Delta^A_{1/m^2} -
\, \sum _{\epsilon _i < \mu}|F_i|^2 \;> \xi_A(\mu)- \Delta^A_{1/m^2}
\label{oursr}
\eeq
due to the positivity of $|F_i|^2$. (For a moment we forget about 
the cubic corrections, $\Delta^A_{1/m^3}$.) 
The perturbative coefficient $\xi _A(\mu )$ is obtained considering 
the
sum rules (\ref{z4}) and (\ref{z5}) in  perturbation theory. To order 
$\as$ \cite{optical}
\beq
\xi_A(\mu)\;=\; 1+
\;2\frac{\as}{\pi}\left[\frac{m_b+m_c}{m_b-
m_c}\ln{\frac{m_b}{m_c}} -
\frac{8}{3} + \frac{1}{3}\left(
\frac{\mu^2}{m_c^2} + \frac{\mu^2}{m_b^2} + \frac{2\mu^2}{3 m_c
m_b}
\right) \right] \;.
\label{z7}
\eeq

\vspace{0.2cm}

The most complicated part of the genuine second-order correction 
in $\xi_A(\mu)$ was calculated in Refs. \cite{czar}. It turned out to be 
small. The two-loop calculation of $\xi_A(\mu)$ was recently completed
\cite{CMU}. 
The BLM-resummation of the one-loop result was carried out in
\cite{blmope} and placed the numerical value of $\xi_A(\mu) $ 
somewhere in between the
tree-level and one-loop estimates, see Fig.~6. This analysis 
suggests
that at $\mu \simeq 0.5\GeV$, the value of $\eta_A(\mu) \equiv
(\xi_A(\mu))^{\frac{1}{2}}$ is $0.99 \pm 0.01$ with the uncertainty 
coming from the 
higher-order corrections and terms $\sim (\mu/m_c)^3$. 

\begin{figure}
\vspace{4.8cm}
\includegraphics{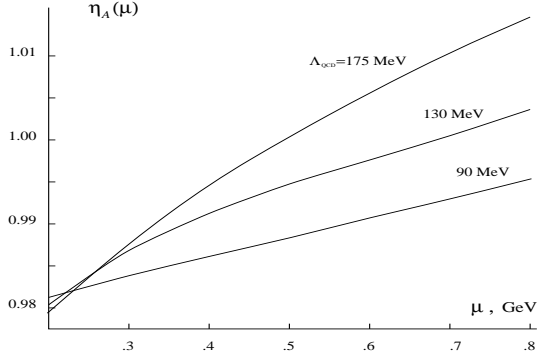}
\caption{The value of $\eta_A(\mu)$ for $\Lam^{\overline{\rm MS}} =90\MeV$,
$130\MeV$ and $175\MeV$. $\;\eta_A(\mu)$ must
be used in the QCD-based calculations of the zero-recoil $B\ra D^*$
formfactor when nonperturbative effects are addressed.}
\end{figure}

The nonperturbative $1/m^2_Q$ power corrections 
$\delta_{1/m^2}$ to $F_{D^*}$, Eq.~(\ref{vpc}), 
are negative. They consist of two terms: the first,  $-\frac{1}{2}
\Delta^A_{1/m^2} $, is known explicitly. The second, the sum over 
the excited states,
also scales as $1/m^2_Q$, it is negative, but is unknown otherwise. It 
cannot be calculated without additional dynamical 
assumptions. $-\delta_{1/m^2} $ exceeds $\mu_G^2/(6m_c^2) 
\simeq 0.035$ in magnitude. The sum over the excited states 
replaces the 
wavefunction overlap 
deficit
of the simple quantum-mechanical  analysis of the transition 
probabilities. We will return to this, the most difficult part of the 
analysis, after 
briefly discussing $1/m_Q^3$ terms in the heavy quark expansion. 

It is not difficult to calculate the $1/m^3_Q$ term in the sum rule. Its 
explicit
form depends on the convention for the $D=5$ operators in 
Eq.~(\ref{z4a}). 
Assuming
that $\mu_\pi^2$ and $\mu_G^2$ there are understood as the 
expectation values of the operators $\bar b (i\vec 
D\,)^2 b$ and
$\bar b\, \frac{i}{2} \sigma G \, b$ over the {\em 
actual}
$B$ mesons rather than the asymptotic states, we get  
\beq
\Delta^A_{1/m^3} \;=\; \frac{\rho_D^3 - \frac{1}{3} 
\rho_{LS}^3}{4m_c^3}\;+\;
\frac{1}{6m_cm_b}\left(\frac{1}{m_c}-\frac{1}{2m_b}\right) 
\,(\rho_D^3 +
\rho_{LS}^3)\, .
\label{z4b}
\eeq
If the asymptotic values of the $D=5$ matrix elements  are used, the 
additional
$1/m_b^3$ parts (see \cite{optical}) come from the expectation 
values in
\eq{z4a}. The $1/m_Q^3$ correction to the {\em inclusive} sum of the 
decay 
probabilities
turns out  to be moderate, and we will not dwell on this issue 
further. 

The excited states in the sum rule are
higher  excitations of $D^*$ and non-resonant $D\pi , ...$ states.
Their contribution in the right-hand side of Eq.
(\ref{oursr}) is negative. What can be said about its absolute value?
Unfortunately, no model-independent answer to this question exists 
at present. 
The best we can do today is to 
assume that the 
sum over the excited states (cut-off at $\mu$) 
is a fraction $\chi$ of the local term
given by $\mu_\pi^2$ and $\mu_G^2$, 
\beq
\frac{1}{2\pi} \:\int_{\epsilon>0}^{\mu} w^A(\epsilon)
\;  d\epsilon
\; = \; \sum _{\epsilon _i < \mu}|F_i|^2 \; = \; \chi\:
\Delta^A_{1/m^2}\, ,
\label{z8}
\eeq
where  on general grounds $\chi \sim 1$.~\footnote{In  perturbation theory 
$\chi=1$ in
the first order, but this relation changes in higher orders.} 
Some idea of the value of 
$\chi$ can be obtained from the 
$D\pi$ contribution which  is readily calculable by virtue of the 
soft-pion technique. The result exhibits the proper scaling behavior 
in all relevant parameters \cite{vcb}. Its absolute value depends on 
the pion coupling to the heavy mesons. The estimate of  this 
constant
was  improved \cite{Khod}. With this improvement, the $D\pi$
contribution to $\chi$ falls in the interval 0.15 to 0.35. It is natural 
to think that the resonant contribution is several times larger. To be 
conservative for numerical estimates we accept $\chi = 0.5\pm 0.5 $. 
Rather arbitrarily we limit  $\chi$ by unity on the upper 
side. 
Note that the larger the value of $\chi$ the stronger the deviation of 
$F_{D^*}(0)$ from unity. 
In this way we arrive at 
\beq
F_{D^*}\;\simeq \; \eta_A(\mu)\;-\;
(1+\chi)\,\left[\frac{\mu_G^2}{6m_c^2} +
\frac{\mu_\pi^2-\mu_G^2}{8}
\left(\frac{1}{m_c^2}+\frac{1}{m_b^2}+\frac{2}{3m_cm_b}\right)
\right] - 
\delta_{1/m^3} 
\, . 
\label{z9}
\eeq

Assembling all pieces together and assuming our standard input
for 
$\mu_\pi^2$ and the low-scale quark masses ($m_c=1.25\GeV$,
$m_b=4.75\GeV$)  we get 
\footnote{The central numerical value here is 
larger by $0.015$ than  that cited in \cite{vcb} for  the same
$\mu_\pi^2$.  $0.01$ is due to a shift in the  perturbative factor 
$\eta_A(\mu)$ ($0.98$ was used in \cite{vcb}). The remaining 0.005  
reflects a different 
definition of $\mu_\pi^2$ at the level of  perturbative corrections.} 
\beq
F_{D^*}\;\simeq \; 0.91 \;-\;0.013\,
\frac{\mu_\pi^2-0.5\GeV^2}{0.1\GeV^2}\;\pm\;
0.02_{\rm excit}\;\pm\;0.01_{\rm pert}\;\pm\;0.025_{1/m^3}\;\;.
\label{z11}
\eeq
Estimates of the uncertainties in the ${\cal O}(1/m_Q^3)$
corrections and the contributions from the higher excitations
are not very firm and reflect a reasonably optimistic viewpoint 
of the convergence of the heavy quark expansion. 
Altogether we get 
\beq
F_{D^*}\;\simeq \; 0.91 \; \pm 0.06\;,
\label{z31}
\eeq
where the  optimistic uncertainty $\pm 0.1 \GeV^2$ is ascribed to 
$\mu_\pi^2$. 
 
The uncertainty can presumably be reduced in the future by one 
percentage point by  accurately measuring $\mu_\pi^2$.  An
additional $0.01$ can be  removed via a dedicated experimental 
study of
the transitions to the excited $P$-wave states.
Using current theoretical technologies,  it seems impossible
to overcome the $5\%$ barrier of the  model-independent theoretical
accuracy in this exclusive decay.

In summary, the 
 QCD-based  analysis  favors a significantly larger deviation of 
$F_{D^*}(0)$ from unity 
than those  {\em en vogue} three years ago. The $1/m_c^2$ shift 
by
$-0.035$ is definitely model-independent. 
Values of $|\delta_{1/m^2}|$ close to the minimal possible
scenario $0.035$
are achievable only if  there are practically no zero-recoil 
transitions $B\ra D\pi, D^{**},...$ up to the mass range  $2.5$ to 
$2.6
\GeV$, and, additionally,  the overall yield of the  lowest $P$-wave 
states in the sum rule for $\mu_\pi^2$ is totally 
suppressed. Needless to say, that this is hardly possible. In any case 
such a distinct pattern 
can also be checked experimentally. Although the present data are 
not
conclusive yet, there is no evidence for such a feature.

\subsubsection{Quantum-mechanical interpretation}

The sum rules (\ref{z4}) have a transparent interpretation
 in the framework of conventional  quantum mechanics \cite{optical}.
From the gluon point of view the  semileptonic
decay of the $b$ quark is an  instantaneous replacement of $b$ by 
$c$
quark. The overall probability of the produced state to hadronize to
some final state is exactly unity, which corresponds to the first term
on the right-hand side of Eq.~(\ref{z4}). The nonperturbative 
corrections in the
sum rule appear since the normalization of the weak current $\bar c
\gamma_\mu \gamma_5 b$ is not unity and depends, in particular, 
on the
external gluon field. Expressing the  current in terms of the
non-relativistic fields used in quantum mechanics  one has, for 
example:
\beq
\bar c \gamma_k \gamma_5 b \leftrightarrow \sigma_k -
\left(\frac{(\vec\sigma
i\vec D)^2\sigma_k}{8m_c^2} +\frac{\sigma_k(\vec\sigma
i\vec D)^2}{8m_b^2}-
\frac{(\vec\sigma i\vec D )\sigma_k(\vec\sigma i\vec D)}{4m_cm_b}
\right)\;+\;
{\cal O}\left(\frac{1}{m^3}\right)\, .
\label{z13}
\eeq
The second term immediately yields the correction to the 
normalization,
which is present on the right-hand side of the sum rule. 
 It is curious to note
that the first  HQET analysis of $1/m^2$ corrections (see, e.g.,
\cite{FN,neubpr}) does not include the first two terms in the brackets 
which
contain the dominant effect
$\sim 1/m_c^2$. They appear, however, in the  approach of the 
Mainz 
group \cite{korner}.

\subsubsection{Digression in the literature} 

In view of the practical importance of the value of $F_{D^*}(0)$ 
a more detailed comment on the literature seems in order. 
We will dwell on a combined analysis \cite{update} often 
cited in this
context as the analysis with the  best  accuracy and reliability 
available today on the theoretical market. Our task is to warn the 
potential reader of hidden assumptions and 
inconsistencies. 

The basic idea of Ref. \cite{update}  was supplementing the sum 
rules 
of \cite{vcb,optical} by  
a certain symmetry relation  obtained previously \cite{FN}.
In this way one gets rid of the sum over the excited states for the
price of 
expressing the  deviation of the formfactor $F_{D^*}$ 
from
unity in terms of two other  unknown
zero-recoil formfactors for the vector current in pseudoscalar and 
vector mesons $\ell_{P,V}$, and an unknown hadronic correlator
$\lambda_G^2$, 
\beq
\delta_{1/m^2}\;=\;-\left(\frac{1}{2m_c}-\frac{1}{2m_b}\right)\,
 \left(\frac{\ell_V}{2m_c}-\frac{\ell_P}{2m_b}\right)\, + 
\, \frac{1}{4m_c m_b} 
\left(-\frac{4}{3}\mu_\pi^2+6\mu_G^2- \lambda_G^2\right)\, .
\label{z20}
\eeq
Here $\ell_P$ and $\ell_V$ are defined via the forward matrix 
elements
$$
\frac{1}{2\sqrt{M_B M_D}}\matel{D}{\bar c \gamma_0 b}{B}\;=\; 
\eta_V \,
\left(1- \left(\frac{1}{2m_c}-\frac{1}{2m_b}\right)^2 \ell_P \right)\, 
,
$$
\beq
\frac{1}{2\sqrt{M_{B^*} M_{D^*}} } \matel{D^*(i)}{\bar c \gamma_0 
b}{B^*(j)}\;=
\; 
\eta_V \,\delta_{ij}\,
\left(1- \left(\frac{1}{2m_c}-\frac{1}{2m_b}\right)^2 \ell_V \right)\, 
,
\label{z21}
\eeq
($\ell_p$ is, {\em theoretically}, observable in the decays $B\ra \tau
\,\ell\nu$). The same sum rules derived in \cite{vcb,optical} 
for $F_{D^{(*)}}$  apply to $\ell_{P,V}$ as well. 
The  parameter $\lambda_G^2$ is a sum of three other unknown
correlators introduced in \cite{FN}.\footnote{As a matter of fact, its 
meaning is seen from relation (\ref{z4}) in the
quantum-mechanical
interpretation at $m_b=m_c$. It represents a sum of the transition 
probabilities
in the spin-flip processes.}
The strategy of Ref.~\cite{update} was to obtain 
constraints on the
introduced unknown parameters following from the sum rules
(\ref{z4})--(\ref{z5}), and, then, in turn, to find the allowed interval 
for $\delta_{1/m^2}$.

Expressing the phenomenologically  relevant  formfactor $F_{D^*}$ in
terms of two other unknown formfactors and one unknown 
correlator
is clearly of no  help unless a new  dynamical input is provided. 
It was provided in a form of an {\em ad hoc} prescription
for the  value of $\lambda_G^2$  which was 
set to 
be very small. This step was crucial since without 
information on
$\lambda_G^2$ the relation (\ref{z20}) provides no constraint.  

Another  crucial element of the analysis was exploiting  the quark 
model to estimate  $\ell_P$ and $\ell_V$.  The quantities
$$
-\left(\frac{1}{2m_c}-\frac{1}{2m_b}\right)^2  \ell_{P,V}
$$
were
identified with the deviations from  unity of the wavefunction 
overlap
between the charm and beauty states, associated with the $c, b$ 
quark mass splitting. The complete relation has, however, the 
form \cite{optical}
$$
\langle\Psi_D | \Psi_B \rangle \;=\;
1-\left(\frac{1}{2m_c}-\frac{1}{2m_b}\right)^2 \ell_P\; + \;
\frac{\mu_\pi^2-\mu_G^2}{2} \left(\frac{1}{2m_c}-
\frac{1}{2m_b}\right)^2
$$
\beq
\langle\Psi_{D^*}^i | \Psi_{B^*}^j \rangle \;=\; \delta_{ij}\;\left[
1-\left(\frac{1}{2m_c}-\frac{1}{2m_b}\right)^2 \ell_V\; + \;
\frac{\mu_\pi^2+\frac{1}{3}\mu_G^2}{2} 
\left(\frac{1}{2m_c}-\frac{1}{2m_b}\right)^2 \right]\;,
\label{z23}
\eeq
where $\Psi_{B^{(*)},D^{(*)}}$ are the normalized wavefunctions for 
charm and beauty.
The terms with $\mu_\pi^2$ and $\mu_G^2$ were missed in 
Ref. \cite{update}. 
These terms emerge due to the Foldy-Wouthuysen 
transformation. 
Thus, even accepting the quark model for guidance, we observe 
inconsistency in evaluating the dominant $1/m_Q^2$ part of the
correction. This led to the surprising statement on the 
insensitivity of the
estimates for $\delta_{1/m^2}$ to the value of kinetic operator.
This insensitivity, together with discarding $1/m_Q^3$  corrections,
was the reason behind a smaller overall uncertainty in $F_{D^*}$ than 
indicated in Eq.~(\ref{z31}).

Another potential problem of this work refers to the treatment of  
radiative 
($\alpha_s^k$)  corrections. The problem arises once one addresses
the issue of higher-order summation ignoring 
the basic aspect  of OPE. Postulating that the perturbative factor 
$\eta_A(\mu)$ in Eq. (\ref{z9})
and its vector counterpart  $\eta_V(\mu)$ coincide with the
sum 
of purely perturbative 
corrections involving all virtual  momenta (including 
those below $\Lam$) we are bound to run into troubles. 
Physically, this is double-counting since the domain below $\mu$ 
was already  included in the matrix elements. It is explicit in this
approach: the sum rules of \cite{vcb,optical} are valid only if the
perturbative coefficients are understood in the Wilsonian sense; they 
definitely {\em do not hold} in the interpretation used in
\cite{update}. Formally 
the resummed series  itself tells us of an 
inconsistency: the resummed values of the perturbative factors 
$\eta_V$, $\eta_A$ cannot be defined at the level  ${\cal O}
(\Lam^2/m_c^2)$; they  develop unphysical 
imaginary parts and so on. This irreducible ambiguity in
thus defined  $\eta_A$  was estimated to be $2.4\%$ \cite{ns}. It is
impossible to get rid of it in the analysis   \cite{update},  since the
hadronic  parameters $\ell_P$,  $\ell_V$  etc. are treated as fixed
positive numbers. This fact was pointed out in \cite{ns} and admitted 
in
\cite{update}. 

A practical manifestation of these theoretical problems was 
later
noted in Ref.~\cite{lig}. It was found that already in the second order
the incorporation of the perturbation theory in the approach of \cite{update}
suffers from large perturbative corrections $\sim 3$-$5\%$.

Combining these observations it is logical to conclude that
the combined analysis \cite{update} loses information rather than 
adds it and worsens the accuracy. It  seems 
unlikely that following this line of reasoning in the future we could  
narrow the 
possible interval for  $F_{D^*}$. Evaluating deviations from the
symmetry limit for 
$\ell_P$ and  $\ell_V$ is neither easier nor more
difficult than evaluating  $F_{D^*}$ itself.  At the moment there 
seems to be no ground for more optimistic attitude. 

\subsubsection{Analyticity and unitarity constraints on the 
formfactor}

Extracting $|V_{cb}|$ from the exclusive decays implies extrapolating 
data to zero recoil. Near zero 
recoil 
statistics in the decays $B\ra D^* \ell\nu$ is
very limited, and the result for the 
differential decay rate at this 
point is  sensitive  to the 
way one extrapolates the experimental data to 
$\vec
q=0$. Most simply  this is 
done through linear extrapolation. 
Noticeable curvature of the 
formfactor
would change  the experimental value for $|V_{cb}F_{D^*}(0)|$.
Under the circumstances it is natural to try to get independent  
theoretical information on
the $q^2$ behavior of the formfactor near zero recoil. 

Some time ago it was emphasized \cite{rafael} that 
additional constraints on the $q^2$ behavior follow from
 analytic properties of the  $B\ra D^*$ 
formfactor considered as
a function of the momentum transfer $q^2$, combined with certain 
unitarity
bounds. In particular, in the dispersion integral 
\beq
F(q^2)\;=\; \frac{1}{\pi}\, \int \; ds \,\frac{\Im F(s)}{s-q^2}
\label{n2}
\eeq
the contribution of the physical $s$-channel domain
$q^2>(M_B+M_{D^{(*)}})^2$  is bound since $|F|^2$ 
describes the exclusive production of $B \bar D$ and cannot exceed 
the
total $b \bar c$ cross section. The integral (\ref{n2}) receives 
important contribution from the domain below the open $b\bar c$
threshold, where $F(q^2)$ has narrow pole-like
singularities corresponding to a few lowest $b\bar c$ bound states.
Introducing a set of unknown residues and making plausible 
estimates of
the positions of these bound states, on the one hand, and adding a 
small 
 non-resonant subthreshold 
contribution, on the other hand,
\res{lebed} suggested an {\em  ansatz }
for the
formfactor in the whole decay domain. Since the residues are 
unknown
and  the momentum transfer in the actual decays 
$B\ra
D^{(*)}\ell\nu $ varies in a rather narrow range, 
the advantages of this parameterization  over the standard  
polynomial fit are not clear at the moment. 

Surprisingly, a much more stringent (compared to what  follows from 
Ref. \cite{lebed})  relation between the slope and the 
curvature of the formfactor (i.e.,
$F'(q^2)$ and $F^{\prime\prime}(q^2)$ at  zero recoil) was found in 
\cite{neubcap}.  This 
relation was hastily  incorporated in some experimental 
analyses. Below we argue that one should be extremely cautious 
in relying on the relation \cite{neubcap} due to a hidden assumption 
which makes the results very vulnerable and unstable. 

The suggestion  of Ref.~\cite{neubcap} was two-fold: consider the 
scalar formfactor 
for the $B\ra D$ transition, and discard the resonant subthreshold 
contribution. In the heavy quark limit the scalar formfactor is 
related to 
usual axial-vector one $F(q^2)$. 
If one neglects  the contribution coming from the  resonances below 
the
open $b\bar c$ threshold (in this channel they are scalar $b\bar c$
relatives of the $\chi_b$ and $\chi_c$ states) and retains only the 
the
non-resonant continuum similar to the one used in \cite{lebed}, 
stringent
bounds do indeed follow from the dispersion  representation.  The
non-resonant continuum,  however,  can be safely disregarded
numerically. It  is $50$ to $100$ times smaller  than what can be
expected from the  resonant  subthreshold  contribution \cite{crad}. 
To
give an idea why it happens  it is sufficient  to recall that in the good  
old
pole-dominance models, the  subthreshold  poles due to $\chi_{bc}$  
saturate the formfactor completely.  Reinstating  the $\chi_{bc}$ 
subthreshold resonances  at appreciable  level suppresses  the
predictive  power of the  dispersion approach,  as is explained in
detail e.g. in Refs. \cite{onTaron}; no advantage compared to a more
careful analysis of \cite{lebed} emerges. 

An additional problem one immediately encounters is the necessity 
of including  $1/m_c$ corrections  translating  the results for the 
scalar current into those referring to the axial
formfactor. Such corrections to the derivatives over the 
velocity
transfer were evaluated \cite{vain}, and turned out to 
be very
significant even for more stable inclusive transitions. For further  
comments see Ref.~\cite{crad}.

The lesson we would like to draw is as follows:  it is not advisable to  
impose the  model-dependent 
relation \cite{neubcap} in  experimental extrapolations to
 zero recoil,  where model dependence is undesirable. 

\section{Challenges in Nonleptonic Beauty Decays}

Concluding the review it is impossible to avoid mentioning some
unsolved mysteries of the heavy quark theory.
Today's question marks carry the seeds of tomorrow's advances.
Basically there are two problems 
where our theoretical understanding is lagging behind.
Both are related to non-leptonic decays. The most placid solution 
(which we would prefer, of course) is experiment evolving towards 
theory, to meet its expectations. If this does not happen, some major 
adjustments in our theoretical  ideas seem to be inevitable. 

\subsection{Semileptonic branching ratio of the $B$ meson and 
$\Gamma (B \ra c \bar c s \bar q)$ }

The theoretical attitude to this problem oscillates with time. 
Twenty years ago it was believed that the parton model would give
a sufficiently accurate prediction,
${\rm BR}(b\ra c \,\ell \nu) \sim 15\%$. Then it was realized that
the $\alpha_s$ radiative corrections were important 
\cite{PETRARCA}.
The present situation is given in Ref.~\cite{BAGAN}. The general 
tendency
associated with the $\alpha_s$ corrections is lowering 
the value of 
${\rm BR}(b\ra c\,\ell\nu )$ down to $\sim 11.5$ to  $13.5\%$. This is still 
noticeably higher than the experimental number
 \cite{PDG96}: 
\begin{equation} 
{\rm BR}(B \to X\,\ell\nu ) = 10.43 \pm 0.24\% \, .
\label{SLBREXP} 
\end{equation} 
Nonperturbative corrections lower the semileptonic 
branching ratio further; yet being of  
order $1/m_b^2$ they are numerically quite small, 
$\Delta_{\rm nonpert}{\rm BR_{\rm sl}}(B) \simeq 0.5\%$, 
which seems to be  insufficient to close the gap between the 
expectation and the data \cite{baffling}. 

The issue of the semileptonic branching ratio
must be considered in conjunction with the charm yield $n_c$,
the number of charm states emerging from $B$ decays.
To measure $n_c$ one assigns charm multiplicity {\em one} 
to $D$, $D_s$, $\Lambda _c$ and $\Xi _c$ and {\em two} to 
charmonia. Zero is assigned to the charmless hadronic final state. 
It is obvious that   
\beq
n_c \simeq 1 +\mbox{BR}(\bar B \ra c \bar c s \bar q)
-\mbox{BR}(\bar B \ra \mbox{no charm})\, .
\label{nc}
\eeq
The experimental situation with $n_c$ is as follows.
The CLEO group finds $n_c = 1.134 \pm 0.043$ whereas 
the ALEPH collaboration reports $n_c = 1.23 \pm 0.07$. 
While both experimental numbers \cite{CASSEL} are consistent with 
each other and consistent with a
``canonical" value 1.15, the 
CLEO number clearly favors lower values ($\sim 1.15$) while ALEPH
does not rule out $n_c$ as large as  1.3. The statistical average is
$n_c = 1.16 \pm 0.04$. 

Further reduction of the theoretical 
prediction for the branching ratio can be provided either by 
higher-order perturbative 
corrections (which are amenable to analysis, at least, in principle)
or by largely uncontrollable duality violations (usual scape-goat). 
In the later case it is natural to suspect  
$\Gamma (\bar B \ra c \bar cs \bar q)$ since the energy release 
in $b \ra c \bar cs$ is not very large.  

With the advent of the heavy quark theory the question 
 of
 the compatibility of the existing theoretical ideas with
the data on BR$(B\ra X_c \ell \nu ) $ and $n_c$ acquired a solid 
footing
\cite{baffling}. While  $n_c\approx 1.15$ came  out naturally,
the excess of BR$(B \ra X_c \ell \nu )$ was an obvious challenge. 
Shortly after, the attitude changed. A natural desire to have all 
problems peacefully settled prevailed, see e.g. the summary talk 
\cite{NEUBERT} where the general conclusion leans towards the 
absence of any problem. A combination of 
two factors was crucial in this respect. Large values
of $\alpha_s$ fashionable two or three years ago
(corresponding to $\alpha_s (M_Z) =0.125$ or even higher)
enhance the non-leptonic width and, hence, suppress
BR$(B \ra X_c\ell \nu )$ down to $11$ or even $10.5 \%$. Simultaneously
$n_c$  jumps up to $0.125$, but since the ALEPH data were newer, 
their significance was overemphasized, and it was tempting to 
close one's eyes on the CLEO data.

The present  perception of the 
issue seems more balanced; it acknowledges the existence of the 
problem, see e.g.
\cite{Dunietz,Kagan}. The only theoretical ingredient one needs to 
reveal the problem is the statement that the $b \ra c \bar c s$ 
channel 
is to blame for the discrepancy in the semileptonic branching ratio. 
It is conceivable that ${\rm BR}(\bar B \ra c \bar c s \bar q)$ 
is actually 
larger than it is usually inferred from the explicit quark-gluon 
calculation, either due to higher order $\alpha_s$ corrections
or due to deviations from duality. Say, if 
BR$(\bar B \ra c \bar cs \bar q) \sim 0.25$, rather than $0.15$,
(which is in line with the fresh CLEO data)
and the charmless modes are negligible,  
this would bring the predicted semileptonic 
branching ratio pretty close to the observed one, but 
$n_c$ becomes at least two standard deviations higher than the 
current average. A possible way out is to assume the charmless
modes (say, $b\ra s$ + gluon) at the level of $10\%$. According
to Eq.~(\ref{nc}), this will bring $n_c$ down to the acceptable value.
Reference \cite{Dunietz} suggests that such high yields of $b\ra s$ + 
gluon are attainable in QCD, which is extremely unlikely, to put it 
mildly. According to Ref.~\cite{Kagan} new physics is supposed to be 
responsible. 

\subsection{Lifetimes of Heavy-Flavor Hadrons}

As stated before, differences between meson and baryon decay 
widths 
arise already in order $1/m_Q^2$. The lifetimes of the various 
mesons 
get differentiated effectively first in order $1/m_Q^3$. A detailed 
review 
can be found in \cite{BELLINI}; here we will comment only briefly 
on the issue. 

Because the charm quark mass is not much larger than typical 
hadronic scales one can expect to make only semi-quantitative 
predictions on the {\em charm} lifetimes, in particular for the 
charm baryons. The agreement of the predictions with the 
data is surprisingly good. (It is quite possible, though, that future 
more 
precise measurements of the $\Xi _c$ and $\Omega _c$ lifetimes 
might reveal serious deficiencies.) 

As far as the {\em beauty} lifetimes are concerned there is much less 
``plausible deniability" when predictions fail. Table \ref{TABLE20} 
contains 
the world averages of published data~\footnote{It is 
unlikely that significant new data will appear before 
the turn of the millennium.} 
together with the predictions. The latter 
were actually made before data (or data of comparable 
sensitivity) became available. 

Data and predictions on the meson lifetimes are completely and 
non-trivially  consistent. Yet even so, a  comment is in order for 
proper  orientation. The numerical prediction is based on the 
assumption of factorization at a typical hadronic scale which is
commonly taken as the one where $\as(\mu_{\rm hadr})\simeq 1$. The related
uncertainty has been emphasized in \cite{NSFACT}, however the recent QCD
sum rule estimates \cite{baek} (though carried out in a simplified
manner) did not support the conjecture of \cite{NSFACT} about the significant 
impact of nonfactorizable contributions. While there is no justification
for factorization at $\mu\sim m_b$, there exists ample
circumstantial evidence in favor of approximate factorization  at a
typical hadronic scale -- from the QCD sum rule calculations, to 
lattice evaluations, to $1/N_c$ arguments. More to  the point,  the
validity of factorization can be probed in semileptonic  decays of $B$
mesons in an independent way, as  was pointed out in \cite{WA}. 

\begin{table}[t]
\caption{QCD Predictions for Beauty Lifetimes  
\label{TABLE20}}  
\begin{center}
\begin{tabular} {|l|l|l|l|}
\hline
Observable &QCD Expectations ($1/m_b$ expansion)& Ref. &
Data from \cite{BELLINI}\\ 
\hline 
\hline 
$\tau (B^-)/\tau (B_d)$ & $1+
0.05(f_B/200\, {\rm MeV} )^2 $ & \cite{mirage} & $1.04 \pm 0.04$ \\
\hline  
$\bar \tau (B_s)/\tau (B_d)$ &$1\pm {\cal O}(0.01)$ & 
\cite{STONE2}  
&  $ 0.97\pm 0.05$ \\ 
\hline 
$\tau (\Lambda _b)/\tau (B_d)$&$\gsim 0.9 $ & \cite{STONE2} & 
$0.77\pm 0.05$ \\
\hline 
\end{tabular}
\end{center} 
\end{table} 

The prediction on $\tau (\Lambda _b)$ versus $\tau (B_d)$ seems to 
be in conflict with the data. However, 
the experimental situation has not been fully settled yet. 
The difference between 
$\aver{\tau (\Lambda _b)/\tau (B_d)}_{\rm exp.} \simeq 0.77$ and 
$\tau (\Lambda _b)/\tau (B_d)|_{\rm theor} \simeq 0.9$ represents a 
large discrepancy.  
A failure of that proportion cannot be rectified 
unless one adopts a new paradigm in evaluating baryonic expectation 
values of the four-fermion operators. 
Two recent papers \cite{boost,NSFACT} have re-analyzed the 
relevant 
quark model calculations and found:    
\begin{equation}
\tau (\Lambda _b)/\tau (B_d) \equiv 1 - \Delta_{(\Lambda_b B)}, \;\; 
\Delta_{(\Lambda_b B)} \sim 0.03\;\mbox{to}\; 0.12 \, .
\label{DEVEST}
\end{equation}  
There are large theoretical uncertainties in
$\Delta_{(\Lambda_b B)}$ since the  baryon lifetimes reflect the 
interplay
of several contributions of  different signs.   Yet  one cannot boost
the size of $\Delta_{(\Lambda_b B)}$ much beyond the $10 \%$ level: to 
achieve the  latter one had to go beyond a usual description of 
baryons when light quarks are ``soft''.
A similar conclusion has 
been
reached by the authors of Ref. \cite{BARI} who analyzed the 
relevant
baryonic matrix elements through QCD sum rules. 

\section{Conclusions and Outlook} 

Heavy quark theory is now a mature branch of QCD.
Many practically important applied problems that defied theoretical
understanding for years, are now tractable. 
At the same time, all natural limitations of QCD take place also
in the heavy quark theory. The infrared part of dynamics is
parameterized rather than solved. Therefore, every new success 
based 
on
the general properties of the quark-gluon interactions, is a precious 
asset.  The most important stages of the success story are heavy 
quark symmetry itself, introduction of the universal Isgur-Wise 
function and  combining  Wilson's approach with the heavy quark 
expansion. The exact inequalities of the heavy quark theory
is another link of the same chain. 

The most clear-cut recent manifestation of the power of the heavy 
quark theory is the framework it provides the determination of
$|V_{cb}|$.
It is remarkable  that the values of $|V_{cb}|$ that 
emerged from exploiting two theoretically complementary  
approaches are  
very  close. The progress was not for free: it became possible 
only
due to essential refinements of the theoretical tools in the  last
several  years, which prompted us, in  particular, that the
zero-recoil $B\ra D^*$ formfactor $F_{D^*}$ is probably close to $0.9$, 
significantly lower than previous expectations. The decrease in 
$F_{D^*}$  
and more accurate experimental data which became 
available shortly after,
reduced the gap between the exclusive and inclusive
determinations of $|V_{cb}|$. There is a hint in 
the experimental data that some discrepancy may still persist: the 
central value of
$|V_{cb}|$ from $B\ra D^*$ decay seems to be somewhat lower than 
that from 
$\Gamma_{\rm sl}(B)$. Both theoretical values, however,
depend to a certain extent on the precise magnitude of $\mu_\pi^2$,
as
is seen from Eqs.~(\ref{w20}) and (\ref{z11}). In the exclusive and 
inclusive formulae the dependence is rather 
similar in 
magnitude but opposite in sign. It is tempting to think that the actual
value of $\mu_\pi^2$ is somewhat larger than the ``canonical" 0.5 
GeV$^2$. Increasing it 
by about $0.2\GeV^2$  makes  the two results much closer. 
We hasten to add, though, that the existing experimental error bars 
are such that any speculations on $\mu_\pi^2$ are premature. 
Moreover,   theoretical uncertainties in the
exclusive formfactor also preclude us from  the above adjustment of 
$\mu_\pi^2$. 
($F_{D^*}$ can
well be, say, $0.87$ even at the canonical value $\mu_\pi^2=0.5 
\GeV^2$). 
Future
accurate measurements will, hopefully, allow one to directly 
measure -- 
through  comparison with $\Gamma_{\rm sl}(B)$ -- the exclusive
formfactor with accuracy better than that achieved by today's 
theory. 
Thus, we  will get  a new source of 
information
on intricacies of the strong  dynamics in a so far rather
poorly known regime.

A large number of applications of the heavy quark theory
are based on duality. Although this notion becomes exact
at asymptotically high energies, at finite energies (momentum 
transfers) certain deviations must be present.
How fast duality sets in and how large are these deviations are 
important questions.
These and similar questions are among most difficult,
with virtually no or very little progress. 
Determinations of $|V_{cb}|$ we discussed rely  on the assumption 
that
 approximate duality between the  actual hadronic amplitudes and 
the
quark-gluon ones sets in already at the excitation energies $\sim 
0.7$ to $1 \GeV$. While there are no 
experimental indications so far that this is not the case (at least, in 
the
semileptonic physics), the proof is not 
 known either. If that is not true, and duality starts only
 above $1\GeV$, most probably one would have to abandon the idea 
of 
accurate determination of $|V_{cb}|$ from the exclusive $B\ra D^{(*)}$
transitions. The only  option still open will  be the inclusive
semileptonic decays where the  energy release is large, 
$\sim 3.5\GeV$. Of course, in
such a pessimistic scenario (which, we believe, is unlikely) the 
theoretical precision in $|V_{cb}|$ will 
hardly  exceed $5\%$. 

What lies ahead? There are problems (e.g. duality violations) which 
we
simply do not know how to attack. Solution of other problems
seem to be possible in the near future. 
A practically 
important problem of this type is perturbation theory in the 
context 
of Wilson's approach, where the soft parts of all diagrams have to 
be removed from the $\alpha_s$ series. This will lead us to more
accurate estimates obtained in the heavy quark expansion.
Another topical problem is constructing a reference model of the 
semileptonic decays where the transition amplitudes will be saturated 
by a minimal set of resonances and satisfy all constraints following
from the heavy quark theory. This list can be continued. We are looking
forward to new exciting developments in the near future.
\vspace*{.4cm}\\
{\bf ACKNOWLEDGMENTS:} \hspace{.4em} The authors are grateful to
R.~Dikeman, 
V.~Braun, A.~Vainshtein and M.~Voloshin for useful discussions.
We thank A.~Leibovich, Z.~Ligeti, I.~Stewart and M.~Wise for drawing our 
attention to the inconsistency in some of the spin-nonsinglet sum
rules in Sect.~4 which are now corrected.
One of the authors (M.S.) thanks M.~Neubert for stimulating discussions
of the virial theorem in QCD. We are grateful to A.~Czarnecki for 
communicating to us recent results prior to their publication. M.S. and 
N.U. thank the CERN Theory Division 
where work on certain parts of this review started, for kind
hospitality. 
We are grateful to B.~Chibisov for assistance with figures. 
This work was supported in part by DOE under the grant number
DE-FG02-94ER40823 and by NSF under the grant
number PHY 92-13313.

\end{document}